\documentclass[aps,prd,reprint,amsmath,amssymb,superscriptaddress,floatfix]{revtex4}
\usepackage{graphicx}
\usepackage{amsfonts}
\usepackage{amssymb}
\usepackage{rotating}
\usepackage{booktabs}
\usepackage{xcolor}
\usepackage{soul}
\usepackage{color}
\usepackage{slashed}
\usepackage{multirow}
\usepackage{makecell}
\usepackage{epsf}
\usepackage{ulem}
\usepackage{cancel}
\usepackage{color,bm}
\usepackage[colorinlistoftodos]{todonotes}

\usepackage[colorlinks=true,citecolor=cyan,urlcolor=blue,bookmarks=true,bookmarks=true,bookmarksopen=true,bookmarksnumbered=true,bookmarksopenlevel=3]{hyperref}

\definecolor{airforceblue}{rgb}{0.36, 0.54, 0.66}
\definecolor{steelblue}{rgb}{0.27, 0.51, 0.71}
\definecolor{amber}{rgb}{1.0, 0.49, 0.0}

\def\comment#1{}

\begin{document}

\title{Photoproduction of $\rm J/\psi$ with forward hadron tagging in hadronic collisions}
\date{\today}
\author{\textsc{Tichouk}}
\author{\textsc{Hao Sun}}
\email{ haosun@mail.ustc.edu.cn\ \ haosun@dlut.edu.cn }
\author{\textsc{Xuan Luo}}
\affiliation{Institute of Theoretical Physics, School of Physics, Dalian University of Technology, \\ No.2 Linggong Road, Dalian, Liaoning, 116024, P.R.China }

\begin{abstract}
We study the photoproduction of $\rm J/\psi$ using the NRQCD formalism with forward hadron tagging at the Large Hadron Collider. We estimate the total cross sections and event rates with and without nuclear shadowing effects in high energy proton-proton, proton-nucleus and nucleus-nucleus collisions. Our results show that the processes which involve $\rm J/\psi$ photoproduction depend on the choice of forward detector acceptances $\xi$. Under some precise cut of $\rm p_{T}^{J/\psi}$ and $\rm z^{J/\psi}$ kinematic variables, we find that the distributions of photoproduction of $\rm J/\psi$ are led by the Fock state $\rm ^{1}S_{0}^{[8]}$ or $\rm ^{3}S_{1}^{[1]}$. The total cross sections and event rates will be smaller if the nuclear shadowing effects are considered. The processes give a crucial photoproduction signature at the LHC with forward detector acceptances. The exploration and the detection of the interactions will be useful for studying the mechanism of heavy quarkonium production.
\end{abstract}
\maketitle
\setcounter{footnote}{0}

\section{Introduction}
\label{Introduction}

Since the discovery of heavy quarkonia in the mid-1970s, their production and decay become reliable tools to improve our understanding of theoretical aspects of quantum chromodynamics (QCD) from the hard region, where the strong interaction is realized with a large momentum transfer, to the soft region, where the strong interaction is carried out with a small momentum transfer. These production and decay of heavy quarkonia are studied under the non-relativistic quantum chromodynamics (NRQCD) due to the large mass of heavy quarks ($\rm m_{c}\sim1.5\ GeV$, $\rm m_{b}\sim4.75\ GeV$). The full description of the theoretical framework was proposed by Bodwin, Braaten and Lepage(BBL) \cite{Bodwin1995}. It factorizes the quarkonium production in terms of short distance QCD cross sections and long distance matrix elements (LDMEs). The former can be pertubatively evaluated in series of the running coupling constant $\rm \alpha_{s}$ while the latter is the probability for a heavy $\rm Q\bar{Q}$ pair  with spin $\rm S$,
orbital angular momentum $\rm L$, total angular momentum $\rm J$, and color multiplicity $\rm a$ = 1 (color-singlet), 8 (color-octet) to evolve into a physical heavy quarkonium state. The value of LDMEs can either be calculated by utilizing non-perturbative methods or extracted phenomenologically from data \cite{Schuler1997}, i.e., QCD lattice simulations or measurements in some production processes. The $\rm Q\bar Q$ pair is generated from the partonic interaction at short distances in color singlet (CS) or in color-octet (CO) states and hadronizes into physical CS observable by emitting soft gluons non-perturbatively. The effect of LDME contribution has a hierarchically ordered scaling with $\rm v$ \cite{Gerhard1997}, $\rm v$ being the nonrelativistic velocity of $\rm Q$ or $\rm\bar{Q}$ in the $\rm Q\bar{Q}$ rest frame. The essence of this theory can now be systematically shortened by the double expansion in powers of $\rm \alpha_{s}$ and $\rm v$.
A lot of phenomenological approaches of this framework are meticulously narrated in Refs \cite{Kniehl2006, Chao2003, SunZhan2015, Nayak2016}, taking into account the complete and spanned structure of the $\rm Q\bar{Q}$ Fock space by the state $\rm n=\,^{2S+1}L_{J}^{(a)}$.

However, the golden age of the NRQCD theory has been to address the limitations of the early proposed models
from which it establishes its coherently solid postulations. The most notable mechanisms for quarkonium production are the color-singlet model (CSM), color octet mechanism (COM), color-evaporation model (CEM) and fragmentation-function approach (FFA). In the CSM \cite{Barbieri1981, AbePrl1997, Brambilla2011, He2010, SunZhan2017, Carlson1976, Nayak2017, Einhorn1975, Berger1981, HeWang2010, Baier1981, Baier1982, Bodwin2008, Ellis1976, Chao1980}, the spin and the color quantum numbers of $\rm Q\bar{Q}$ pair don't change during the hadronization due to the absence of gluon emission, it is restricted by sorting out the infrared divergences in P-wave and D-wave decay widths of heavy quarkonium, and it is also incapable to handle the high transverse momentum spectrum of $\rm J/\psi$ in experiments. However, the spin and the color quantum numbers of $\rm Q\bar{Q}$ pair do change during the hadronization with the emission of gluons \cite{Klasen2002} in COM. The CEM depends more on statistical factors and is unable to explain the ratio of the $\rm P_{T}$ distribution of $\rm J/\psi$ and $\rm \chi_{CJ}$, both in photo- and hadroproduction from experiments \cite{Bodwin2005, Ben2017, Ma2016, Kang2005}. In the FFA \cite{Kniehl1997, Ma2014}, the cross section can be written in terms of convolution of the partonic production cross section with light-cone fragmentation functions at large transverse momentum.

More recently, the accessibility of experimental data including the feed-down and non-prompt contributions has provided a very good description of $\rm J/\psi$ photoproduction in a wide energy range \cite{Yuan2000, Cacciari1996, Klasen2002} in investigating color-singlet and color-octet contributions. Though there are many studies of $\rm J/\psi$ production within NRQCD factorization approach have been carried out in,
for example, $\rm \gamma N$ \cite{Amundson1997}, $\rm \gamma p$ \cite{Ko1996}, $\rm e^+e^-$ \cite{Chen2013}, forward rapidity $\rm pp$ \cite{Costa2017}, $\rm p\bar{p}$ and $\rm ep$ \cite{Krmer2001} collisions, the theoretical tools for the explanation of decay and production quarkonium at existing colliders are still under intense debate. For that reason, we should predict paths to examine more heavy quarkonium production and decay for the future colliders.

A plan \cite{Albrow2009} which is suggested by the FP420 R\&D Collaboration in 2009 allows to study forward physics at the LHC such as the study pomeron-induced processes, low-$\rm x$ QCD physics, hadronic models of ultra high energy cosmic rays and photon-photon interactions, etc \cite{Staszewski2011}. To reach this new realm of interest, detectors in the LHC tunnel need to be readjusted so as to precisely measure very forward hadrons. The setting-up of forward detectors at LHC may permit to predict the single diffraction and double pomeron exchange cross sections \cite{Sun2017} which are important for future measurements at the LHC, may endow the likelihood to open a new window to investigate new physics even up to TeV scale \cite{Sun2014, Inan2014, Kepka2008, Chapon2010} with extremely clean environment, and may enable to tag  forward hadron or nucleus in photon-hadron  and photon-nucleus collisions at high rapidity regions. It may also offer knowledge about unexplored phase space areas. The acceptance of the forward hadron detector system is defined as the region of the interaction and dependent on the mass of the centrally produced system. The interval of collision is classified as follows: $0.1<\xi_{1}<0.5$, $0.0015<\xi_{2}<0.5$ for CMS-TOTEM forward detector, and $0.015<\xi_{3}<0.15$ for AFP-ATLAS forward detector \cite{Albrow2009}. The probe of coherent interactions in existing and future colliders with the installation of additional detectors at low scattering angle and with the outgoing particle remaining intact after collision is feasible \cite{Vysotsky2018}.

\begin{figure}[h]
\centering
\includegraphics[width=3.0in]{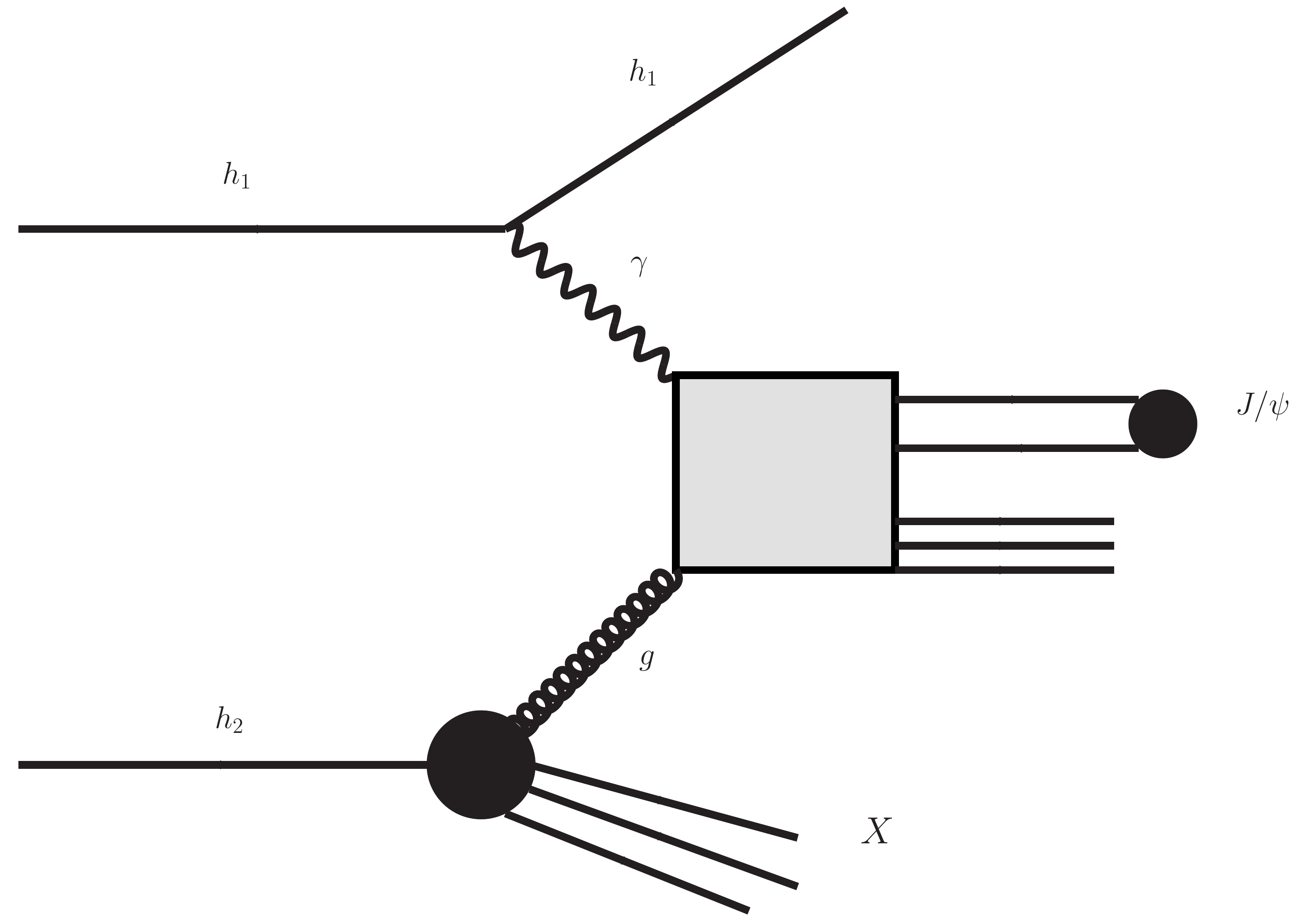}
\caption{General diagram for the mechanism of heavy quarkonium production in coherent hadron-hadron interactions at the CERN LHC:
 $\rm h_{1}h_{2}\rightarrow h_{1}\gamma h_{2}\rightarrow h_{1}\mathcal{Q}(+g)+X$.
\label{fig1}}
\end{figure}

In this present work, we estimate the predictions of the cross sections of single $\rm J/\psi$ photoproduction for the coherent processes in proton-proton, proton-nucleus, nucleus-nucleus interactions including both 2$\rightarrow$1 and 2$\rightarrow$2 subprocesses. The illustrative diagram of $\rm J/\psi$ meson photoproduction by photon induced process at LHC is shown in Fig.\ref{fig1}.
A number of phenomenological studies on photoproduction have been achieved  in \cite{Monfared2016, Sahin2012, Cho2015, Sun2014PRD, Sun2015JHEP} and references therein. The aim of this paper is to study in detail the inclusive photoproduction of single heavy quarkonium in the NRQCD factorization formalism at the LHC with forward detector acceptances, which is characterized by one rapidity gap in the final state related to the photon exchange, in order to explain the unsuccessfully known inclusive quarkonium mechanism. The study of $\rm \gamma proton$ and $\rm \gamma Pb$ as well as $\gamma\gamma$ interactions at the LHC could offer valuable information on the heavy quarkonium production at high energies and more clean backdrop compared to normal hadronic $\rm pp$, $\rm Pb Pb$ or $\rm Pbp$ interactions.

The paper is structured into three sections including the introduction in Section \ref{Introduction}. The detailed description of the hadron tagging devices and the formalism framework for the leading order (LO) cross section of $\rm J/\psi$ production at the LHC are clearly described in Section \ref{FRAMEWORK}. The input parameters and discussed numerical results are shown in Section \ref{NUMERICAL}. A summary is briefly given in Section \ref{SUMMARY}.

\section{SETUP OF THE CALCULATION PROCEDURE}
\label{FRAMEWORK}

\subsection{Forward hadron tagging}

The produced $\rm J/\psi$ meson in $\rm pp$, $\rm PbPb$ and $\rm Pbp$ collisions is measured in the central detector accompanied by gluon and $\rm X$. However, the scattered lead and proton remaining intact after collision will not have high transverse momentum and will be difficult to detect with central detector. For that reason, the central detector needs to be complemented by the installation of the forward hadron detectors which are equipped with charged particle trackers to tag the intact hadrons. The goal of these forward tagging hadron detectors is to catch the undissociated lead and proton after the collision events. They are symmetrically positioned in the region 220 m (420 m) away from the ATLAS and/or CMS interaction points to be 2 mm (5 mm) from the beam \cite{Cox2007, Royon2013Forwarda, Royon2013Forwardb, Albrow2009}. These regions cover the range where proton and lead are either both detected at 420 m (symmetric tagging) or one is detected at 220 m and one at 420 m (asymmetric tagging). The forward tagging hadron detectors are characterized by their acceptance, resolution  and  ability to measure the time-of-flight from interaction point. The resolution and the acceptance of the hadron tagging detector are fixed by the LHC high-lunimosity beam optics. The hadron tagging detectors use the FPtract program \cite{Chaichian2009, Royon2014, Royon2010} to track the path of the hadrons through the LHC lattice so as to completely find out the acceptance. The acceptance of the forward detectors is controlled by the distance of the active edge of the detector from the beam, which fixes the smallest measurable energy loss of the outgoing hadrons, and the space of the LHC beam elements between the interaction point and the forward detector. The hadrons that lose too much momentum will be absorbed by beam elements, enforcing an upper limit on the measurable momentum loss of hadrons. The hadron tagging technique immediately measures the hadrons surviving the coherent emission.

The suitable forward kinematic variables in hadronic and nuclear reactions are the transverse momentum $\rm p_{T}$, the longitudinal momentum $\rm p_{L}$, the rapidity $\rm y$ and the polar angle $\theta$. The detector can directly measure the four-momentum of a given final-state produced in the interactions. These momenta and rapidity depend on polar angle with respect to beam axis. The rapidity can be thought of as the relativistically invariant measure of the longitudinal velocity. Pseudorapidity and rapidity are taken equal for negligible masses or highly relativistic particles. Forward rapidity region is the interval where the forward tagging hadron can be detected and is defined beyond $\rm \left\vert\rm y\right\vert \approx 3$. The produced system can be detected by the central detector at the midrapidity $\rm y \approx 0$. The rapidity gap is large region of pseudorapidity completely devoid of particles, where an intact particle moves at a polar angle $\theta$ with respect to the beam \cite{Staszewski2015, Khachatryan2015} and is tagged by forward detector. Moreover, rapidity gaps can also emerge due to fluctuations of the distance between neighbouring particles. A gap in rapidity is present between hadrons and central produced system and one or both interacting hadron stay intact due to the coherent and colorless emission of photon over the hadron. One expects a one-rapidity pattern in the interactions due to the emission of a virtual photon by one of the hadron emerging as undissociated outgoing hadrons at very low angles (few $\mu$rad) \cite{Nurse2016}. The pseudorapidity difference $\rm \Delta y$ in an event between the intact hadron and the produced system is given as $\rm \Delta y \simeq-log \xi$. $\rm \xi$ is the fractional longitudinal momentum loss of the outgoing intact hadron \cite{Nurse2016}.

The advantages of forward detectors added to the central detector are to allow the whole reconstruction of the $\rm pp$, $\rm PbPb$ and $\rm Pbp$ collision events by covering a large range of rapidity, to remove the ambiguity of a rapidity gap tag which suffers from background due to multiplicity events, to veto events (the final state particles in the collision) in which there are no forward rapidity gaps \cite{Cox2005, Enterria2008, Akiba2016}.

\subsection{Cross Section Formulas}

We refer to the heavy quarkonium $\rm J/\psi$ as $\rm \mathcal{Q}$ whereas $\rm h_{1}h_{2}$ is symbolized by $\rm pp, \rm PbPb$ and $\rm Pbp$ or $\rm pPb$ . The cross sections for the $\rm h_{1}h_{2}\to h_{1}\gamma h_{2}\to h_{1}\mathcal{Q}(+g)+X$ process can be expressed as:
\begin{equation}
\rm \sigma (h_{1}h_{2}\to h_{1}\gamma h_{2}\to h_{1}\mathcal{Q}(+g)+X)=
\int d\xi dx \sum_{n} \widehat{\sigma}(\gamma g\to Q\overline{Q}[n](+g)+X)\rm \langle 0|\mathcal{O}_{1,8}^{J/\psi }[n]|0\rangle[f_{\gamma /h_{1}}(\xi)
G_{g/h_{2}}(x,\mu _{f}) + h_{1}\leftrightarrow h_{2}],
\end{equation}
where $\rm \langle 0|\mathcal{O}_{1,8}^{J/\psi }[n]|0\rangle$ is the long-distance matrix element which describes the hadronization of the $\rm Q\overline{Q}$ heavy pair into the physical observable quarkonium state $\rm J/\psi$.
The $\rm \widehat{\sigma }(\gamma g\rightarrow Q\overline{Q}[n](+g))$ denotes the short-distance cross section
for the partonic process $\rm \gamma g\rightarrow Q\overline{Q}[n](+g)$, which is found by operating the covariant projection method. The Fock state $\rm n$ is given as follows: $\rm ^{1}S_{0}^{[8]},^{3}P_{0}^{[8]},^{3}P_{2}^{[8]}$ for $\rm \gamma g\rightarrow Q\overline{Q}[n]$ partonic 2$\to$1 subprocess, and $\rm ^{3}S_{1}^{[1]},^{1}S_{0}^{[8]},^{3}S_{1}^{[8]},^{3}P_{0}^{[8]},^{3}P_{1}^{[8]},^{3}P_{2}^{[8]}$ for $\rm \gamma g\to Q\overline{Q}[n]+g$ partonic 2$\rightarrow$2 subprocess. The $\rm G_{g/h_{2}}(x,\mu_{f})$ stands for the gluon parton density function while $\rm x$, Bjorken variable, is the momentum fraction of the hadron (proton and lead) momentum carried by the gluon.
The nuclear gluon distribution is given by $\rm x\rm G_{g/Pb}(x,\mu_{f})$=$\rm A\cdot R_{g}\cdot x\rm G_{g/p}(x,\mu_{f})$, where $\rm R_{g}$ is the nuclear modification factor for gluon taking into account the nuclear shadowing effects \cite{Armesto2006} as given by the EPS09 parametrization \cite{Eskola2009}, which is based on a global fit of the current nuclear data. $\rm R_{g}=1$ disregards the shadowing corrections. $\rm x G_{g/p}(x,\mu_{f})$ is the proton-gluon distribution.

A considerable fraction of $\rm pp$, $\rm PbPb$ and $\rm Pbp$  collisions at the LHC will include quasi-real (low-$\rm Q^2$) photon interactions, which is called the photon induced processes. The photon induced interactions have three  processes at the LHC with forward detector: (1) photon-photon process, (2) photon-proton process, (3) photon-lead process. In these processes, the emitted photon from the electromagnetic field of one of the two colliding hadrons can interact with one photon of the other proton or lead (photon-photon process) or directly with the other hadrons (photon-proton process or photon-lead process). This photon  from the hadron will have energy $\rm E_{\gamma}$.
Consequently, the hadron which emits the photon should have some momentum fraction loss $\xi$, which is defined as the forward detector acceptance. The ratio between scattered low-$\rm Q^{2}$ photons $\rm E_{\gamma}$ and incoming energy $\rm E$ is symbolized by $\rm \xi =E_{\gamma}/E$. The fundamental concept is that the electromagnetic interaction is supposed to dominate at large impact ($\rm b>R_{A}+R_{B}$, where $\rm R_{A}$ and $\rm R_{B}$ are the hadron radii) and ultrarelativistic energies. In the heavy ion collisions, the heavy nuclei generate strong electromagnetic fields as a result of the coherent action of all protons in the nucleus, which can interact with each other. The analytic estimate for the equivalent photon flux of a hadron, $\rm Pb$ or $\rm p$, can be evaluated considering the requirement that photoproduction is not followed by hadronic interaction (ultra-peripheral collisions), which is given by \cite{Vysotsky2018}
\begin{eqnarray}
\mathrm{f_{\gamma /h}(\xi)=\frac{Z^{2}\alpha _{em}}{\pi \xi}[(4a+1)\ln (1+}\frac{%
1}{a}\mathrm{)-}\frac{24a^{2}+42a+17}{6(a+1)^{2}}],
\end{eqnarray}
where $\rm h$ can be either $\rm p$ ($\rm Z=1$) or $\rm Pb$ ($\rm Z=82$), $\rm a=(\omega /\Lambda _{h}\gamma )^{2}$,
$\rm \omega =\frac{\sqrt{s}}{2}\xi$ with $\rm \Lambda_{p} \approx 0.20\times e^{\frac{17}{12}}$ GeV and $\rm \Lambda_{Pb}=80$ MeV \cite{Jentschura2009}.
\begin{figure}[htp]
\centering
   \includegraphics[height=1.3cm,width=2.5cm]{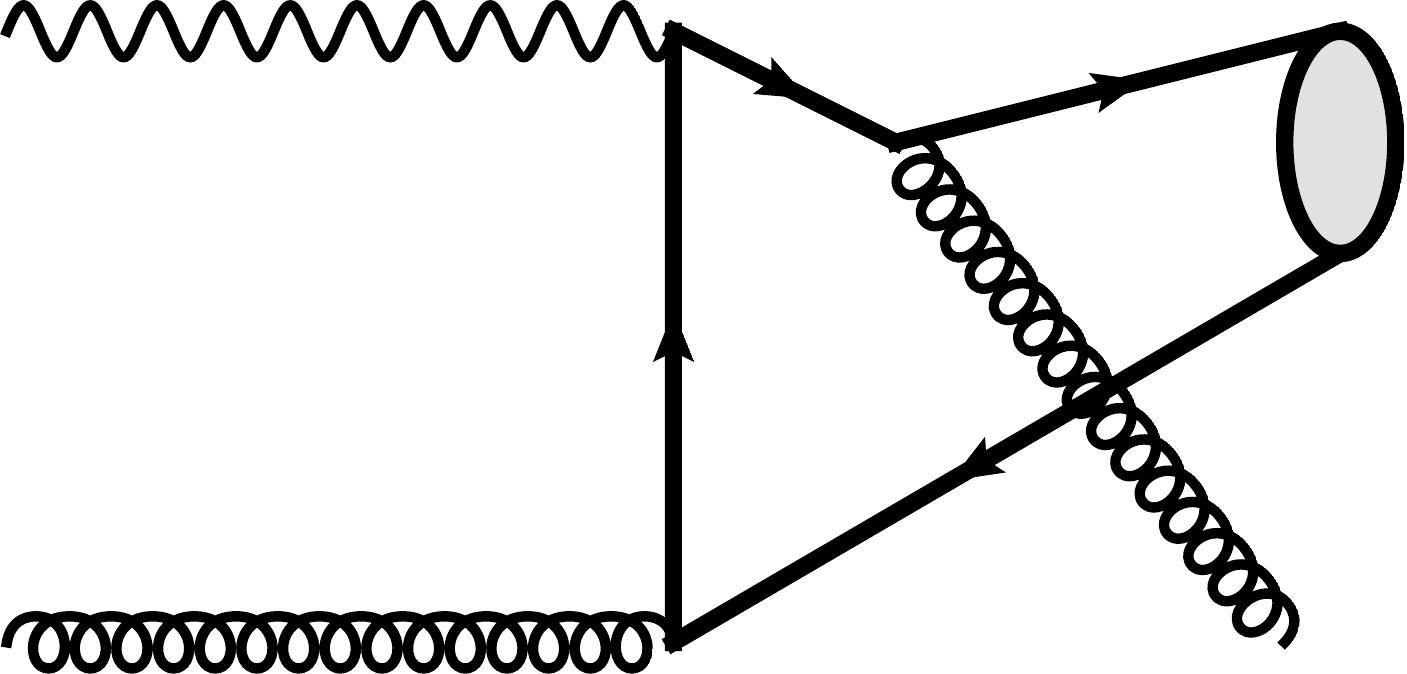}
   \includegraphics[height=1.3cm,width=2.5cm]{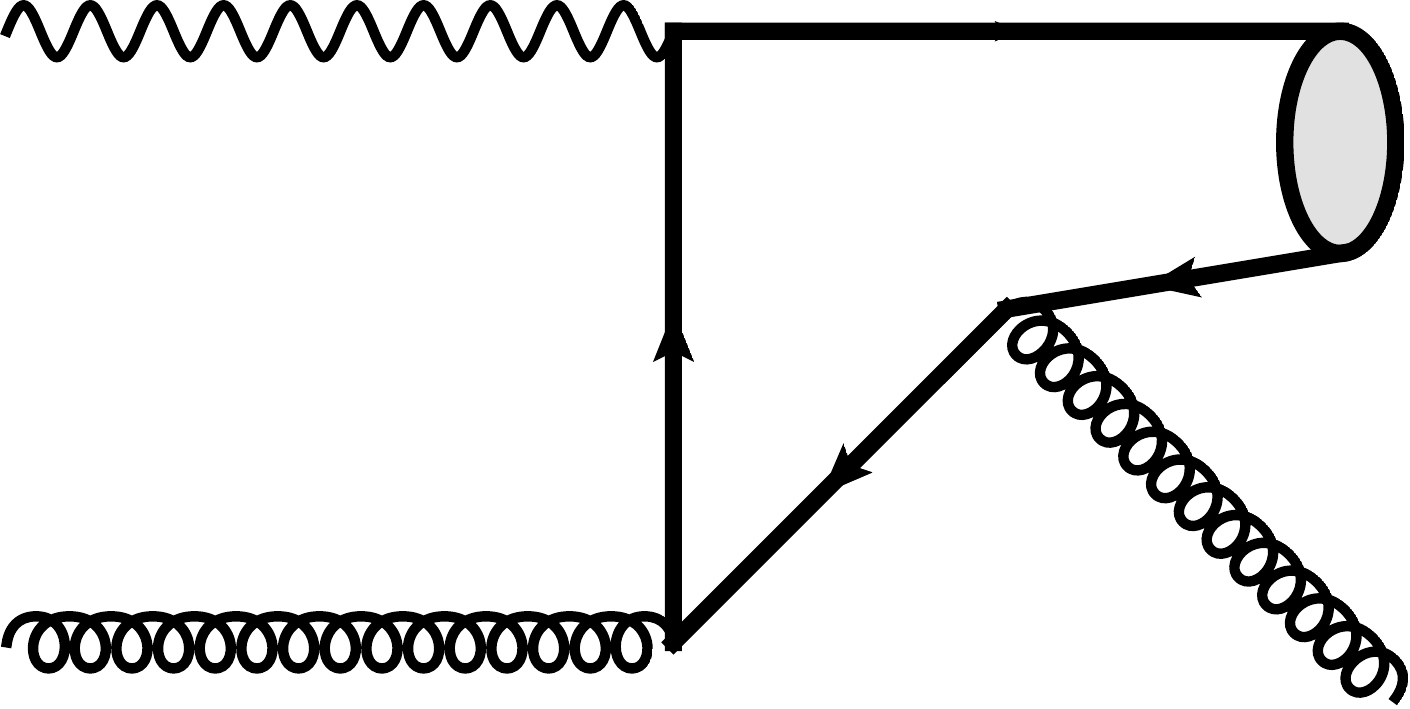}
   \includegraphics[height=1.3cm,width=2.5cm]{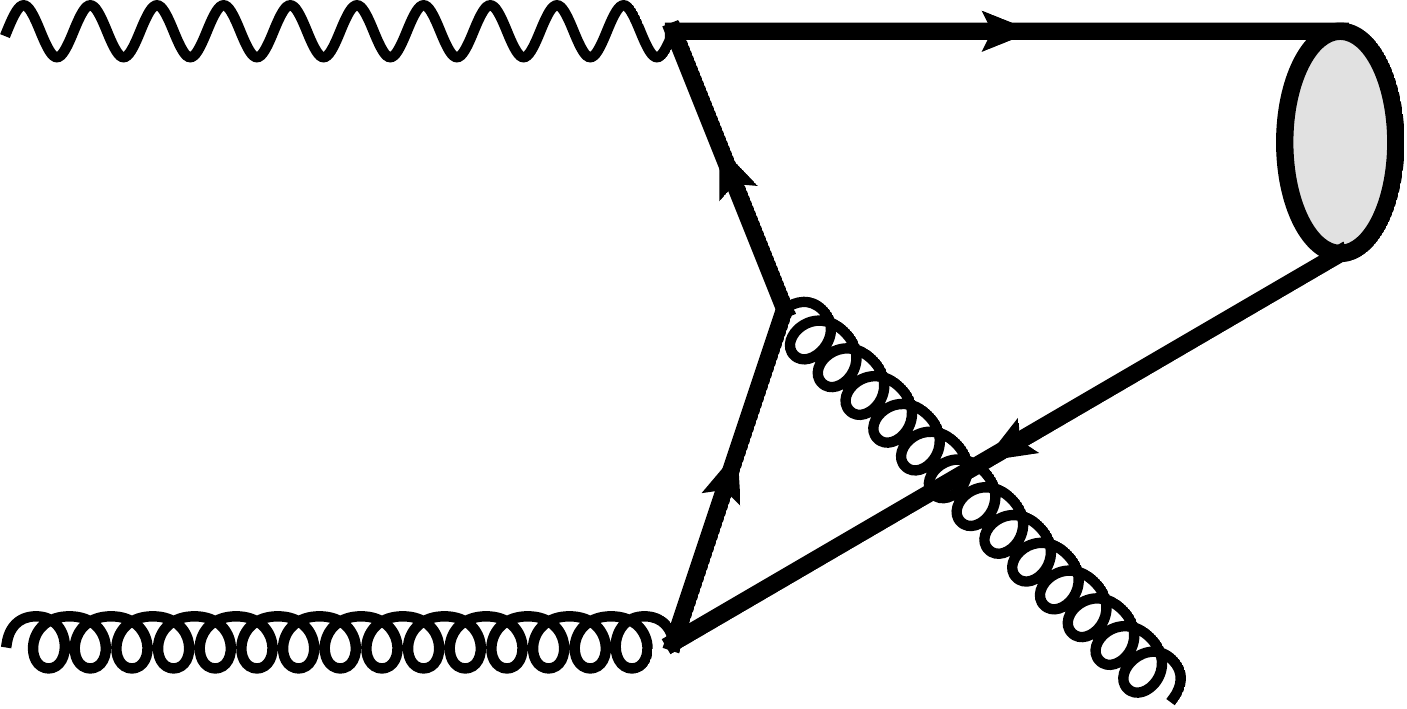}
   \includegraphics[height=1.3cm,width=2.5cm]{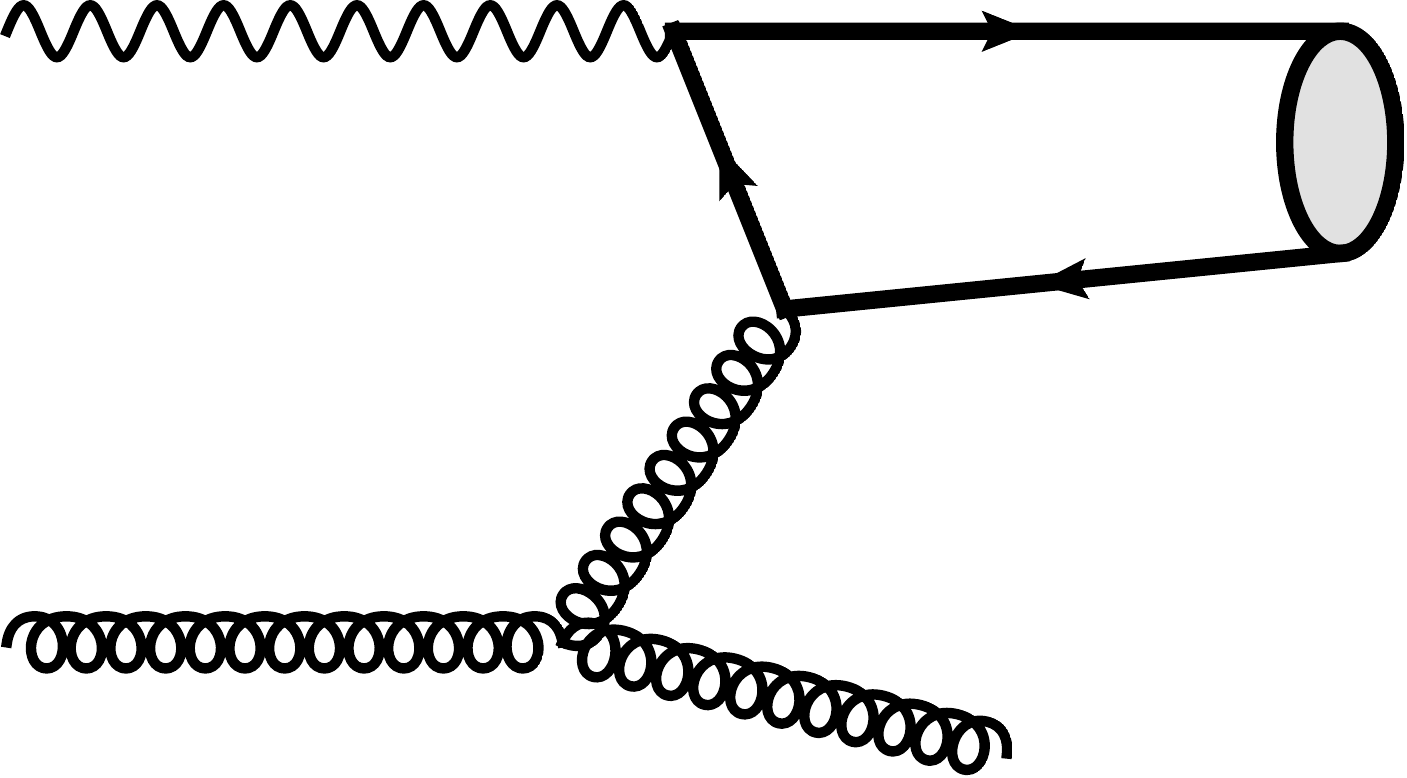}
   \includegraphics[height=1.3cm,width=2.5cm]{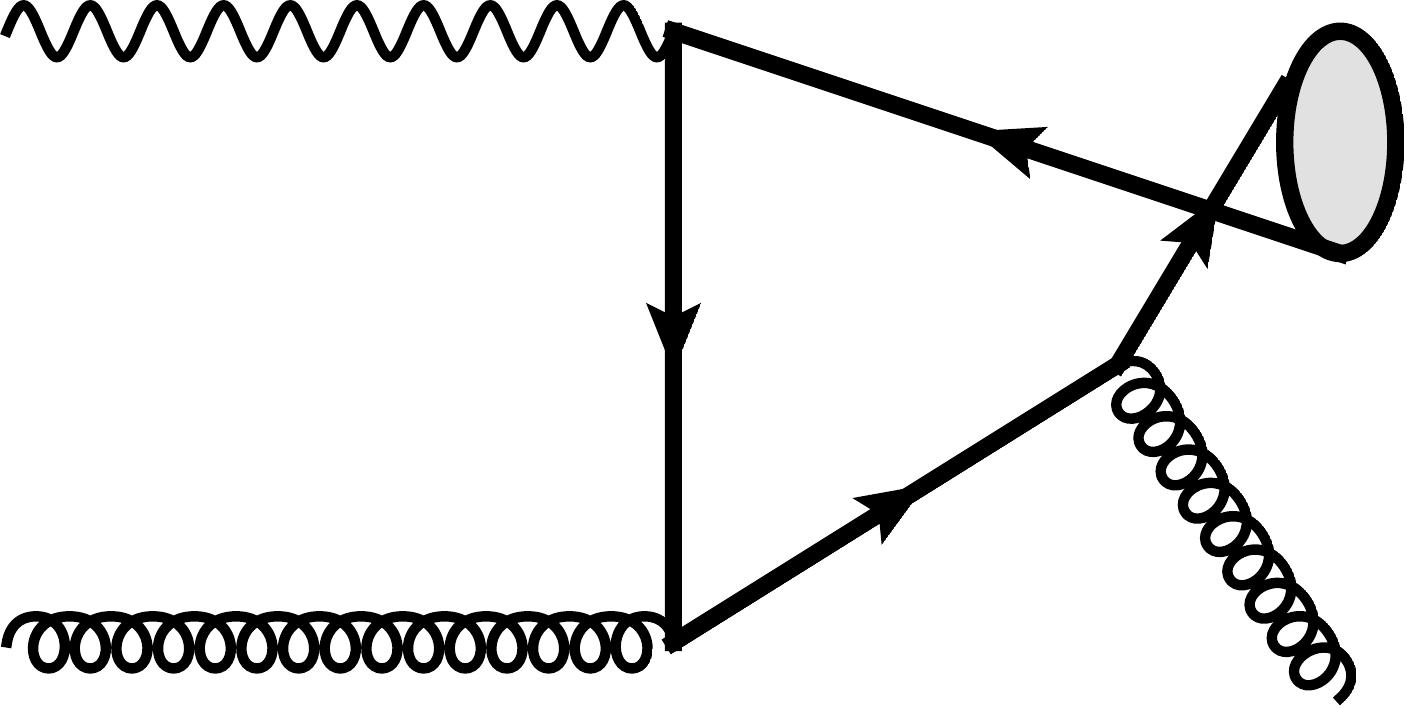}\\
   \includegraphics[height=1.3cm,width=2.5cm]{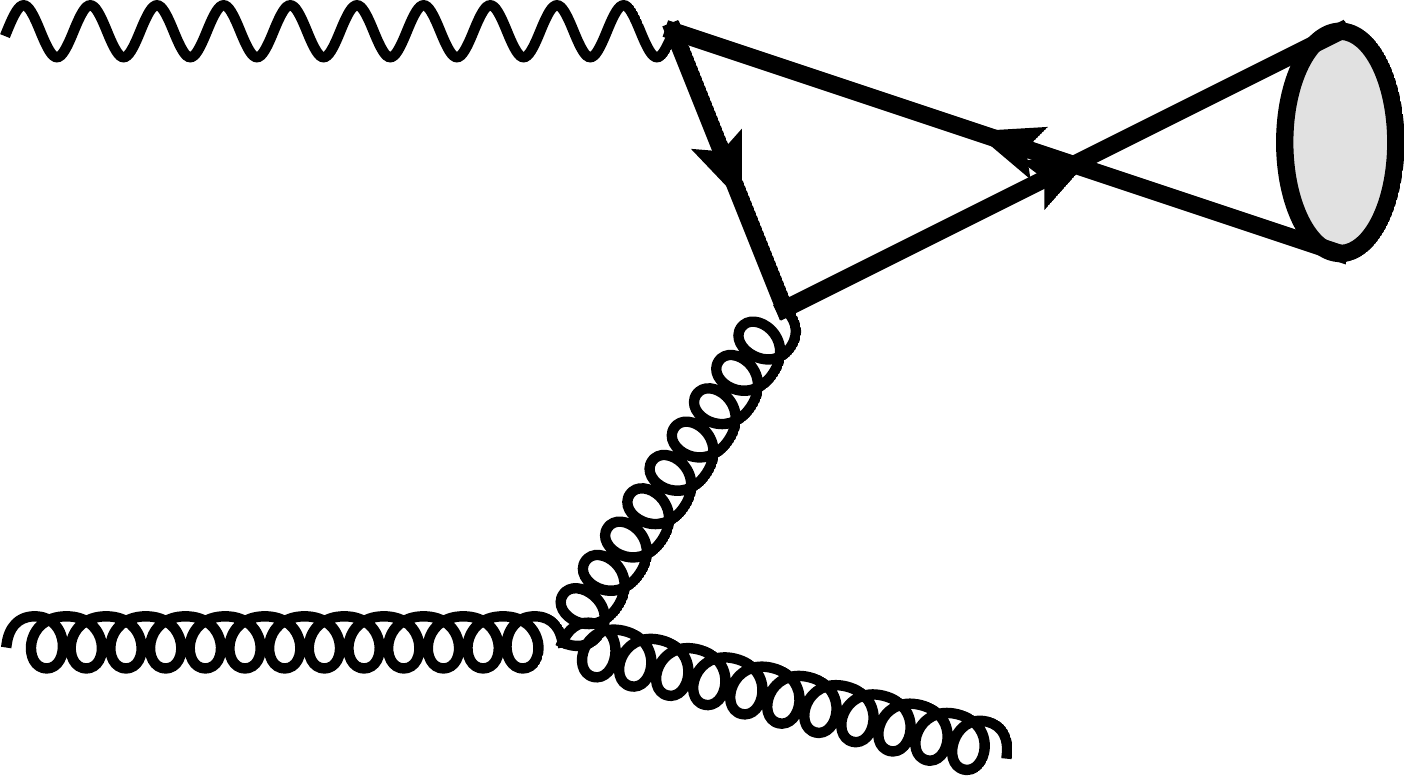}
   \includegraphics[height=1.3cm,width=2.5cm]{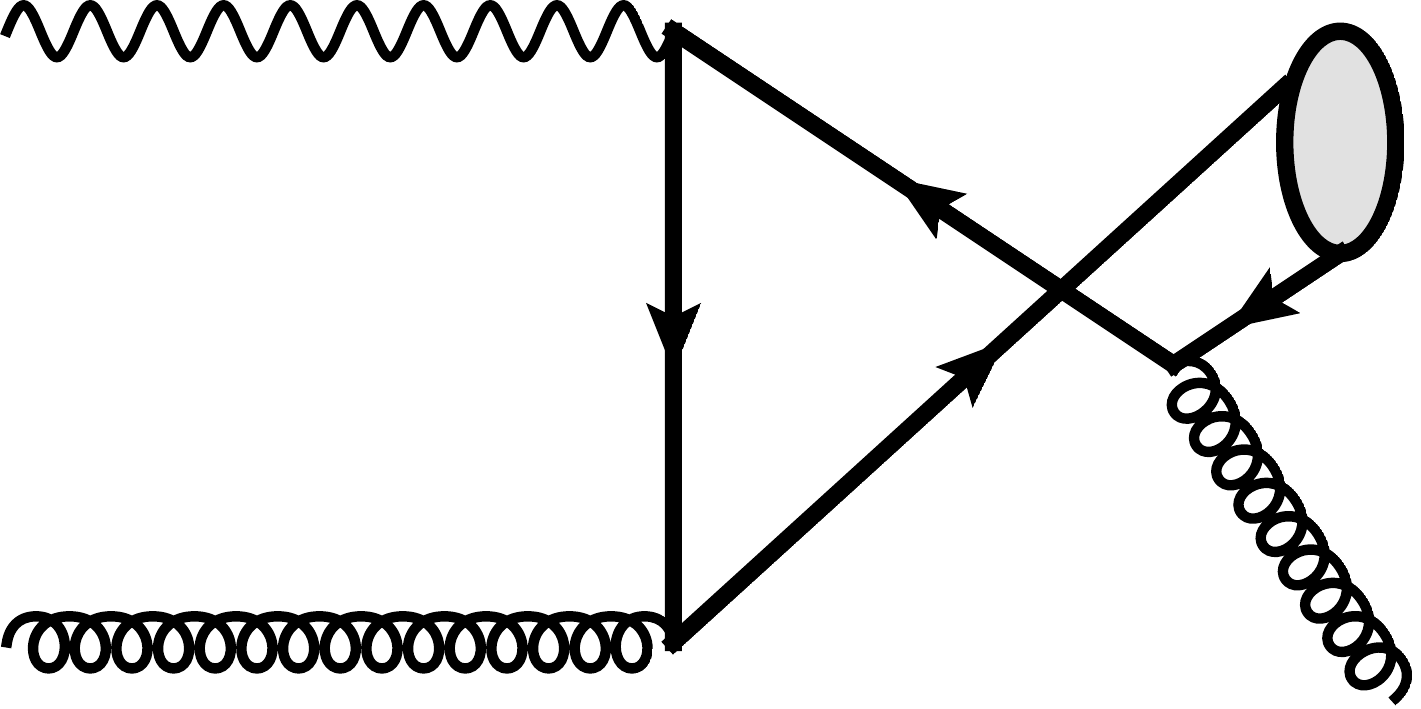}
   \includegraphics[height=1.3cm,width=2.5cm]{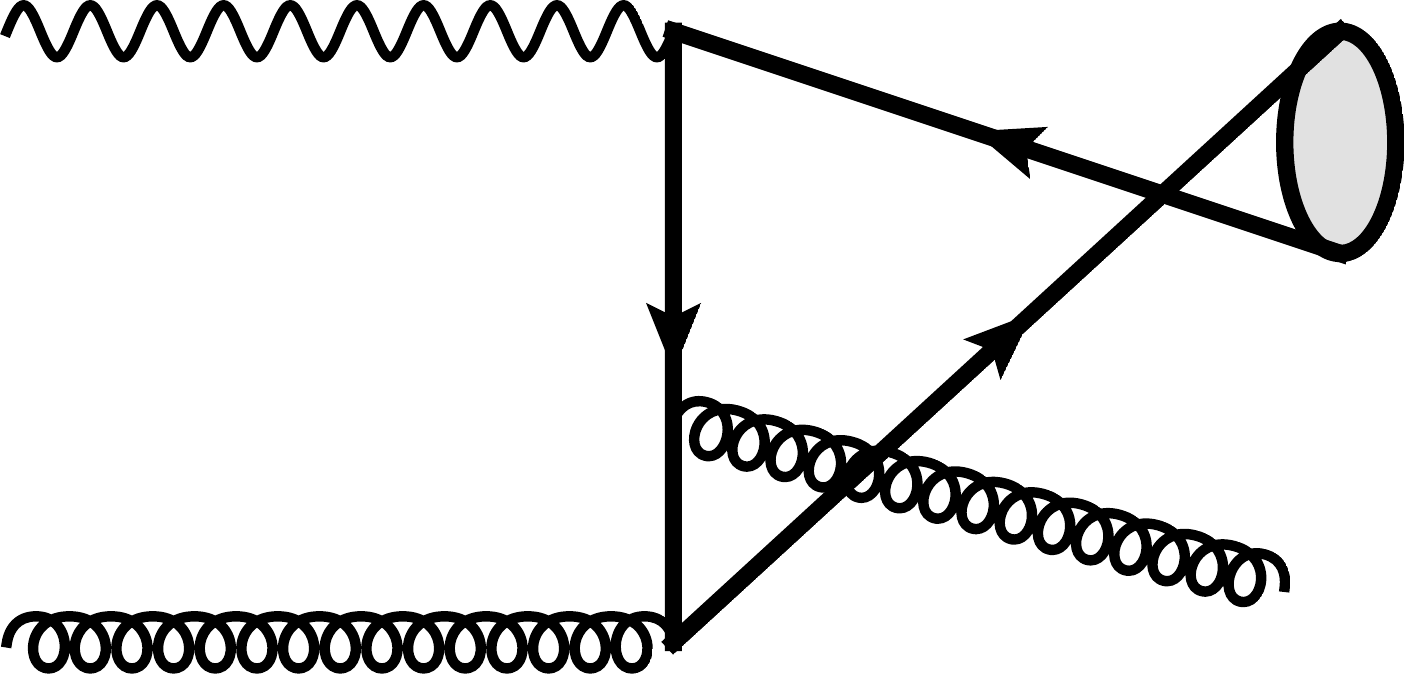}
   \includegraphics[height=1.3cm,width=2.5cm]{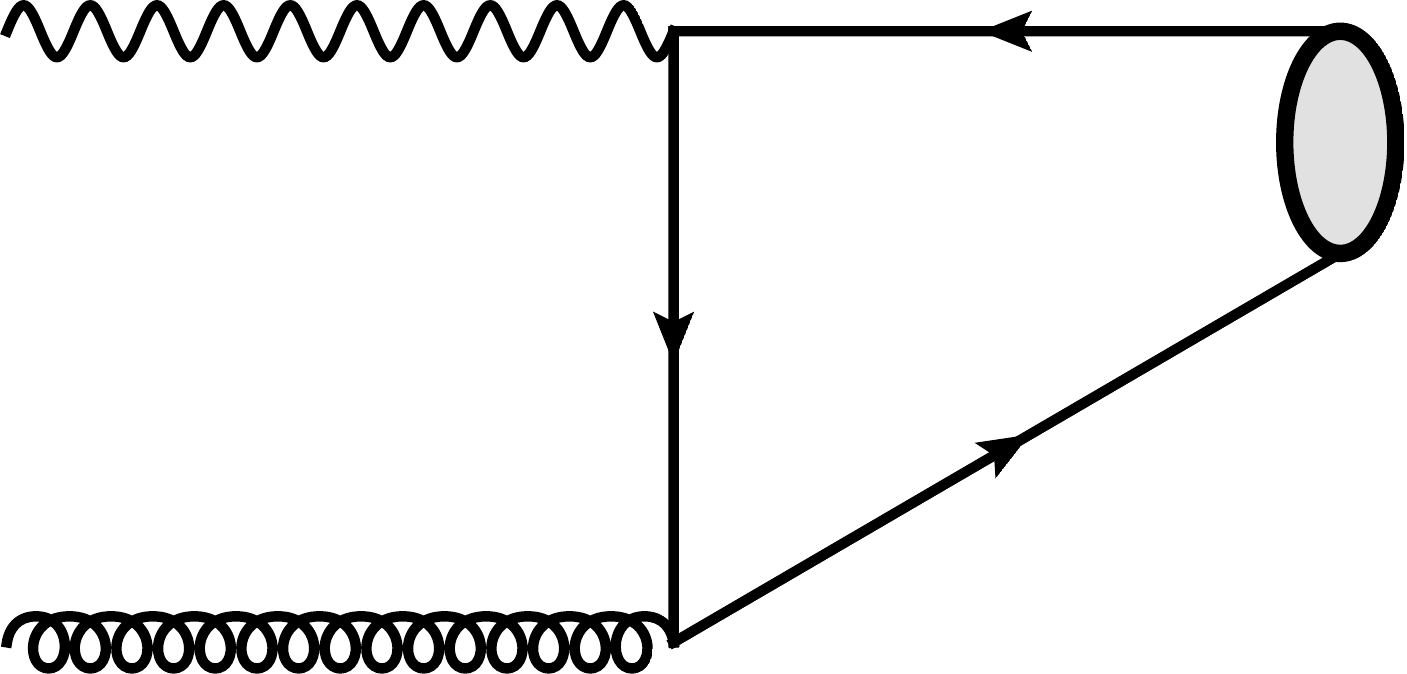}
   \includegraphics[height=1.3cm,width=2.5cm]{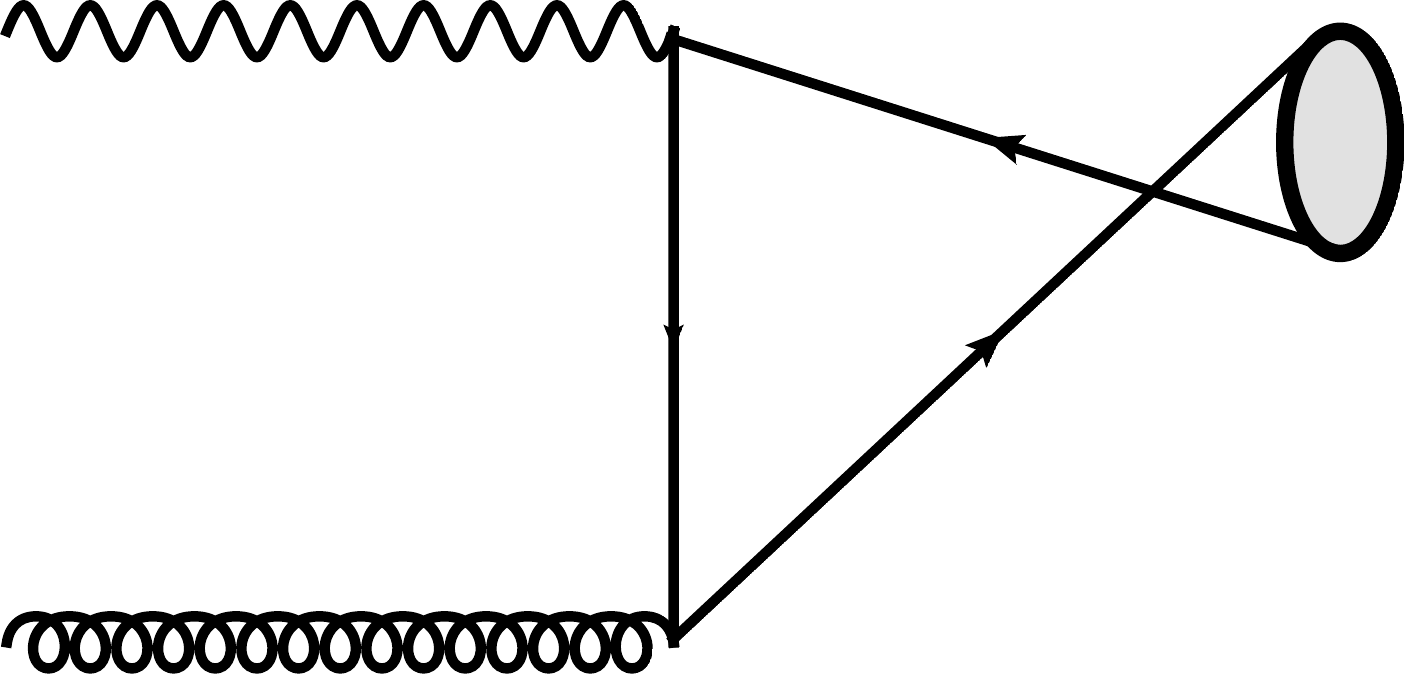}
      \caption{Feynman diagrams for partonic processes $\rm \gamma+g\rightarrow Q\overline{Q}[n]+g$ (a) and $\rm \gamma+g \rightarrow Q\overline{Q}[n]$ (b).
\label{fig2}}
\end{figure}
In considering the partonic processes $\rm \gamma(p_{1})+g(p_{2})\rightarrow Q\overline{Q}[n](p_{3})(+g(p_{4}))$, there are eight diagrams for $\rm \gamma g \to Q\overline{Q}[n]+g$ and the last two ones are for $\rm \gamma g \to Q\overline{Q}[n]$. The Feynman diagrams for both processes are shown in Fig.\ref{fig2}.

In $\rm pp$, $\rm PbPb$ and $\rm pPb$ interactions, the total cross section for 2$\to$2 and 2$\to$1 subprocesses can be factorized in terms of the equivalent flux of photons into the hadron ($\rm p$ or $\rm Pb$) projectile and the photon-target production cross section:
\paragraph{$\rm h_{1}h_{2}=pp\ or\ PbPb$.}
For the  $\rm h_{1}h_{2}\rightarrow h_{1}\gamma h_{2}\rightarrow h_{1}\mathcal{Q}(+g)+X$ process where the incident particles are similar, the photon is emitted by one of the incoming particle $\rm p (or\ Pb)$ and the gluon by the other incoming particle $\rm p (or\ Pb)$. The final total cross section for $\rm h_{1}h_{2}\rightarrow h_{1}\gamma h_{2}\rightarrow h_{1}\mathcal{Q}+g+X$ process can be now written as:
\begin{eqnarray}\label{2to2a} \nonumber
\rm \sigma (h_{1}h_{2}\rightarrow h_{1}\gamma h_{2}\rightarrow h_{1}\mathcal{Q}g+X) &=& \rm
\int
\frac{1}{16\pi^2 x\xi s^2} \frac{p_{T}^{J/\psi }}{\xi-\frac{m_{T}^{J/\psi}}{\sqrt{s}}e^{-y^{J/\psi }}}
\frac{1}{N_{col}N_{pol}} \overline{\sum }\left\vert\mathcal{ A}_{S,L}\right\vert ^{2} \\
&&\rm
[f_{\gamma/h_{1}}(\xi) G_{g/h_{2}}(x,\mu_{f}) + (h_{1}\leftrightarrow h_{2})] dp_{T}^{J/\psi } d\phi^{J/\psi} dy^{J/\psi } d\xi,
\end{eqnarray}
with x fixed by $\rm x=\frac{1}{\sqrt{s}}(m_{T}^{J/\psi}e^{y^{J/\psi }}+m_{4T}e^{y_4})$ where $\rm y_4 = -log(\frac{\sqrt{s}\xi}{m_{T}^{J/\psi}}-e^{-y^{J/\psi }})$. The final total cross section for $\rm h_{1}h_{2}\rightarrow h_{1}\gamma h_{2}\rightarrow h_{1}\mathcal{Q}+X$ process can also be expressed as:
\begin{eqnarray}\label{2to1a}
\rm \sigma (h_{1}h_{2}\rightarrow h_{1}\gamma h_{2}\rightarrow h_{1}\mathcal{Q}+X) &=& \rm
\int
\frac{\pi p_{T}^{J/\psi }}{s^2 x\xi^2} \frac{1}{N_{col}N_{pol}} \overline{\sum }\left\vert\mathcal{ A}_{S,L}\right\vert ^{2}
[f_{\gamma/h_{1}}(\xi) G_{g/h_{2}}(x,\mu_{f}) + (h_{1}\leftrightarrow h_{2})] dp_{T}^{J/\psi } d\phi^{J/\psi} d\xi,
\end{eqnarray}
with x given by $\rm x=(m_{T}^{J/\psi})^2/(s\xi$).
\paragraph{$\rm h_{1}h_{2} = Pbp\ or\ pPb$.}
For the $\rm h_{1}h_{2}\rightarrow h_{1}\gamma h_{2}\rightarrow \mathcal{Q}(+g)h_{1}+X$ process where the incident particles are different, the final total cross section is split into two cases: the first case the photon comes from lead and the proton emits gluon (denoted as $\rm \gamma p$ collision); the second case takes place with the exchange of $\rm Pbp$ to $\rm pPb$, the photon originates from proton and lead emits gluon (denoted as $\rm \gamma Pb$ collision). The expression of cross section is arranged in the same way as defined in Eq.\ref{2to2a}(Eq.\ref{2to1a}).

In these formulas, $\xi$ is integrated in the region $ \xi_{\min} < \xi < \xi_{\max}$ and $\xi_{\min}(\xi_{\max})$ is the lower(upper) limit of forward detector acceptance. The $\rm s$ and  $\rm m_{Q}$ are respectively the square of center-of-mass energy of collider and the mass of heavy quark. The heavy quark represents the charm quark in our study. $\rm p_{T}^{J/\psi}$, $\rm y^{J/\psi }$, $\rm m_T^{J/\psi}$, $\phi ^{J/\psi}$ are the transverse momenta, rapidity, transverse mass and azimuthal angle of the charmnium  where $\rm m_{T} ^{J/\psi}=\sqrt{\left( p_{T}^{ {J/\psi}}\right) ^{2}+(M_{J/\psi})^{2}}$. $\rm m_{4T}$ and $\rm y_4$ are the transverse mass and rapidity of gluon, respectively. Similarly as in photoproduction induced by electron proton collisions, we can define the $\rm \rm z^{J/\psi }$ parameter $ \rm z^{J/\psi } = P_{h}\cdot P^{J/\psi}/P_{h}\cdot q_{\gamma} $ where $\rm P_{h}$, $\rm q_{\gamma}$, $\rm P^{J/\psi}$ are the momenta of the hadron,
the virtual photon and charmonium, respectively. The data are somehow taken in either inelastic or elastic regime.
The inelastic regime is commonly considered to be the area where $\rm z$ is below 0.8 or 0.9 while the elastic regime is considered to be the area near $\rm z^{J/\psi }=1$ \cite{Amundson1997}. The gluon shadowing is interpreted in the nuclear infinite momentum frame as a result of fusion of gluons originating from different bound nucleons \cite{Alvioli2010}. The fundamental form of gluon shadowing is gluon saturation \cite{Gribov1981}. The nuclear shadowing effect manifests itself in $\rm PbPb$ and $\rm \gamma Pb$ interactions where the nucleus emits the gluon while $\rm pp$ and $\rm \gamma p$ interactions don't reveal this effect.

The summations in Eqs.(\ref{2to2a}) and (\ref{2to1a}) are taken over the spins and colors of initial and final states,
and the bar over the summation denotes averaging over the spins and colors of initial parton. The Mandelstam variables are defined as $\rm \widehat{s}=(p_{1}+p_{2})^{2}$, $\rm \widehat{t}=(p_{1}-p_{3})^{2}$ and $\rm \widehat{u}=(p_{1}-p_{4})^{2}$. $\rm N_{col}$ and $\rm N_{pol}$ refer to as the numbers of colors and polarization of states $\rm n$, separately. The QCD amplitudes with amputated heavy-quark spinors read as follows \cite{Petrelli1998}:
\begin{equation}
\begin{split}
\rm \mathcal{A}_{Q\overline{Q}}[^{1}S_{0}^{(1/8)}]=&\rm Tr[\mathcal{C}_{(1/8)}\Pi _{0}\mathcal{A}]_{q=0},\\
\rm \mathcal{A}_{Q\overline{Q}}[^{3}S_{1}^{(1/8)}]=&\rm \epsilon _{\alpha }Tr[\mathcal{C}_{(1/8)}\Pi_{1}^{\alpha }\mathcal{A}]_{q=0},\\
\rm \mathcal{A}_{Q\overline{Q}}[^{1}P_{1}^{(1/8)}]=&\rm \epsilon _{\beta }\frac{d}{dq_{\beta }}Tr[\mathcal{C}_{(1/8)}\Pi _{0}\mathcal{A}]_{q=0},\\
\rm \mathcal{A}_{Q\overline{Q}}[^{1}P_{J}^{(1/8)}]=&\rm \epsilon_{\alpha \beta } ^{(J)}\frac{d}{dq_{\beta }}Tr[\mathcal{C}_{(1/8)}\Pi _{1}^{\alpha }\mathcal{A}]_{q=0},
\end{split}
\end{equation}
where $\rm \mathcal{A}$ denotes the QCD amplitude with amputated heavy-quark spinors,
the lower index $\rm q$ represents the momentum of the heavy-quark in the $\rm Q\overline{Q}$ rest frame.
$\rm \Pi _{0/1}$ are spin projectors onto spin singlet and spin triplet states stated as
\begin{equation}
\begin{split}
\rm \Pi _{0}          =&\rm \frac{1}{\sqrt{8m_{Q}^{3}}}(\frac{\slashed{P}}{2}-\slashed{q}-m_{Q})\gamma _{5}(\frac{\slashed{P}}{2}+\slashed{q}+m_{Q}),\\
\rm \Pi _{1}^{\alpha }=&\rm \frac{1}{\sqrt{8m_{Q}^{3}}}(\frac{\slashed{P}}{2}-\slashed{q}-m_{Q})\gamma ^{\alpha }(\frac{\slashed{P}}{2}+\slashed{q}+m_{Q}),
\end{split}
\end{equation}
where $\rm P$ is the total momentum of heavy quarkonium and $\rm q$ is the relative momentum between the $\rm Q\overline{Q}$ pair.
$\rm \mathcal{C}_{1/8}$ are color factor projectors onto the color-singlet and color-octet states and can be expressed as follows:
\begin{equation}
\begin{split}
\rm C_{1} =&\rm \frac{\delta _{ij}}{\sqrt{N_{c}}}\\
\rm C_{8} =&\rm \sqrt{2}T_{ij}^{c},
\end{split}
\end{equation}
where $\rm N_{c}$ is the number of color, and $\rm T_{ij}^{c}$ is the generator of $\rm SU(N_{c})$, and $\rm i$, $\rm j$ are color indices for the heavy quarks.
The summation over the polarization is given as:
\begin{equation}
\begin{split}
\rm \sum_{J_{z}}\varepsilon _{\alpha }\varepsilon _{\alpha ^{^{\prime }}}^{\ast
} =&\rm \Pi _{\alpha \alpha ^{^{\prime }}},\\
\rm \sum_{J_{z}}\varepsilon _{\alpha \beta }^{0}\varepsilon _{\alpha ^{^{\prime
}}\beta ^{^{\prime }}}^{0\ast } =&\rm \frac{1}{3}\Pi _{\alpha \beta }\Pi
_{\alpha ^{^{\prime }}\beta ^{^{\prime }}},\\
\rm \sum_{J_{z}}\varepsilon _{\alpha \beta }^{1}\varepsilon _{\alpha ^{^{\prime
}}\beta ^{^{\prime }}}^{1\ast } =&\rm \frac{1}{2}(\Pi _{\alpha \alpha
^{^{\prime }}}\Pi _{\beta \beta ^{^{\prime }}}-\Pi _{\alpha \beta ^{^{\prime
}}}\Pi _{\alpha ^{^{\prime }}\beta }),\\
\rm \sum_{J_{z}}\varepsilon _{\alpha \beta }^{2}\varepsilon _{\alpha ^{^{\prime
}}\beta ^{^{\prime }}}^{2\ast } =&\rm \frac{1}{2}(\Pi _{\alpha \alpha
^{^{\prime }}}\Pi _{\beta \beta ^{^{\prime }}}+\Pi _{\alpha \beta
^{^{\prime }}}\Pi _{\alpha ^{^{\prime }}\beta })-\frac{1}{3}\Pi _{\alpha
\beta }\Pi _{\alpha ^{^{\prime }}\beta ^{^{\prime }}},
\label{tesfin}
\end{split}
\end{equation}
where $\rm \varepsilon_{\alpha}$ ($\varepsilon_{\alpha \beta }$) represents the polarization vector (tensor)
of the $\rm Q\overline{Q}$ states, and $\rm \Pi_{\alpha\beta}=-g_{\alpha\beta}+\frac{P_{\alpha}P_{\beta}}{M _{J/\psi}^{2}}$, and $\rm M _{J/\psi}$ is the $\rm J/\psi$ mass. The amplitude squares multiplied by long distance matrix elements are listed in the appendix \ref{factorization formalism}.

\section{NUMERICAL RESULTS AND DISCUSSION}
\label{NUMERICAL}

In this section, we discuss the numerical results of the photoproduction of $\rm J/\psi$ by using some physical parameters. $\rm M_{p}=0.94$ GeV is the mass of proton. The mass of the heavy quark is chosen as $\rm m_{Q}$=1.5 GeV where $Q$ represents the charm quark in our case. The mass of $\rm J/\psi $ is literally put at $\rm M _{J/\psi}=2m_{Q}$. The colliding energies used in this paper are $\sqrt{s}$=14 TeV for $\rm pp$, and $\sqrt{s}$=5 TeV for  $\rm PbPb$ and $\rm Pbp$. CTEQ6L1 \cite{Pumplin2002CTEQ} is used for the proton-gluon distribution which is probed at the factorization scale chosen as $\rm \mu_{f}=m_{T}^{J/\psi}$, where $\rm m_{T}^{J/\psi}=\sqrt{\left( p_{T}^{J/\psi}\right) ^{2}+ M_{J/\psi}^{2}}$ is the transverse mass. The minimum value of the transverse momentum of the quarkonium is chosen at $\rm {p_{T}^{\mathcal{Q}}}_{min}=1$ GeV. The range of
$\rm z^{J/\psi}$ is selected from zero to the area near one ($\rm {z^{J/\psi }< 1}$) and the predictions are not strongly dependent on the inferior limit of the integration $\rm z^{J/\psi }_{min}$ \cite{Goncalves2014}. Numerical calculations are carried out by in-house monte carlo generator.
\begin{table}[h]
\centering
\begin{tabular}{ll}
\hline
$\rm \langle 0|\mathcal{O}_{1,8}^{J/\psi }(^{2S+1}L_{J})|0\rangle$      & Numerical value \\
\hline
$\rm \langle 0|\mathcal{O}_{1}^{J/\psi }(^{3}S_{1})|0\rangle $        &1.16   GeV$^3$ \\
$\rm \langle 0|\mathcal{O}_{8}^{J/\psi }(^{3}S_{1})|0\rangle$      &  $0.3\times 10^{-2}$  GeV$^3$ \\
$\rm \langle 0|\mathcal{O}_{8}^{J/\psi }(^{1}S_{0})|0\rangle$      &  $8.9\times 10^{-2}$  GeV$^3$ \\
$\rm \langle 0|\mathcal{O}_{8}^{J/\psi }(^{3}P_{0})|0\rangle$      & $1.26\times 10^{-2}$  GeV$^5$ \\
%$\langle 0|Q_{8}^{J/\psi }(^{3}P_{2})|0\rangle$       &  $g_{0}/2\pi$\\[1ex]
%%$\langle 0|Q_{8}^{J/\psi }(^{3}P_{2})|0\rangle $     &  $\mathcal{L}$ \\[1ex]
\hline
\end{tabular}
\caption{\label{mous} Numerical values of LDMEs.}
\end{table}
The choice of the LDMEs for $\rm J/\psi $ is taken from \cite{Chao2012} as shown in Table \ref{mous}.
For $\rm \langle 0|\mathcal{O}_{8}^{J/\psi }(^{3}P_{J})|0\rangle $ with $\rm J$=0,1,2,
and following the heavy-quark spin symmetry, we get the relations:
\begin{equation}
\rm \langle 0|\mathcal{O}_{8}^{J/\psi }(^{3}P_{J})|0\rangle =(2J+1)\langle 0|\mathcal{O}_{8}^{J/\psi}(^{3}P_{0})|0\rangle.
\end{equation}

\begin{table*}[htbp]
\begin{center}
\begin{tabular}{|c | c | c | c | c | c | c | c | c |}
\hline
\multirow{2}{*}{$\rm \sigma[nb] \diagdown \xi_{i}$} & \multicolumn{2}{c|}{$0.015<\xi_{3}<0.15$} & \multicolumn{2}{c|}{$0.0015<\xi_{2}<0.5$}
& \multicolumn{2}{c|}{$0.1<\xi_{1}<0.5$} & \multicolumn{2}{c|}{$0<\xi<1$} \\
\cline{2-9} & with-s & no-s & with-s & no-s & with-s & no-s & with-s & no-s \\
\hline
$\rm \sigma^{pp}_{2\to 1}$   &-&4.71$\times$10$^{1}$&-&1.20$\times$10$^{2}$&-&1.17$\times$10$^{1}$&-&2.45$\times$10$^{2}$ \\
$\rm \sigma^{pbpb}_{2\to 1}$ &1.92$\times$10$^{6}$  &2.75$\times$10$^{6}$ &1.73$\times$10$^{7}$ &2.41$\times$10$^{7}$
& 2.58$\times$10$^{3}$  &3.79$\times$10$^{3}$  & 6.41$\times$10$^{7}$ & 7.93$\times$10$^{7}$\\
$\rm \sigma^{\gamma Pb}_{2\to 1}$ &4.45$\times$10$^{3}$  &6.43$\times$10$^{3}$ &1.13$\times$10$^{4}$ &1.60$\times$10$^{4}$
&1.09$\times$10$^{3}$  &1.61$\times$10$^{3}$   & 2.20$\times$10$^{4}$ & 2.88$\times$10$^{4}$ \\
$\rm \sigma^{\gamma p}_{2\to 1}$ &-&1.32$\times$10$^{4}$ &-&1.16$\times$10$^{5}$ &-&1.82$\times$10$^{1}$  &-&3.81$\times$10$^{5}$ \\
\hline
$\rm \sigma^{pp}_{2\to 2}$ &-&4.56$\times$10$^{1}$ &-&1.07$\times$10$^{2}$ &-&1.23$\times$10$^{1}$&-&1.66$\times$10$^{2}$ \\
$\rm \sigma^{pbpb}_{2\to 2}$ &1.56$\times$10$^{6}$ &1.98$\times$10$^{6}$ &1.20$\times$10$^{7}$ & 1.46$\times$10$^{7}$
&2.39$\times$10$^{3}$  &3.15$\times$10$^{3}$  &2.78$\times$10$^{7}$ & 3.12$\times$10$^{7}$ \\
$\rm \sigma^{\gamma Pb}_{2\to 2}$ & 3.78$\times$10$^{3}$ & 4.88$\times$10$^{3}$ &8.68$\times$10$^{3}$
&1.09$\times$10$^{4}$ &1.03$\times$10$^{3}$ &1.38$\times$10$^{3}$  &1.25$\times$10$^{4}$ &1.50$\times$10$^{4}$ \\
$\rm \sigma^{\gamma p}_{2\to 2}$ &-&9.52$\times$10$^{3}$ &-&7.01$\times$10$^{4}$ &-&1.52$\times$10$^{1}$&-&1.50$\times$10$^{5}$ \\
\hline
\end{tabular}
\end{center}\caption{\label{tab:0b}
The total cross section (nb) for the $\rm J/\psi$ photoproduction with nuclear shadowing effect ($\rm with-s$) and without nuclear shadowing effect ($\rm no-s$) effect at LHC with forward detector acceptances.}
\end{table*}
In Table \ref{tab:0b}, we present our estimates for the total cross sections of $\rm J/\psi$ photoproduction including both 2$\rightarrow$1 and 2$\rightarrow$2 subprocesses at LHC with forward detector acceptances. We consider the $\rm pp$, $\rm PbPb$ and $\rm Pbp$ collisions where $\rm Pbp$ means gluon can be emitted from proton ($\rm \gamma p$) or lead ($\rm \gamma Pb$). The nuclear shadowing effect is included only when gluons are emitted from nucleus. Our predictions show that the  total cross sections with $\xi_{2}$ and $\xi_{3}$ are close and  keep  more contribution to $\rm J/\psi$ photoproduction than $\xi_{1}$. The 2$\rightarrow$1 subprocess produces only color octet channels. The 2$\rightarrow$2 subprocess generates  color octet channels as well as the color singlet channel due to the emission of gluon.
The color singlet  and color octet  $\rm ^{1}S_{0}^{[8]}$ channels have dominantly contributed to the total cross section. These two contributions have enhanced the total cross section of the 2$\rightarrow$2 subprocess. However, its total cross sections are small in most cases with reference to 2$\rightarrow$1 because of the two-body final state phase space suppression. Notice here we have applied a $\rm {p_{T}^{\mathcal{Q}}}_{min}=1$ GeV cut for 2$\rightarrow$2 production. It has been found that the total cross sections with nuclear shadowing effect are lower than those without nuclear shadowing effect. The reduction is due to nuclear modification of the parton distribution functions \cite{Stahl2017,Andronic2015} and the decrease of the gluon density in the nucleus at low depletion of Bjorken variable at $x\lesssim 0.1$ \cite{Alvioli2010}. In certain case, the production of charmonium can be altered by nuclear matter effect such as parton energy loss due to multiple scattering in the nucleus \cite{Peigne2008}.  It has also been shown that the cross section of $\rm PbPb$ interactions is larger than that of $\rm Pbp$ and $\rm pp$ interactions, $\rm Pbp$ interactions is larger than that of $\rm pp$, which can be accounted for by the enhancement of the photon flux from nuclei and the gluon distribution from lead , i.e., proportional to $\rm Z^2$  and $\rm A$.

\begin{table*}[htbp]
\begin{center}
\begin{tabular}{|c | c | c | c | c |}
\hline
\multirow{2}{*}{$\rm \sigma[nb]$} & \multicolumn{4}{c|}{$0.0015<\xi_{2}<0.5$}  \\
\cline{2-5} & $\rm pp$ &  $\rm PbPb$
& $\rm \gamma p$ & $\rm \gamma Pb$ \\
\hline
$\rm 2\to 1\ total$ &1.20$\times$10$^{2}$ & 1.73$\times$10$^{7}$ & 1.16$\times$10$^{5}$& 1.13$\times$10$^{4}$ \\
\hline
\hline
$\rm 2\to 2\ total$ &1.07$\times$10$^{2}$  &1.20$\times$10$^{7}$ &7.01$\times$10$^{4}$ &8.68$\times$10$^{3}$  \\
\hline
$\rm ^{3}S_{1}^{[1]}$ &4.15$\times$10$^{0}$ &5.30$\times$10$^{5}$ &3.36$\times$10$^{3}$ & 3.57$\times$10$^{2}$  \\
$\rm ^{1}S_{0}^{[8]}$ &7.66$\times$10$^{1}$ &8.57$\times$10$^{6}$ & 4.99$\times$10$^{4}$&6.23$\times$10$^{3}$ \\
$\rm ^{3}S_{1}^{[8]}$ &0.02$\times$10$^{0}$&2.57$\times$10$^{3}$ &1.63$\times$10$^{1}$ &1.73$\times$10$^{0}$ \\
$\rm ^{3}P_{0}^{[8]}$  &9.25$\times$10$^{0}$&1.04$\times$10$^{6}$ & 6.10$\times$10$^{3}$&7.55$\times$10$^{2}$ \\
$\rm ^{3}P_{1}^{[8]}$  &3.69$\times$10$^{0}$ &4.01$\times$10$^{5}$ &2.30$\times$10$^{3}$ &2.97$\times$10$^{2}$  \\
$\rm ^{3}P_{2}^{[8]}$  &1.29$\times$10$^{1}$ &1.44$\times$10$^{6}$ &8.36$\times$10$^{3}$ &1.05$\times$10$^{3}$ \\
\hline
\end{tabular}
\end{center}
 \caption{\label{tab:1b}
The total cross sections of $\rm pp$, $\rm PbPb$  and $\rm Pbp$ collisions where $\rm Pbp$ means gluon can be emitted from proton ($\rm \gamma p$) or lead ($\rm \gamma Pb$), and the cross sections of different color octet and color singlet subprocesses contributions with and without nuclear shadowing effect for the $\rm J/\psi$ photoproduction at the LHC with forward detector acceptances $\xi_2$.}
\end{table*}
In Table \ref{tab:1b}, we predict the dominant total cross sections of $\rm pp$, $\rm PbPb$ and $\rm Pbp$ collisions and the cross sections of different color octet and color singlet contributions including both 2$\to$1 and 2$\to$2 with and without shadowing effect for the $\rm J/\psi$ photoproduction at the LHC with forward detector acceptances $0.0015<\xi_{2}<0.5$. We see that the cross section of color-octet channel subprocess contribution $\rm ^{3}S_{1}^{[8]}$ ( $\rm ^{1}S_{0}^{[8]}$) has a lower (larger) contribution to the total ones.

\begin{table*}[htbp]
\begin{center}
\begin{tabular}{|c | c | c |c|c| }
\hline
\multirow{2}{*}{$\rm N \diagdown \xi_{i}$}    & \multirow{2}{*}{$\rm \mathcal{L}_{LHC}$[$\rm mb^{-1}s^{-1}$]}& \multirow{2}{*}{T[s]} & \multicolumn{2}{c|}{$0.0015<\xi_{2}<0.5$}  \\
\cline{4-5}& & &with-s & no-s  \\
\hline
$\rm N^{pp}_{2\to 1}$  & $10^{7}$&10$^{7}$  &-&1.2$\times$ 10$^{10}$ \\
$\rm N^{pbpb}_{2\to 1}$  & 0.42&10$^{6}$   &7.23$\times$10$^{6}$ &1.01$\times$10$^{7}$ \\
$\rm N^{\gamma Pb}_{2\to 1}$&150& 10$^{6}$&1.70$\times$10$^{6}$ &2.40$\times$10$^{6}$ \\
$\rm N^{\gamma p}_{2\to 1}$&150&10$^{6}$ &-&1.74$\times$10$^{7}$  \\
\hline
$\rm N^{pp}_{2\to 2}$ &$10^{7}$& 10$^{7}$&-&1.07$\times$10$^{10}$   \\
$\rm N^{pbpb}_{2\to 2}$ & 0.42&10$^{6}$ &5.04$\times$10$^{6}$ &6.13$\times$10$^{6}$  \\
$\rm N^{\gamma Pb}_{2\to 2}$&150& 10$^{6}$ & 1.30$\times$10$^{6}$&1.64$\times$10$^{6}$ \\
$\rm N^{\gamma p}_{2\to 2}$& 150&10$^{6}$ &-&1.05$\times$10$^{7}$  \\
\hline
\end{tabular}
\end{center}
 \caption{\label{tab:2b}
The production rates of $\rm pp$, $\rm PbPb$ and $\rm Pbp$ interactions of $\rm J/\psi$ photoproduction with nuclear shadowing effect ($\rm with-s$) and without nuclear shadowing effect ($\rm no-s$) effect at the LHC with forward detector acceptances $0.0015<\xi_{2}<0.5$.}
\end{table*}
In Table \ref{tab:2b}, we present our estimates for production rates of $\rm pp$, $\rm PbPb$ and  $\rm Pbp$ interactions by assuming the design integrated luminosities $\rm \mathcal{L}_{LHC}^{pp}$, $\rm \mathcal{L}_{LHC}^{PbPb}$, $\rm \mathcal{L}_{LHC}^{Pbp}$ and run times (T) \cite{Goncalves2013}. Our predictions show that the event rates  of $\rm pp$ interactions dominate over the event rates of $\rm PbPb$ and  $\rm Pbp$ interactions, due to its considerable design  integrated luminosity and running time. The design integrated luminosity of  $\rm Pbp$ interactions, which are two order of magnitude larger than for $\rm PbPb$, lessens the suppression for the event rates. Nonetheless, the resulting production rates  are still minute in comparison to the $\rm pp$ results. Although the values of the cross sections are considerable, the clean topology of coherent interactions involves a higher signal to background ratio. Hence, the experimental detection is evidently possible. The main drawback is that the signal is supposed to be downgraded due to the event pileup. An option to rate coherent events at the LHC is
by tagging the intact hadron in the final state.

\begin{table*}[htbp]
\begin{center}
\begin{tabular}{|c | c | c |c|c| }
\hline
\multirow{2}{*}{$\rm N \diagdown \xi_{i}$} & \multirow{2}{*}{$\rm \mathcal{L}_{LHC}$[$\rm mb^{-1}s^{-1}$]}& \multirow{2}{*}{T[s]} & \multicolumn{2}{c|}{$0.0015<\xi_{2}<0.5$} \\
\cline{4-5} & & & with-s & no-s  \\
\hline
$\rm N^{\gamma Pb}_{2\to 1}$ &$10^{4}$&10$^{7}$&1.13$\times$10$^{9}$&1.60$\times$10$^{9}$ \\
$\rm N^{\gamma p}_{2\to 1}$&$10^{4}$&10$^{7}$ &-&1.16$\times$10$^{10}$   \\
\hline
$\rm N^{\gamma Pb}_{2\to 2}$&$10^{4}$&10$^{7}$ & 8.68$\times$10$^{8}$&1.09$\times$10$^{9}$ \\
$\rm N^{\gamma p}_{2\to 2}$&$10^{4}$&10$^{7}$ &-&7.01$\times$10$^{9}$  \\
\hline
\end{tabular}
\end{center}
 \caption{\label{tab:3b}
The production rate of  $\rm  Pbp$ interactions of  $\rm J/\psi$ photoproduction
with nuclear shadowing effect ($\rm with-s$) and without nuclear shadowing effect ($\rm no-s$) effect at upgraded LHC with forward detector acceptances $0.0015<\xi_{2}<0.5$.}
\end{table*}
As for  $\rm Pbp$ interactions, an enhanced setup was suggested in Ref.\cite{dEnterria2010} and the design integrated luminosity $\rm \mathcal{L}_{LHC}^{Pbp}$ and  the running time (T) were upgraded. The upgraded production rates are arranged in Table \ref{tab:3b} and our estimate reveals that the main upgraded production rates hail from $\rm \gamma p$ interactions because of the large cross sections. Therefore, the experimental analysis can be made possible with the upgraded event rates of  $\rm \gamma p$ interactions due to the fact that the numbers are reasonable and those collisions are supposed  to trigger on and perform the measurement with almost no pileup. For that reason, the upgraded $\rm \gamma p$ scenarios give one of the best feasibilities to detect the inelastic $\rm J/\psi$  photoproduction in coherent processes.

\begin{figure}[htp]
\centering
   %\begin{minipage}[t]{4.0cm}
   \includegraphics[height=4.8cm,width=4.4cm]{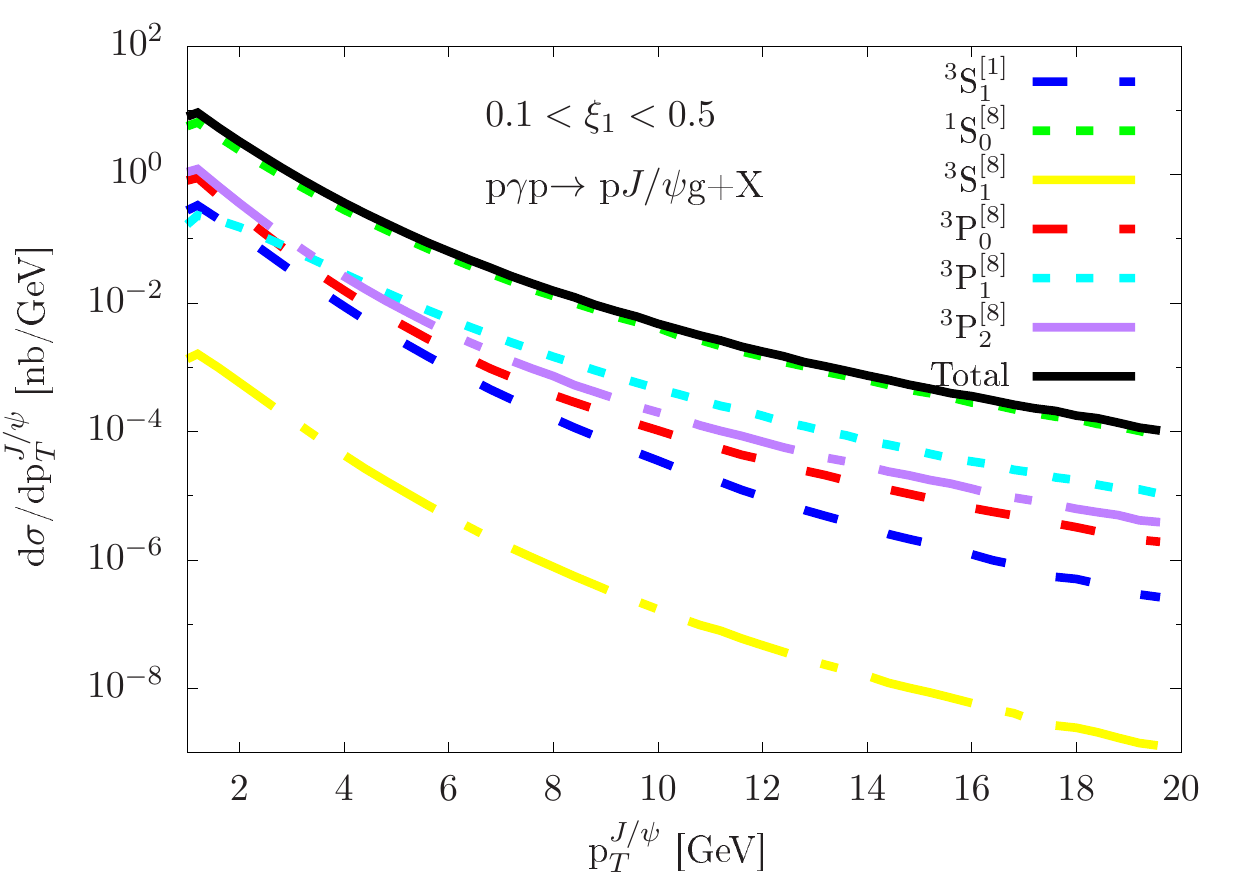}
   \includegraphics[height=4.8cm,width=4.4cm]{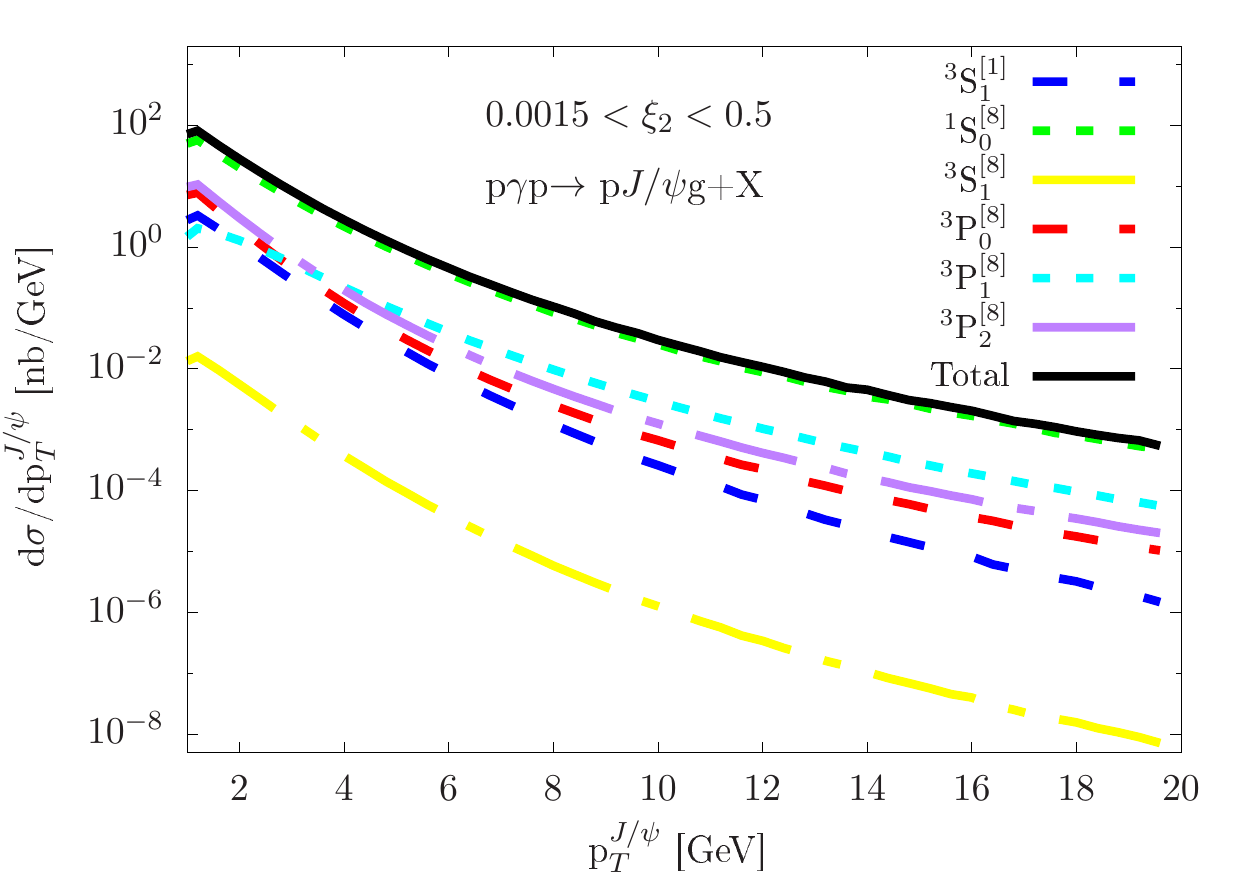}
   \includegraphics[height=4.8cm,width=4.4cm]{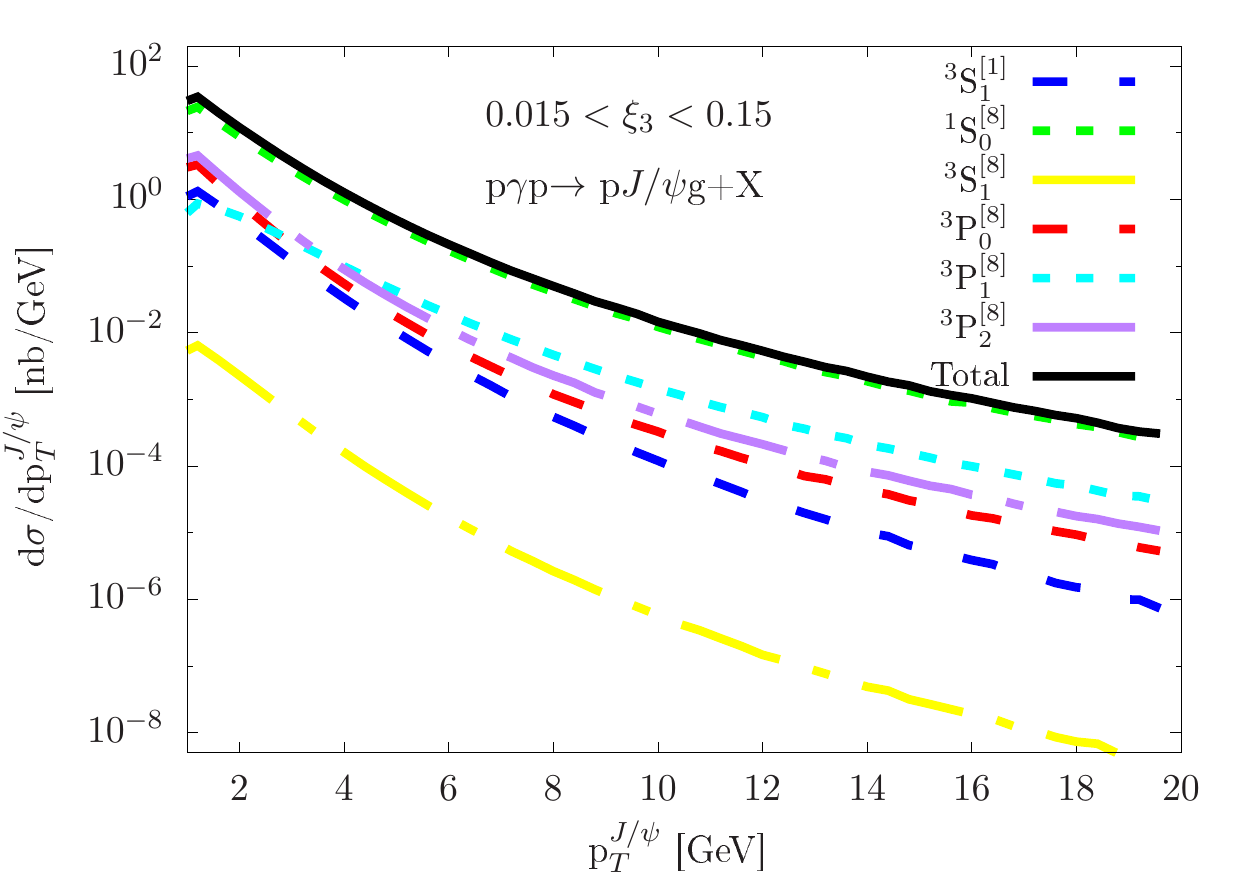}
   \includegraphics[height=4.8cm,width=4.4cm]{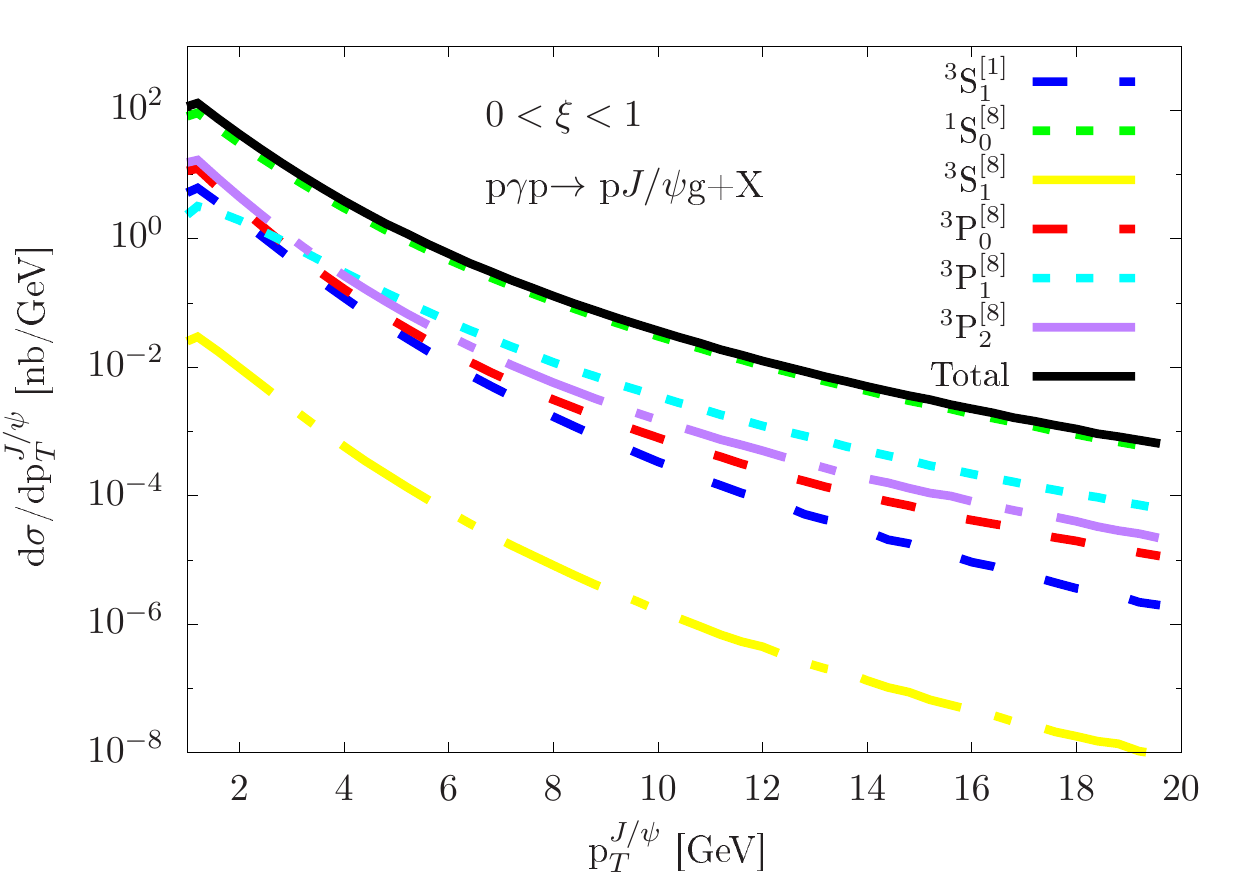}
   \includegraphics[height=4.8cm,width=4.4cm]{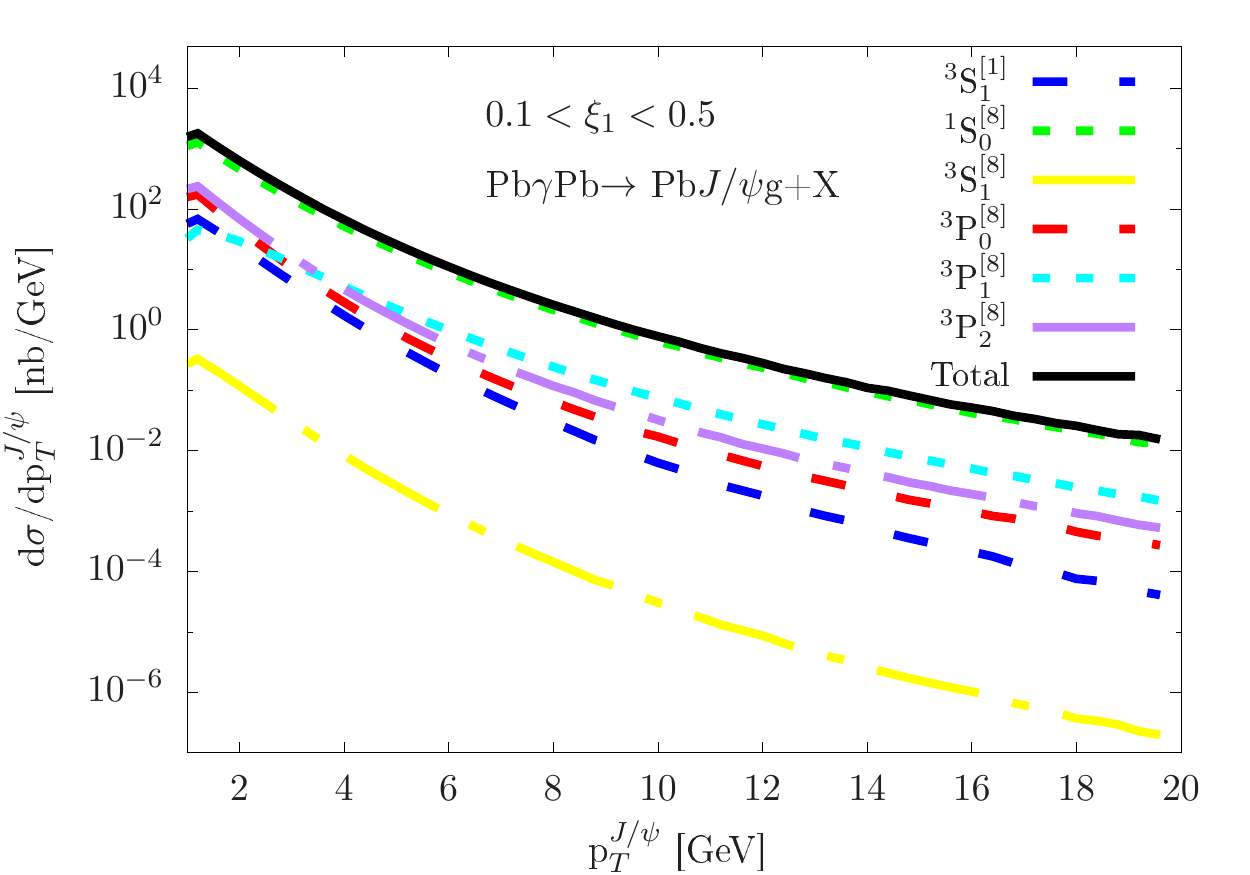}
   \includegraphics[height=4.8cm,width=4.4cm]{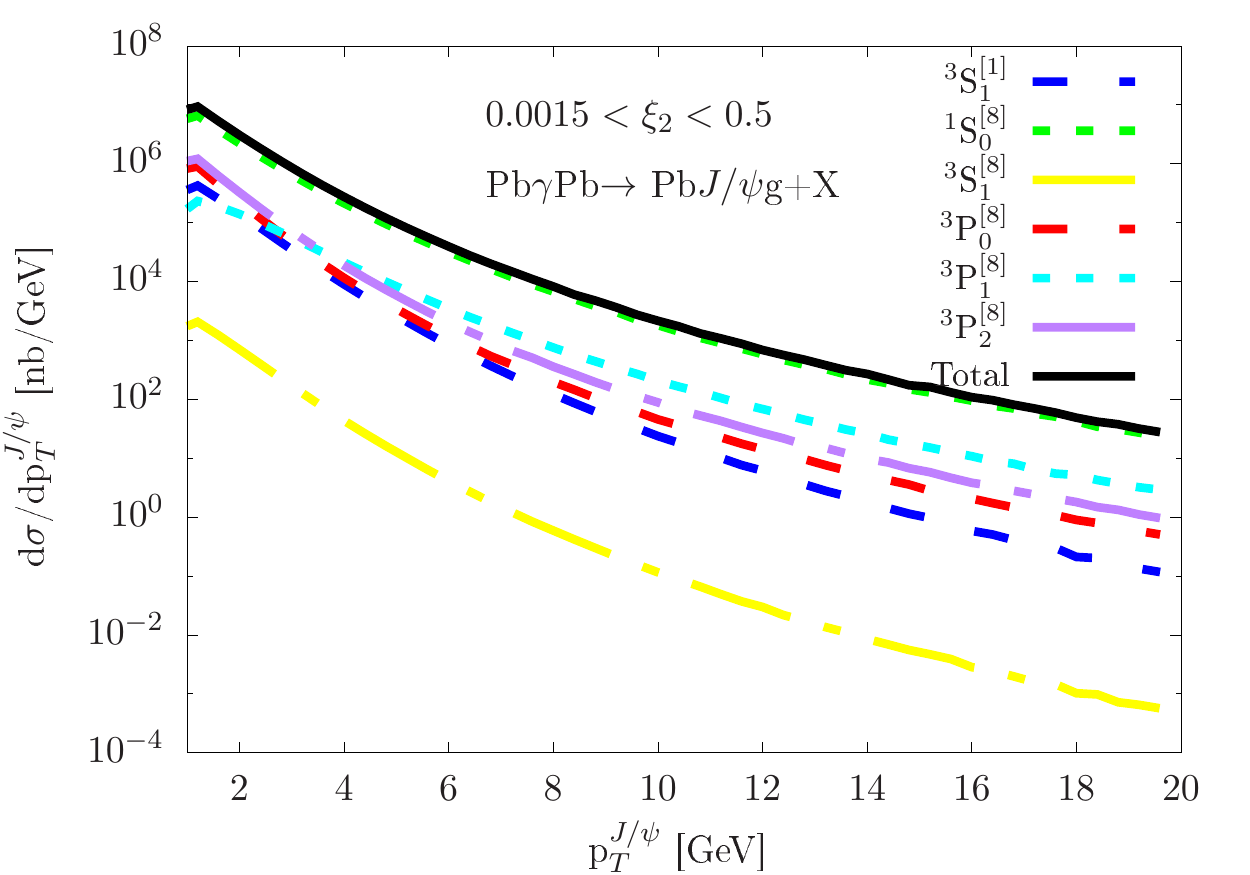}
   \includegraphics[height=4.8cm,width=4.4cm]{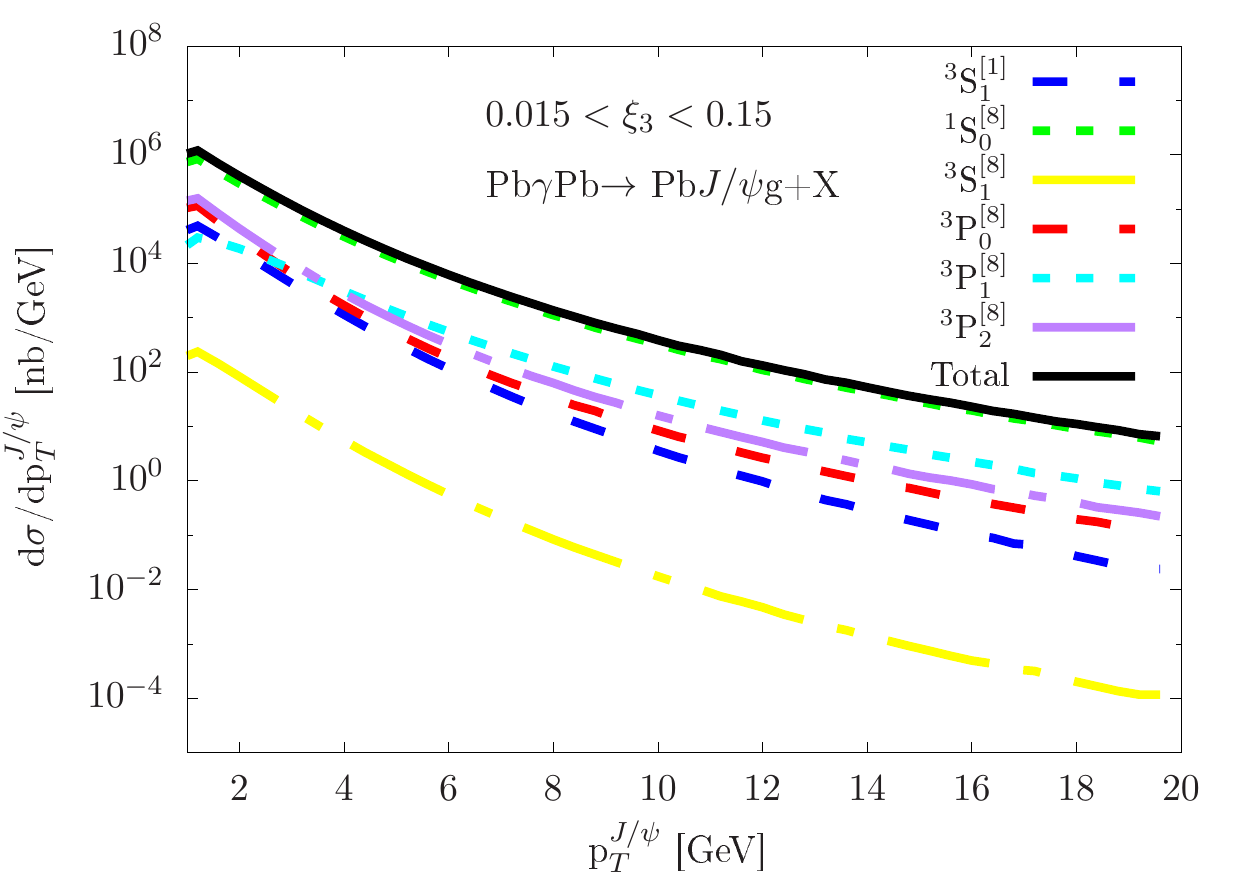}
   \includegraphics[height=4.8cm,width=4.4cm]{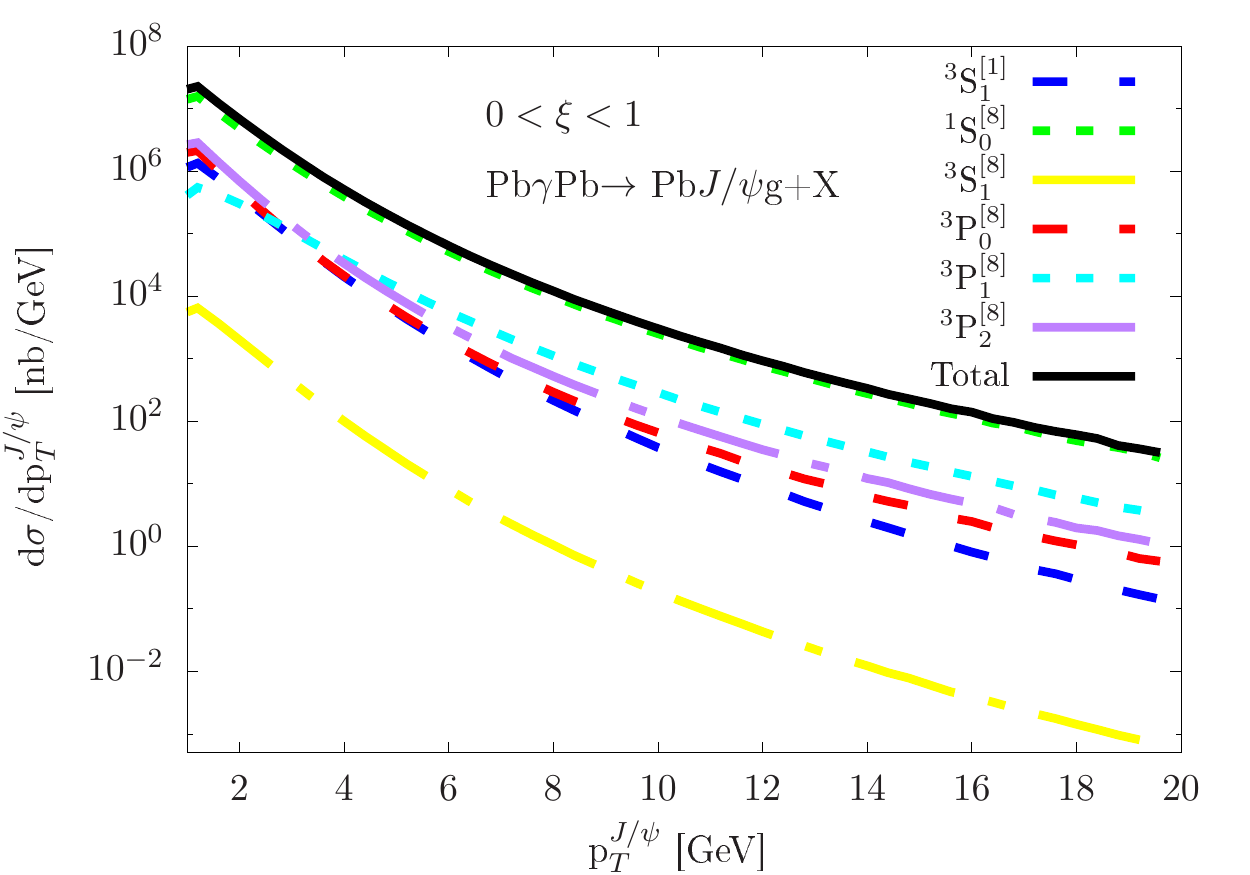}
   %\end{minipage}
   \caption{ \normalsize (color online)
   The $\rm p_{T}^{J/\psi}$ distributions  for the
   $\rm p\gamma p\rightarrow p\mathcal{Q}g+X$ (top panel) and $\rm Pb\gamma Pb\rightarrow Pb\mathcal{Q}g+X$  (bottom panel) processes
   , and the contributions of the $\rm ^{3}S_{1}^{[1]}$ (blue dashed line),
   $\rm ^{1}S_{0}^{[8]}$ (green dotted line), $\rm ^{3}S_{1}^{[8]}$ (yellow dash dotted line),
   $\rm ^{3}P_{0}^{[8]}$ (red dashed line), $\rm ^{3}P_{1}^{[8]}$ (cyan dotted line), $\rm ^{3}P_{2}^{[8]}$ (purple dash dotted line) channels and
   Total (black solid line).}
\label{fig3:limits}
\end{figure}
\begin{figure}[htp]
\centering
  % \begin{minipage}[t]{4.0cm}
   \includegraphics[height=4.8cm,width=4.4cm]{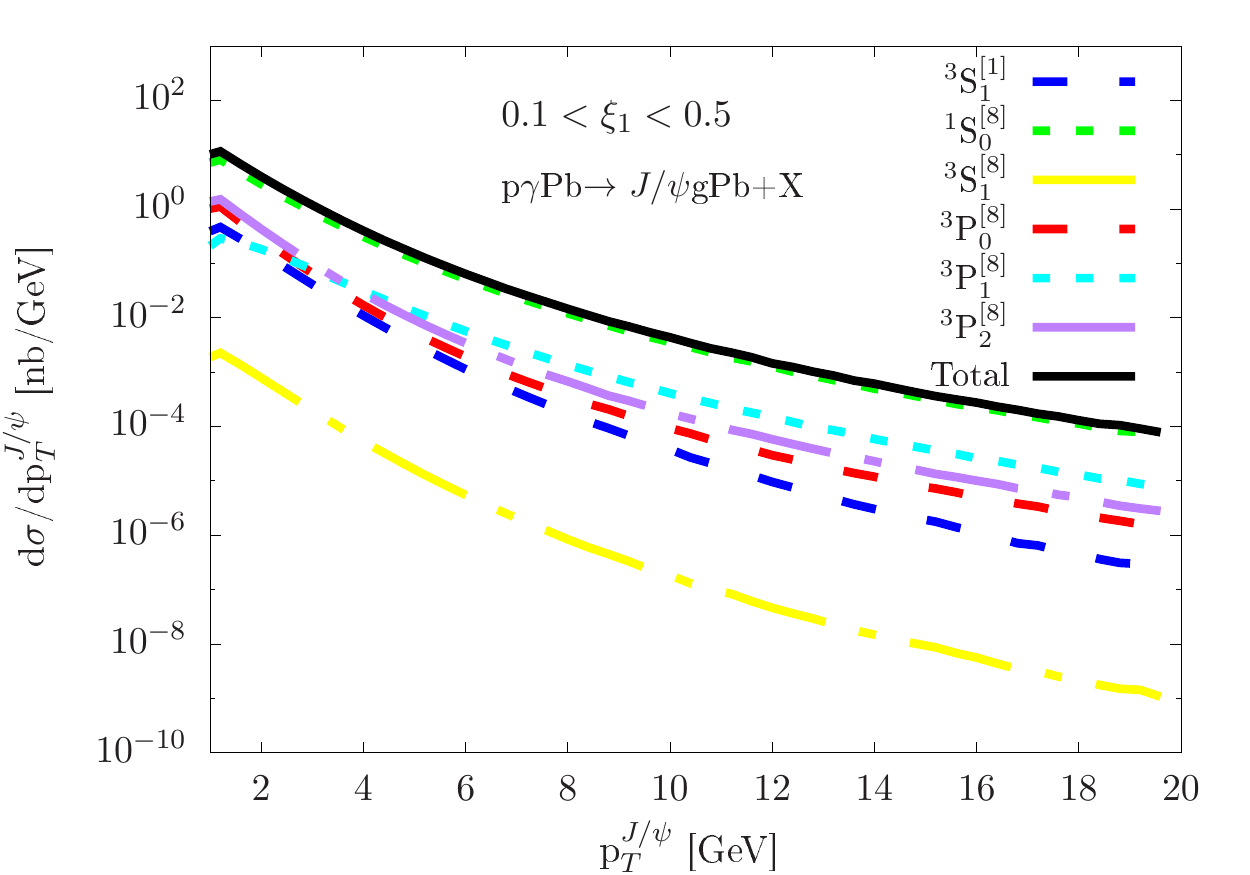}
   \includegraphics[height=4.8cm,width=4.4cm]{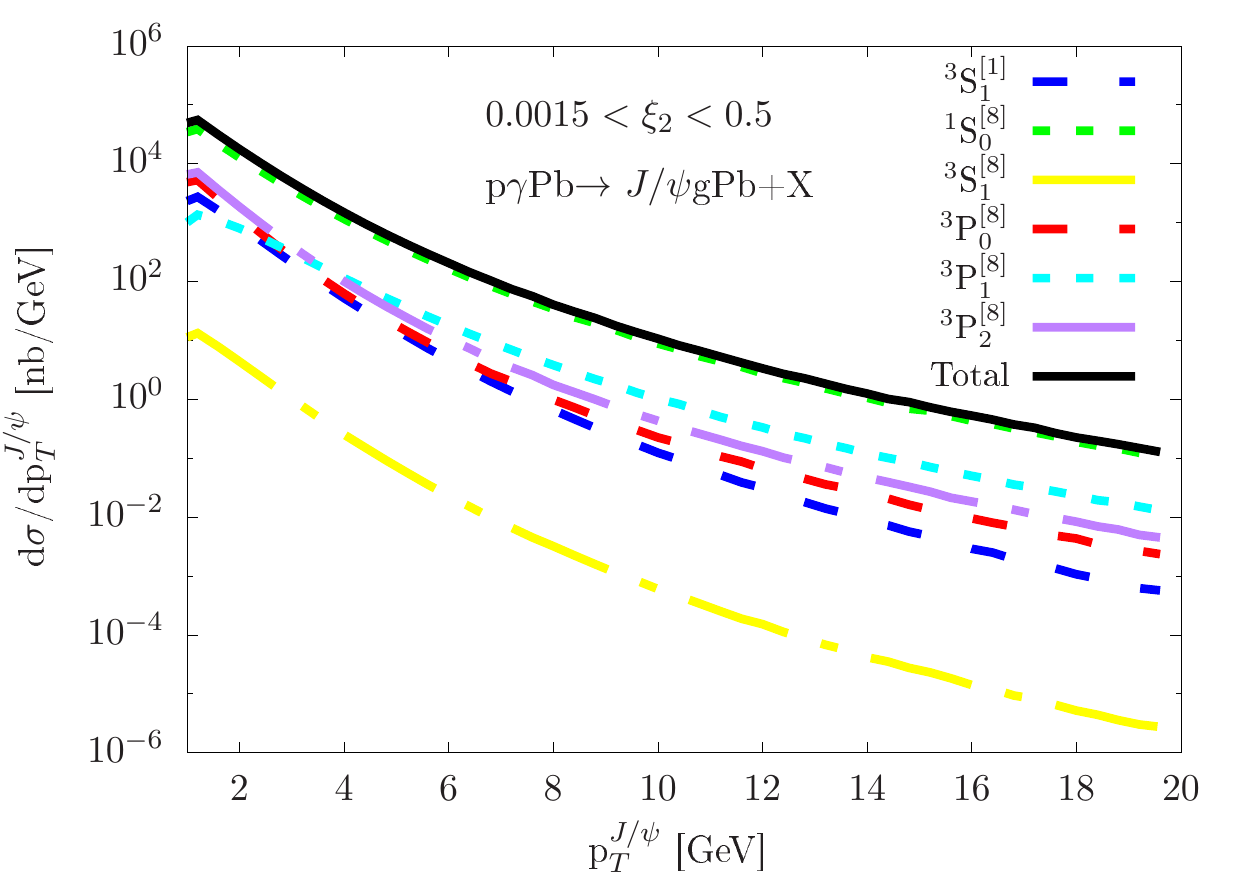}
   \includegraphics[height=4.8cm,width=4.4cm]{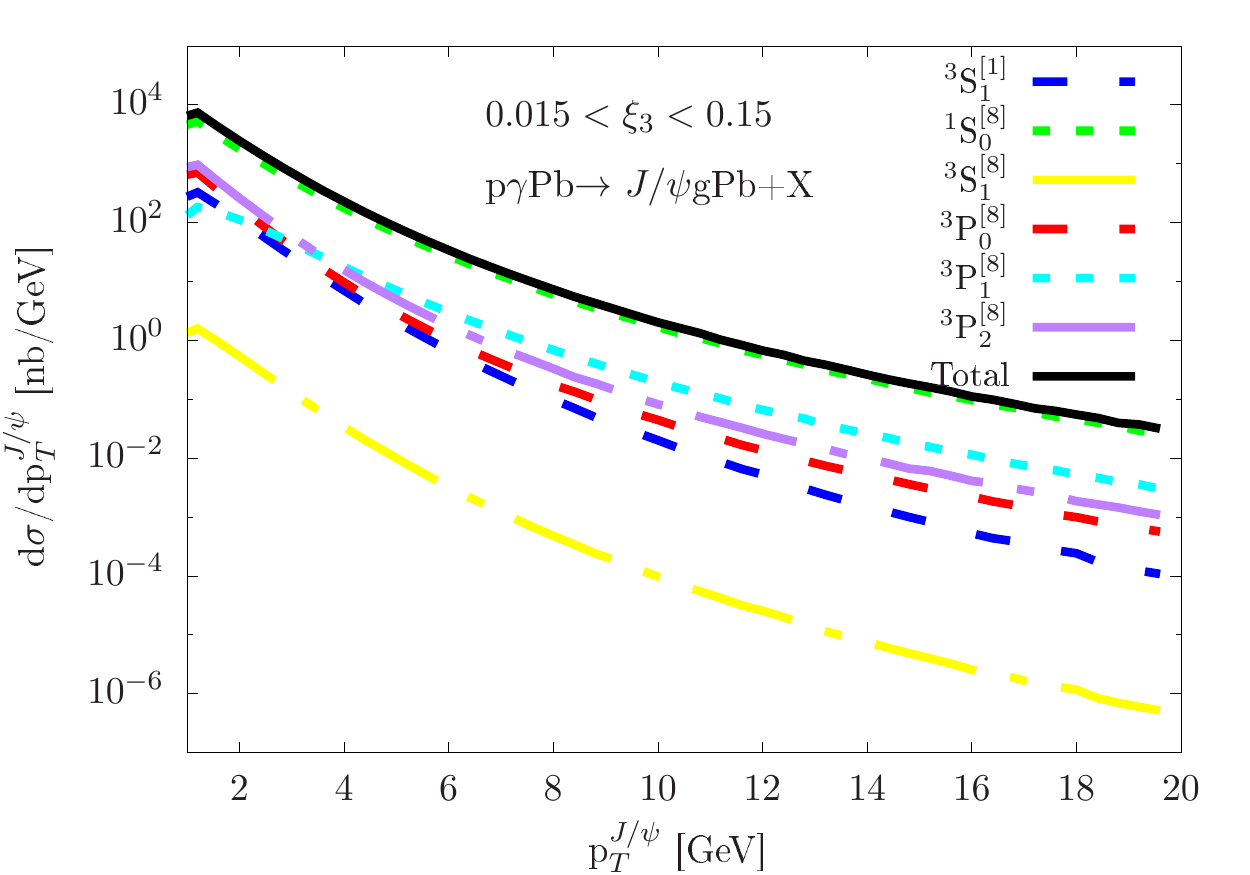}
   \includegraphics[height=4.8cm,width=4.4cm]{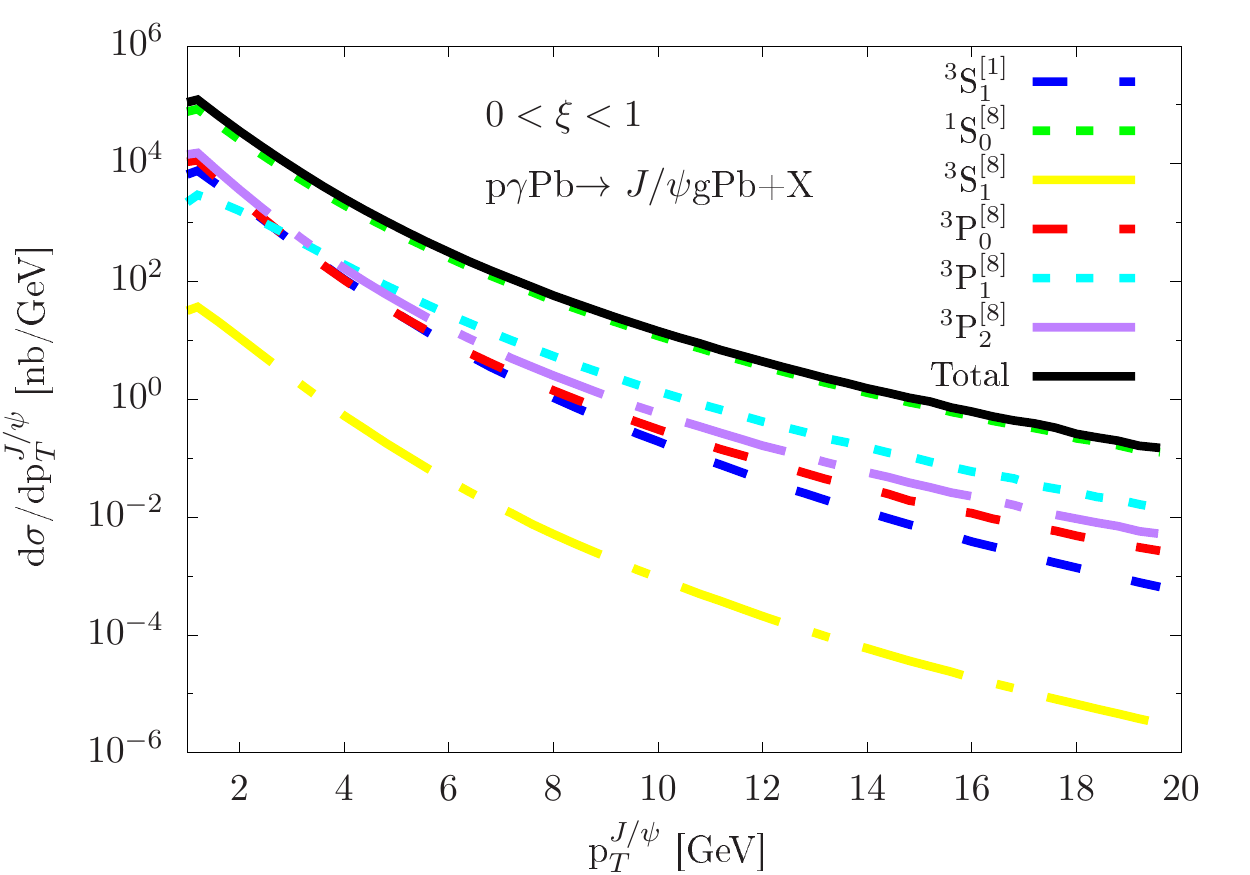}
   \includegraphics[height=4.8cm,width=4.4cm]{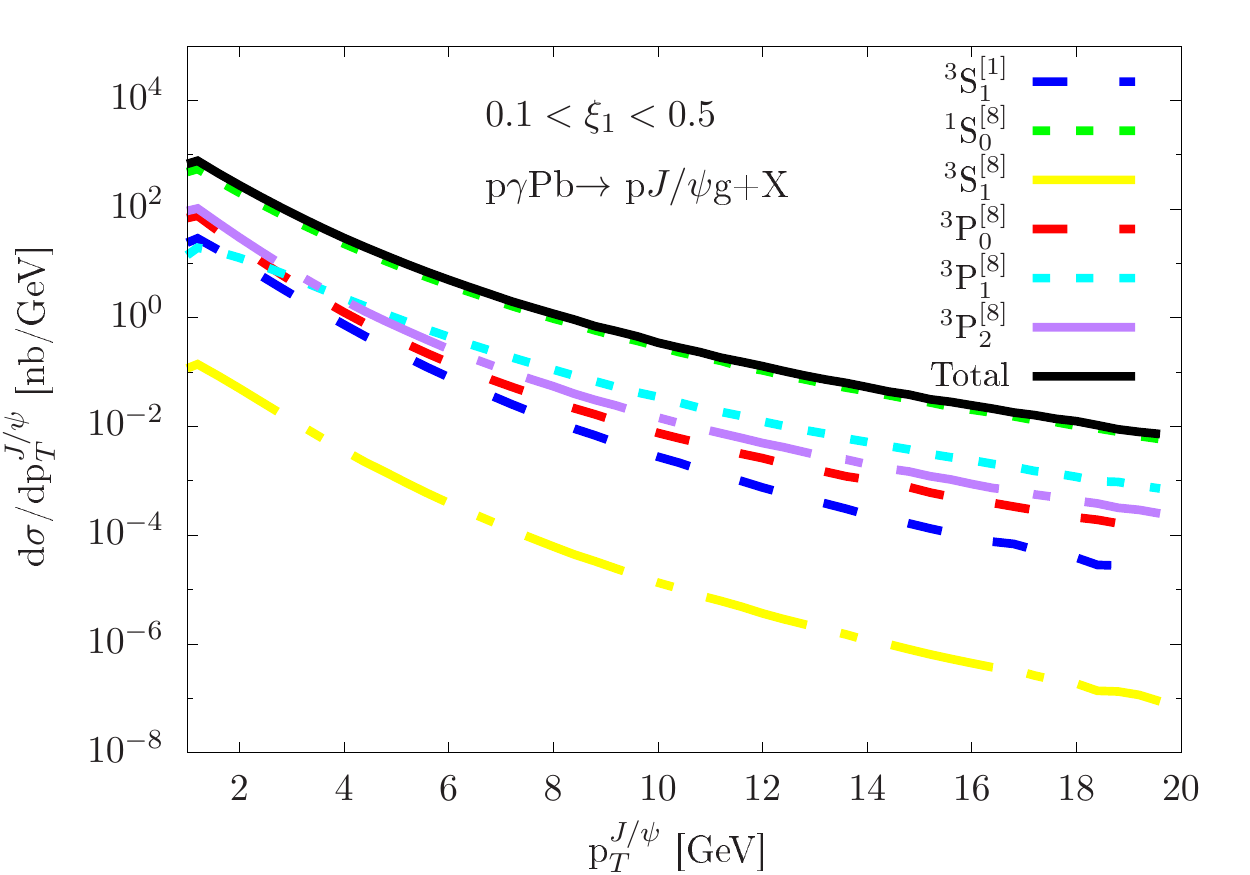}
   \includegraphics[height=4.8cm,width=4.4cm]{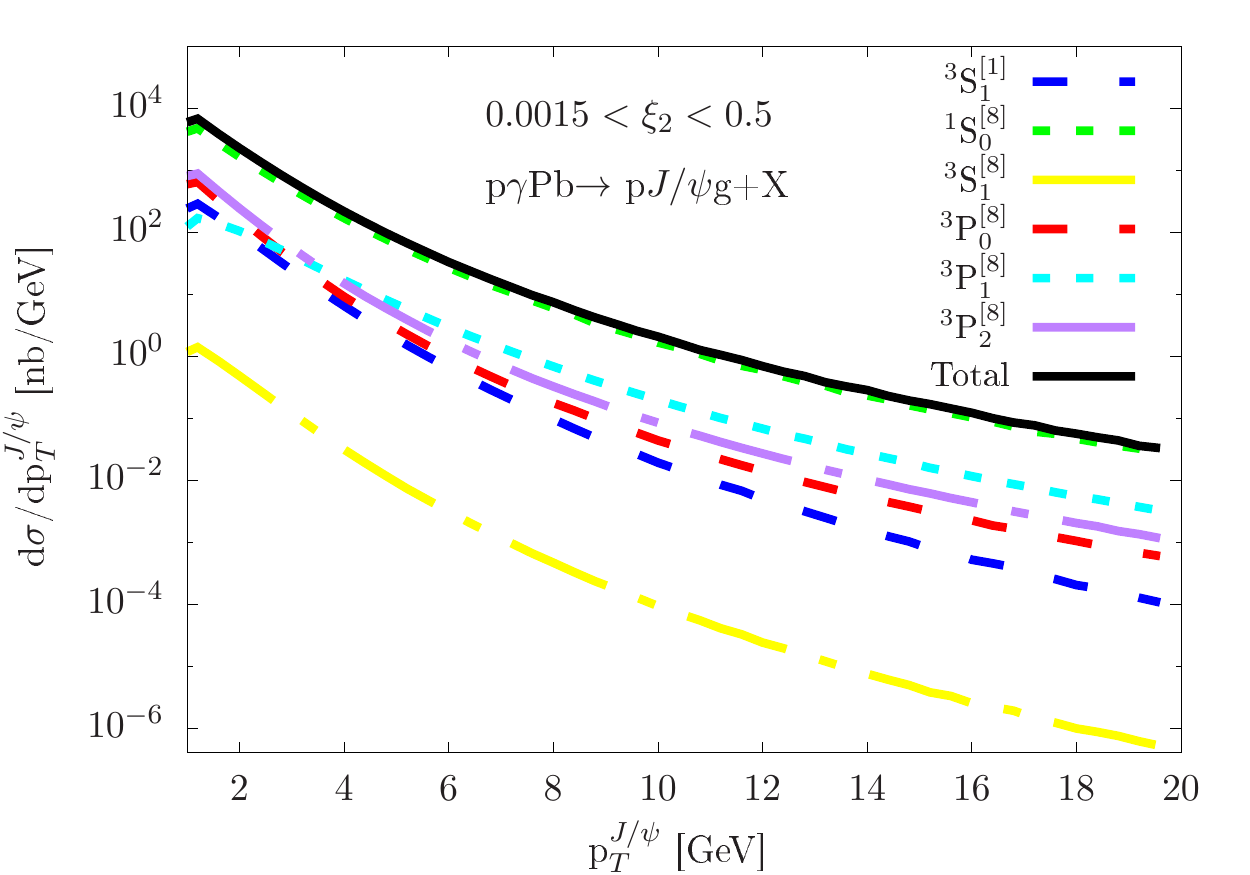}
   \includegraphics[height=4.8cm,width=4.4cm]{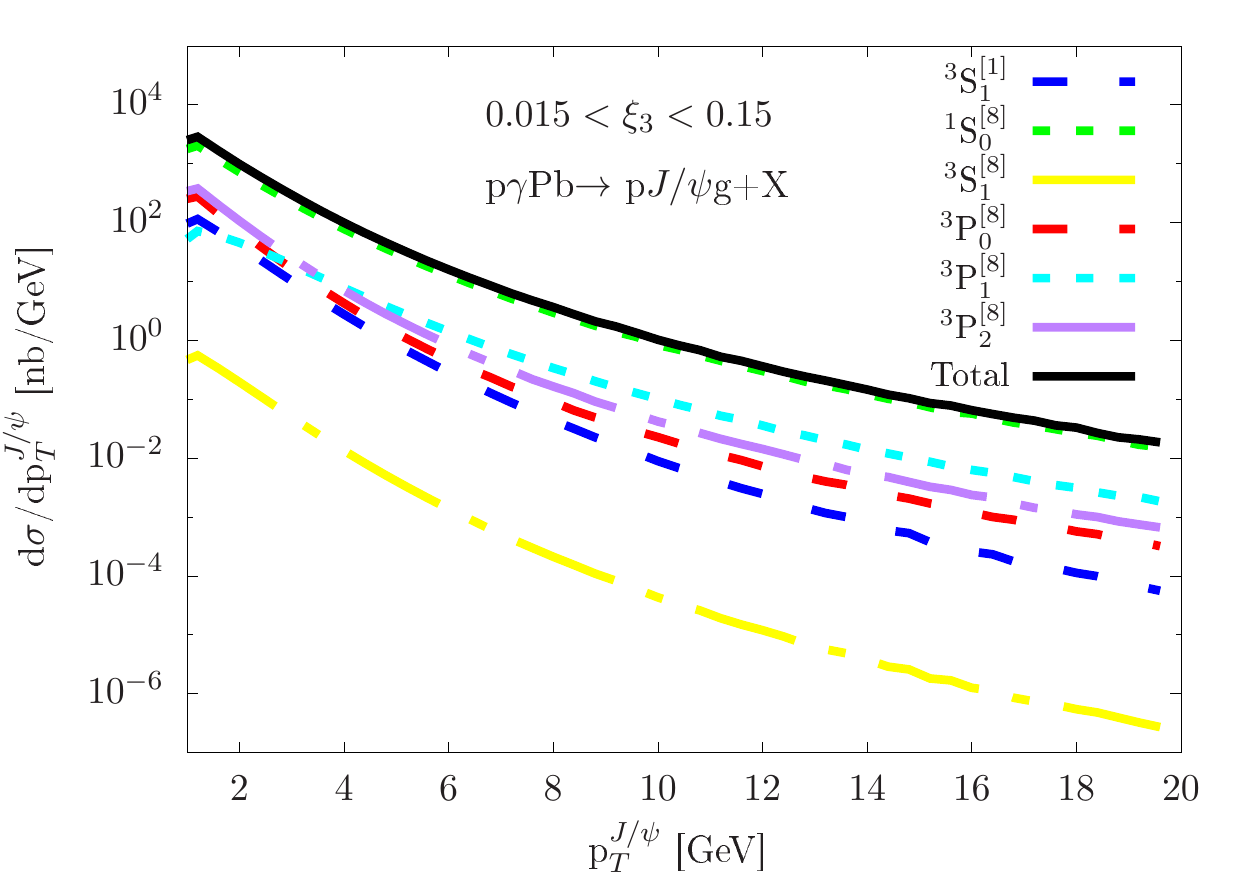}
   \includegraphics[height=4.8cm,width=4.4cm]{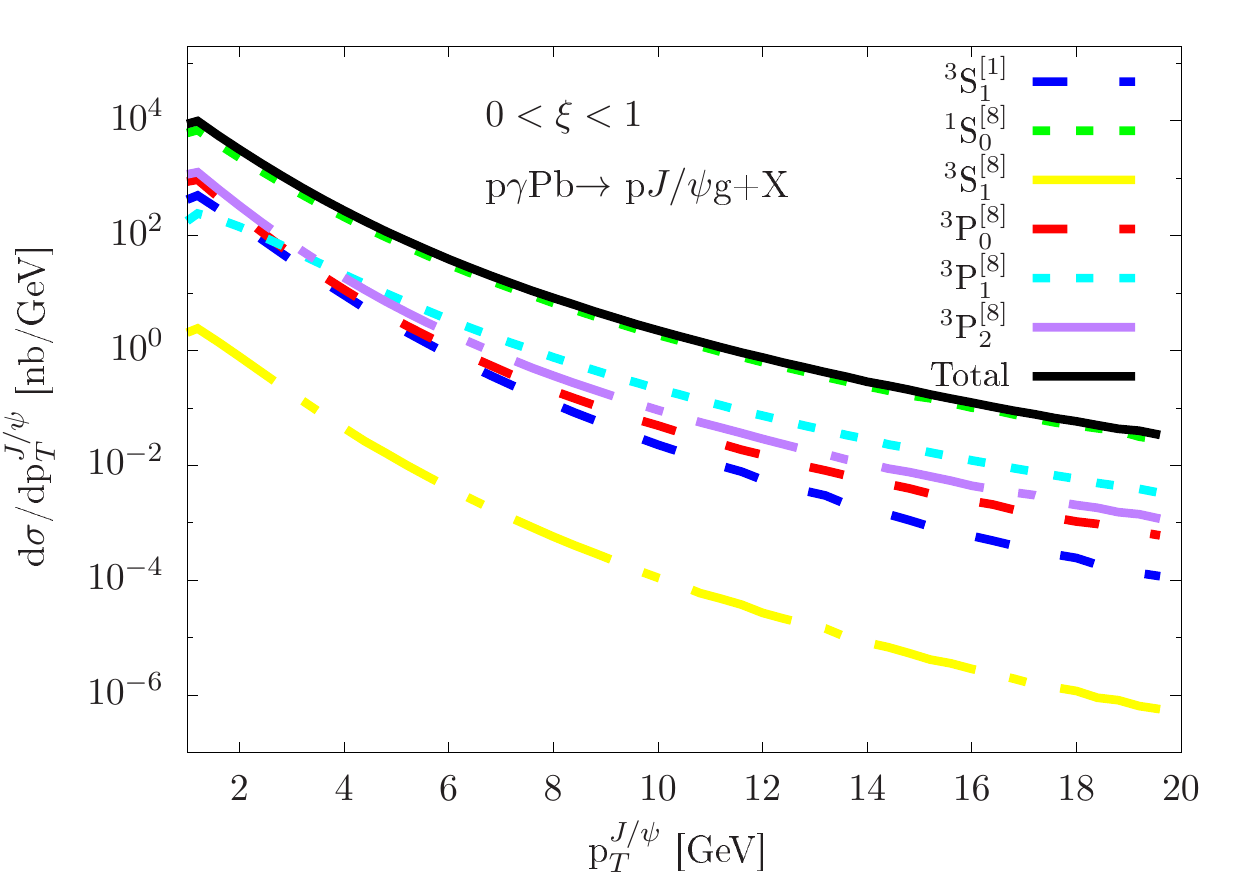}
   %\end{minipage}
   \caption{ \normalsize (color online) The $\rm p_{T}^{J/\psi}$  distributions  for the
   $\rm p\gamma Pb\to \mathcal{Q}gPb+X$  (top panel)  and $\rm p\gamma Pb\to p\mathcal{Q}g+X$  (bottom panel)  processes,
   and the contributions of the $\rm ^{3}S_{1}^{[1]}$ (blue dashed line),
   $\rm ^{1}S_{0}^{[8]}$ (green dotted line), $\rm ^{3}S_{1}^{[8]}$ (yellow dash dotted line),
   $\rm ^{3}P_{0}^{[8]}$ (red dashed line), $\rm ^{3}P_{1}^{[8]}$ (cyan dotted line), $\rm ^{3}P_{2}^{[8]}$ (purple dash dotted line) channels and
   Total (black solid line).}
\label{fig4:limits}
\end{figure}
The values of scaling variable  $\rm p_{T}^{J/\psi }$ can be used to choose the appropriate models of quarkonia production.
At small transverse momentum, the CSM is usually used to explain the $\rm J/\psi$ photoproduction.
However, the CO mechanism and the fragmentation approach are preferentially selected at large transverse momentum to address the quarkonia production. In Figs.\ref {fig3:limits} and \ref{fig4:limits}, we plot the predictions of the $\rm p_{T}^{J/\psi }$ distributions  in $\rm pp$, $\rm PbPb$ and  $\rm Pbp$ collisions with $0.1<\xi_{1}< 0.5$ for CMS-TOTEM forward detector, $0.0015 <\xi_{2}< 0.5$ for CMS-TOTEM forward detector, $0.015 <\xi_{3}< 0.15$
for AFP-ATLAS forward detector and $ 0 <\xi <1$  for different Fock state contributions.
The $\rm pPb$ collision is split into $\rm p\gamma Pb\to p J/\psi(+g)+X$ which signifies  the  initial photon is emitted from proton while this proton stays undissociated in the final state ($\rm \gamma Pb$ interaction) and $\rm p\gamma Pb\to J/\psi Pb(+g)+X$ which signifies the initial photon is emitted from lead while this lead stays undissociated ($\rm \gamma p$ interaction).
The forward $\rm p_{T}^{J/\psi }$  contribution is only considered in this case study, meanwhile $\rm \gamma Pb$ and $\rm \gamma p$ interactions are separately described. The total contribution will be two times larger if backward is included in $\rm PbPb$ and $\rm pp$ interactions, and sum of the two in  $\rm Pbp$ interaction. The $\rm p_{T}^{J/\psi}$ distributions show that the color-octet and color-singlet channels have a similar peak  in small-$\rm p_{T}^{J/\psi}$ region at $\rm p_{T}^{J/\psi}\approx m_{Q}$.
From that point, they start decreasing and dropping down logarithmically. Indeed, at moderate transverse momentum $\rm p_{T}^{J/\psi}\geq m_{Q}$, the $\rm J/\psi $ photoproduction has much milder slope. The color-octet channels  $\rm ^{3}S_{1}^{[8]}$,  $\rm ^{3}P_{0}^{[8]}$, $\rm ^{3}P_{2}^{[8]}$ and the color-singlet channel  $\rm ^{3}S_{1}^{[1]}$ drop much faster than the color-octet channels $\rm ^{1}S_{0}^{[8]}$ and $\rm ^{3}P_{1}^{[8]}$ with the increase of $\rm p_{T}^{J/\psi}$. The slopes of $\rm ^{3}P_{1}^{[8]}$and $\rm ^{1}S_{0}^{[8]}$ are small  different from others. From our description, it has been observed that the $\rm p_{T}^{J/\psi }$  distribution decreases with $\rm p_{T}^{J/\psi}$ following a power law behavior proportional to $\rm 1/{(P_{T}^{J/\psi })^{n}}$ where $\rm n$ being energy dependent and different for each meson stands for the effective power. The color-octet channel  $\rm ^{1}S_{0}^{[8]}$ dominates over the other color-octet  and color- singlet contributions in small and large $\rm p_{T}^{J/\psi}$ regions which provides the main contribution to the $\rm p_{T}^{J/\psi}$ distribution of photoproduction processes, while the color-octet channel $\rm ^{3}S_{1}^{[8]}$ gives the low one. Certain processes for example diffraction  \cite{Butenschoen2010} and fragmentation \cite{Braaten1994,Rajesh2018} can also contribute to $\rm p_{T}^{J/\psi }$  distribution
 at small $\rm p_{T}^{J/\psi}$ and with very small total cross section, respectively. These two processes have not been taken into account and will be studied in our future work.

\begin{figure}[htp]
\centering
  % \begin{minipage}[t]{4.0cm}
   \includegraphics[height=4.8cm,width=4.4cm]{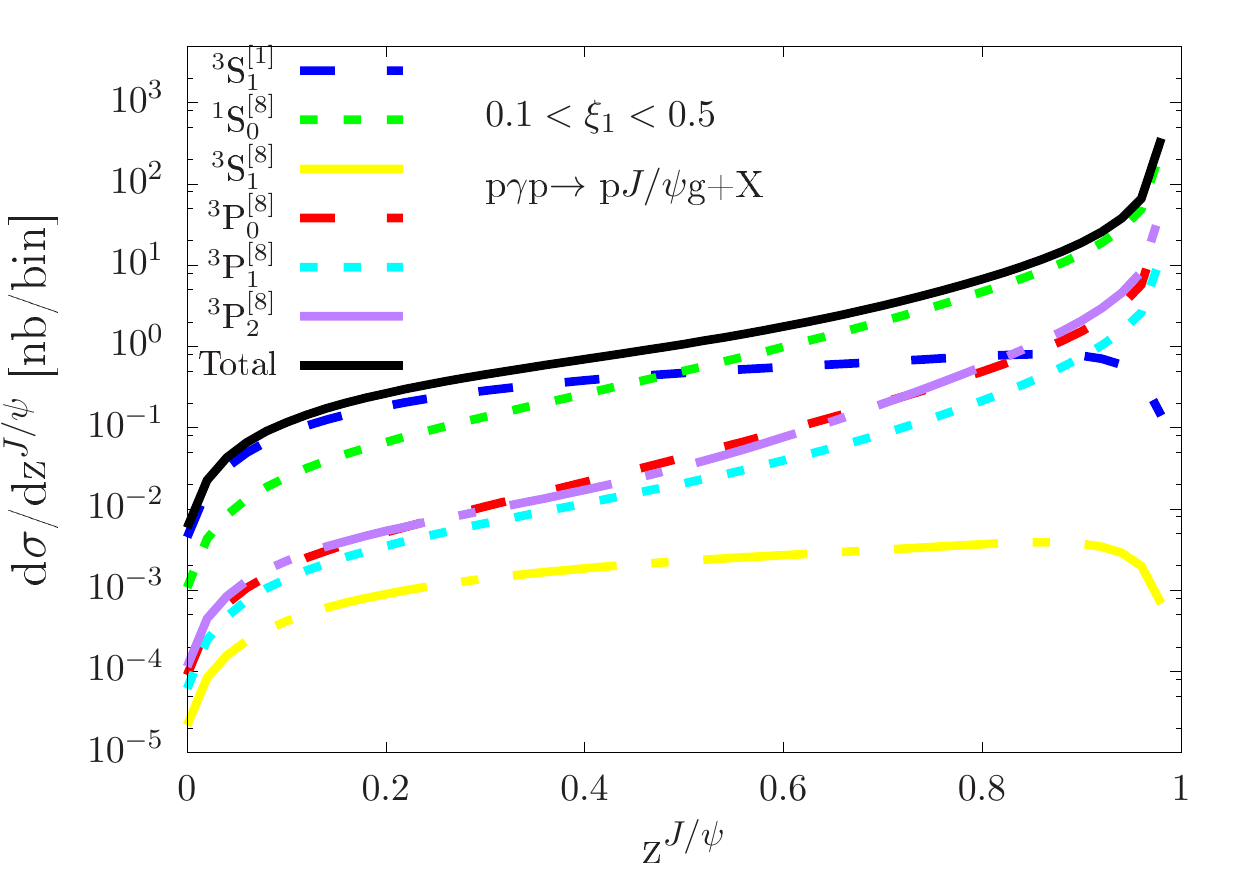}
   \includegraphics[height=4.8cm,width=4.4cm]{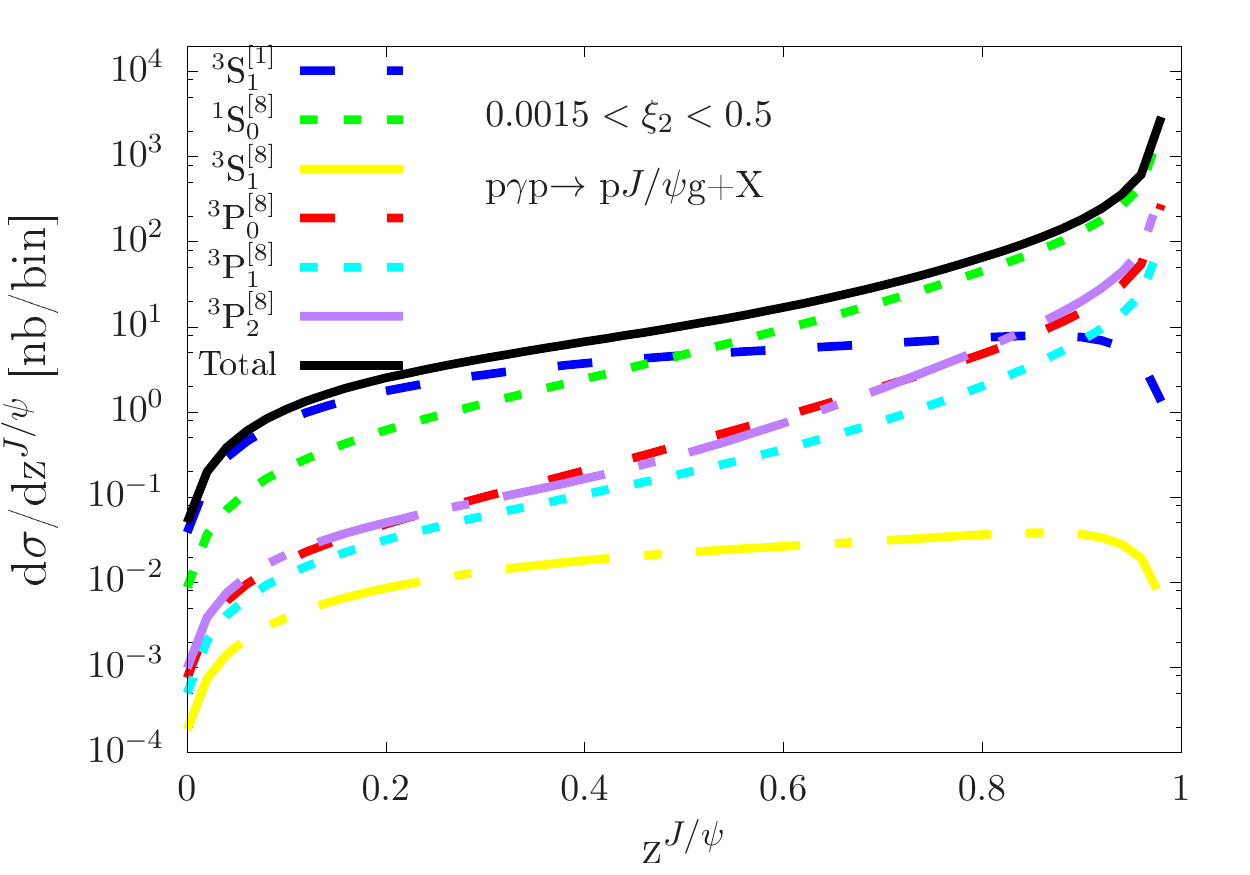}
   \includegraphics[height=4.8cm,width=4.4cm]{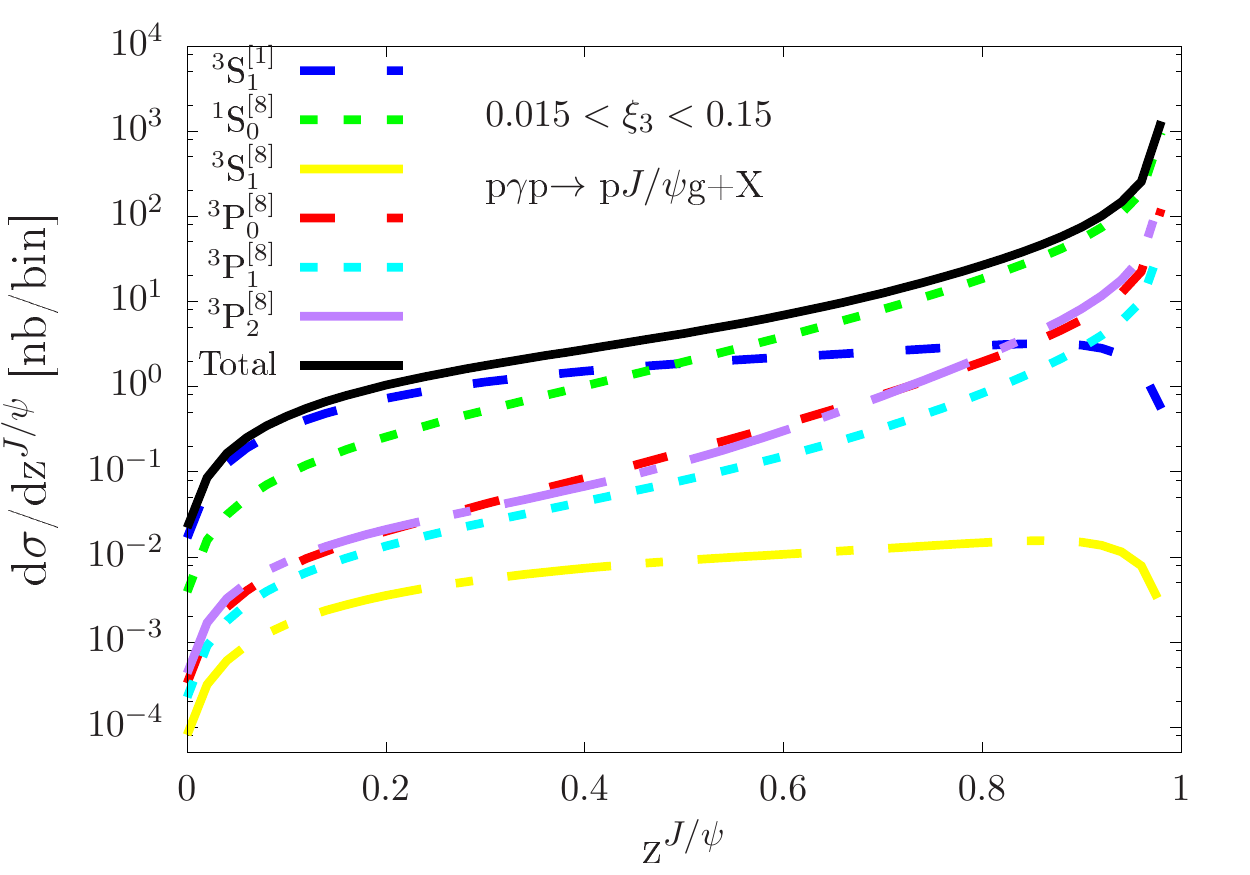}
   \includegraphics[height=4.8cm,width=4.4cm]{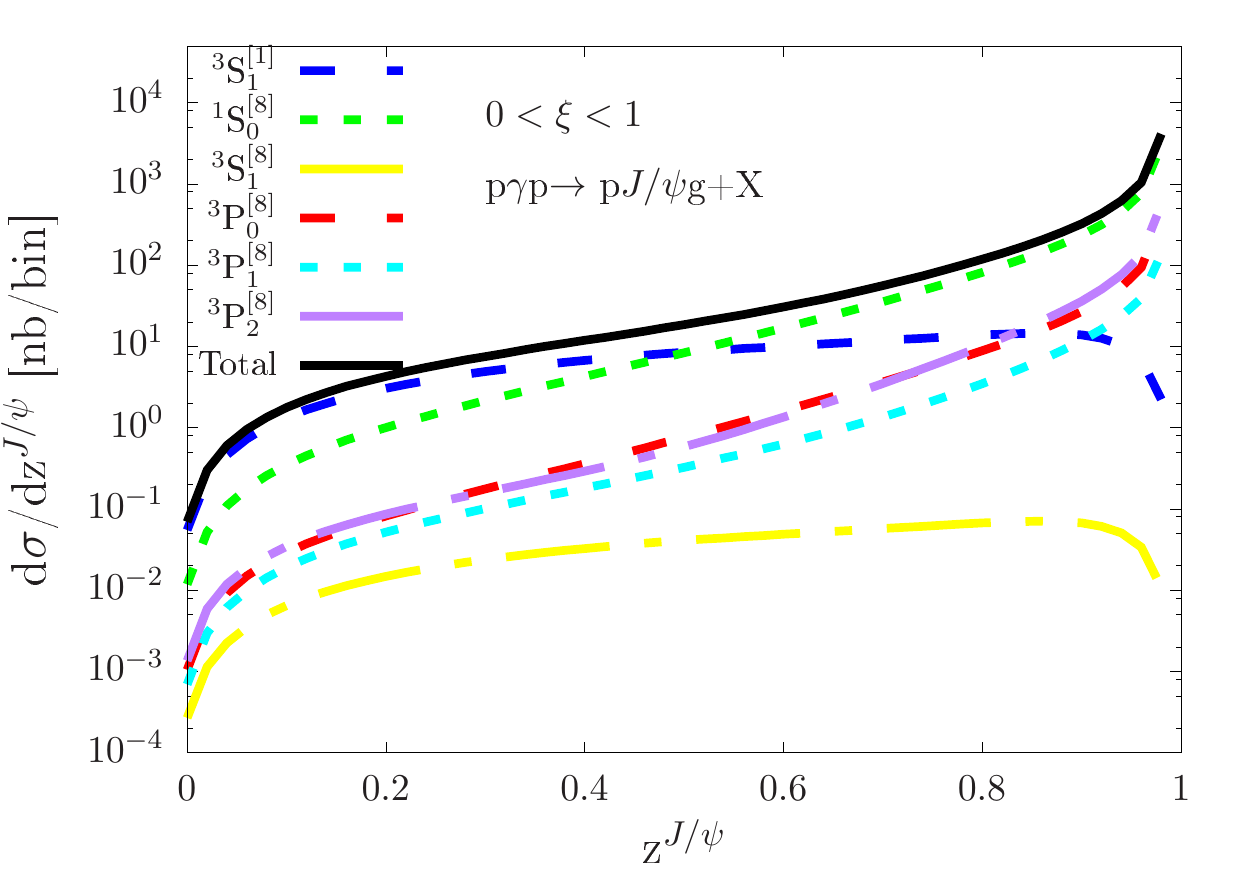}
   \includegraphics[height=4.8cm,width=4.4cm]{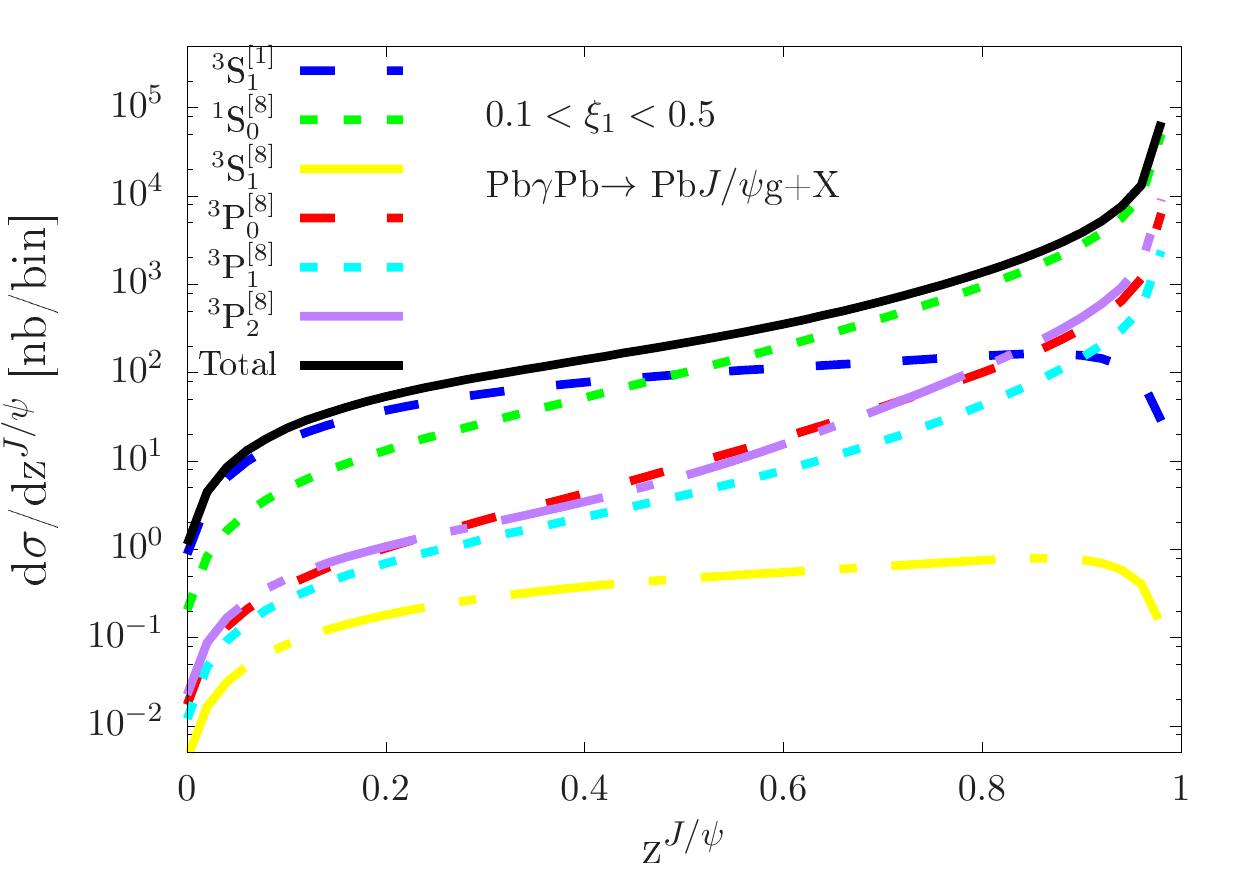}
   \includegraphics[height=4.8cm,width=4.4cm]{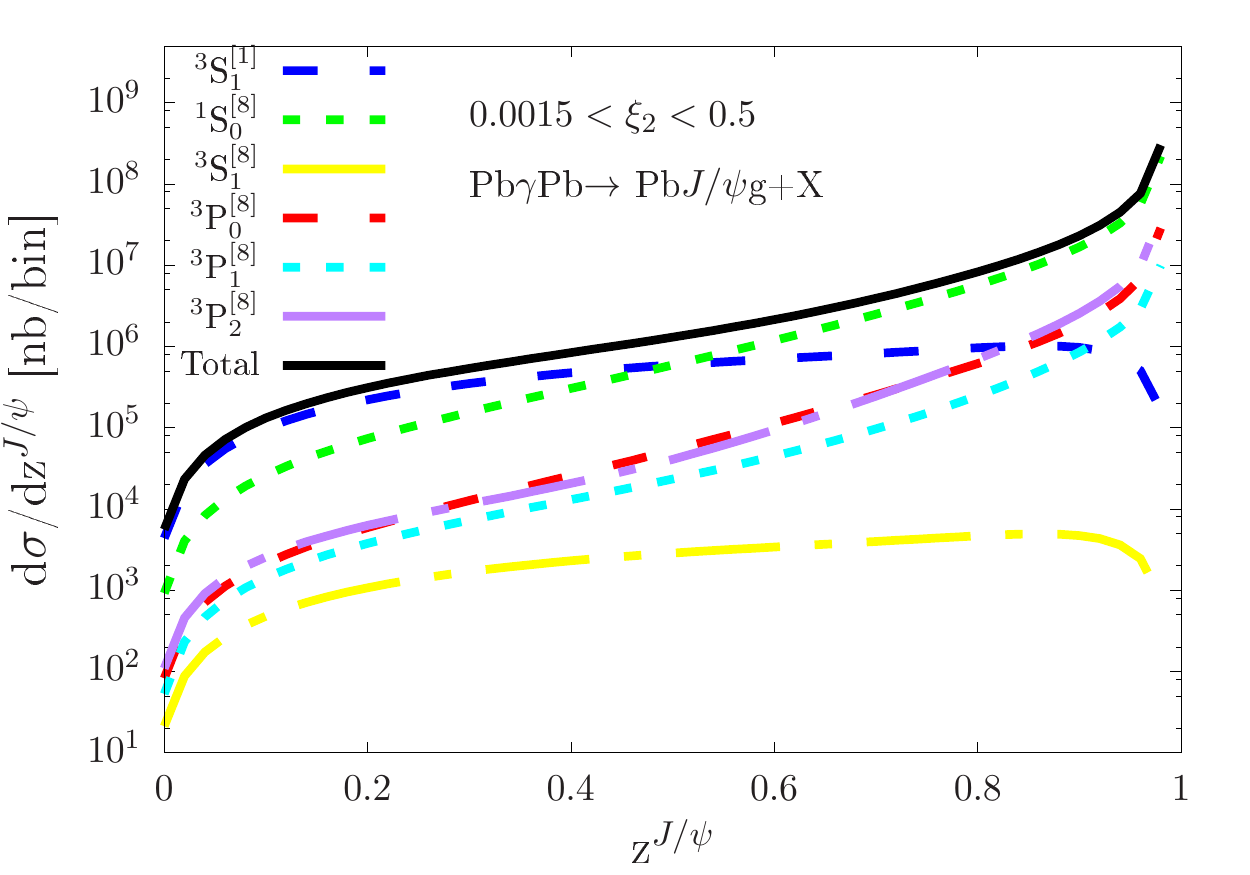}
   \includegraphics[height=4.8cm,width=4.4cm]{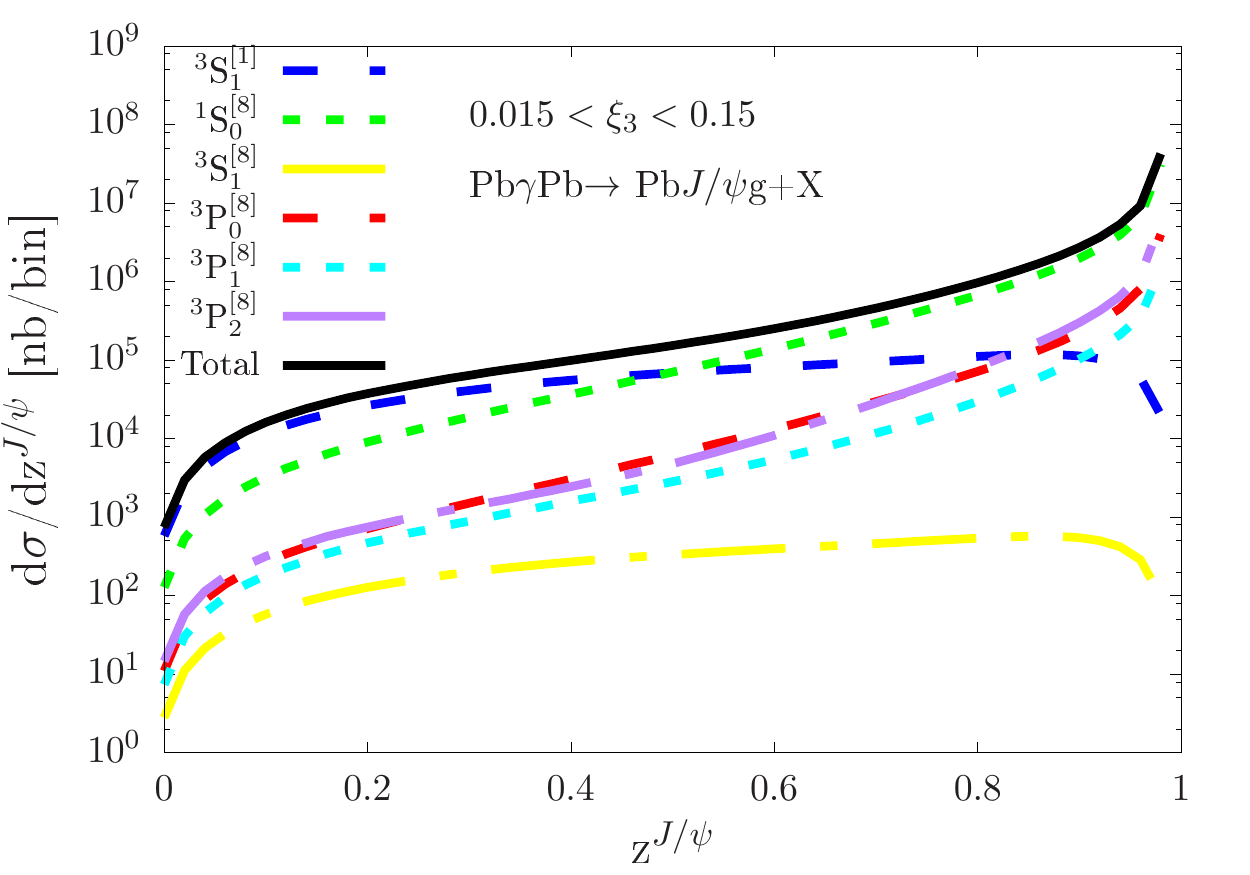}
   \includegraphics[height=4.8cm,width=4.4cm]{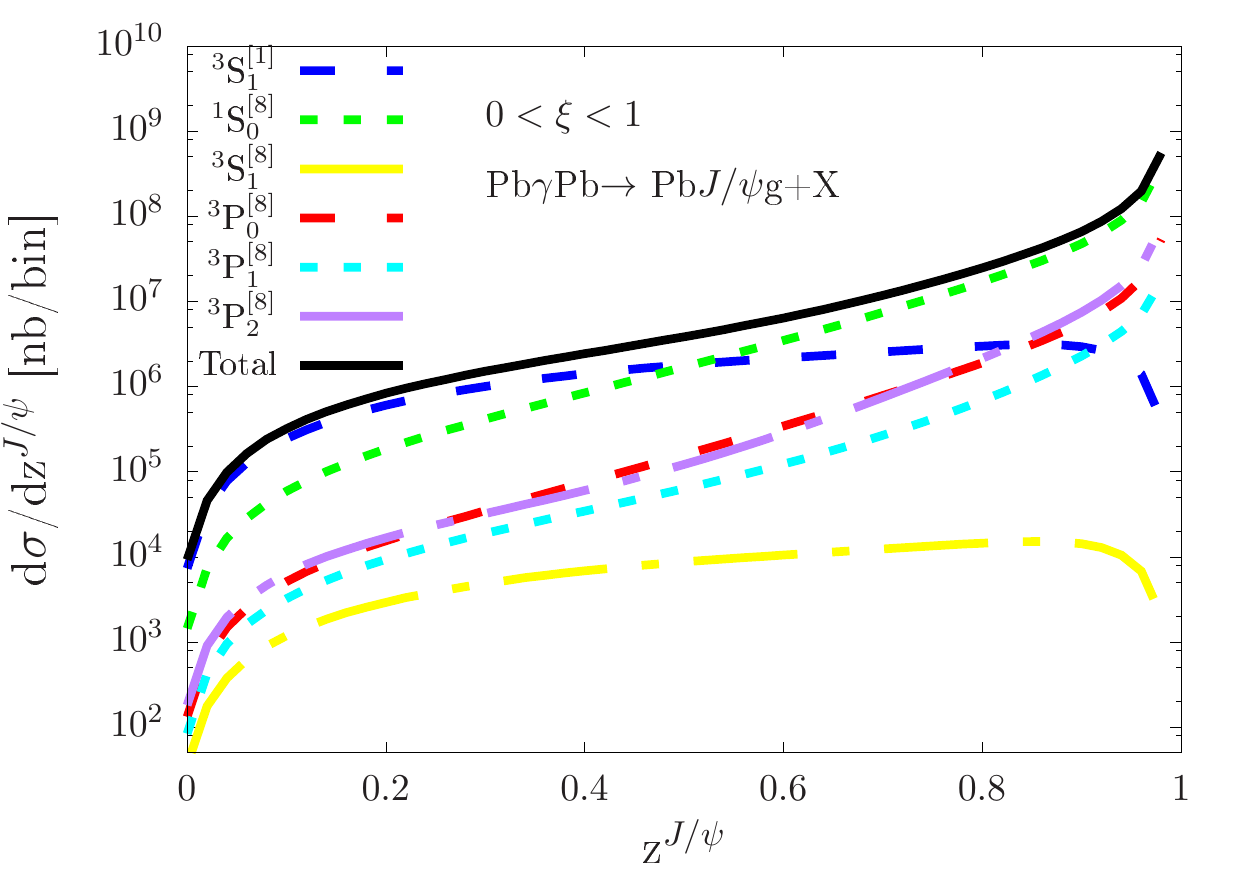}
   %\end{minipage}
   \caption{ \normalsize (color online) The $\rm z^{J/\psi }$ distributions  for the
   $\rm  p\gamma p\to p\mathcal{Q}g+X$  (top panel)  and $\rm Pb\gamma Pb\rightarrow Pb\mathcal{Q}g+X$  (bottom panel)  processes,
    and the contributions of the $\rm ^{3}S_{1}^{[1]}$ (blue dashed line),
   $\rm ^{1}S_{0}^{[8]}$ (green dotted line), $\rm ^{3}S_{1}^{[8]}$ (yellow dash dotted line),
   $\rm ^{3}P_{0}^{[8]}$ (red dashed line), $\rm ^{3}P_{1}^{[8]}$ (cyan dotted line), $\rm ^{3}P_{2}^{[8]}$ (purple dash dotted line) channels and
   Total (black solid line).}
\label{fig5:limits}
\end{figure}
\begin{figure}[htp]
\centering
  % \begin{minipage}[t]{4.0cm}
   \includegraphics[height=4.8cm,width=4.4cm]{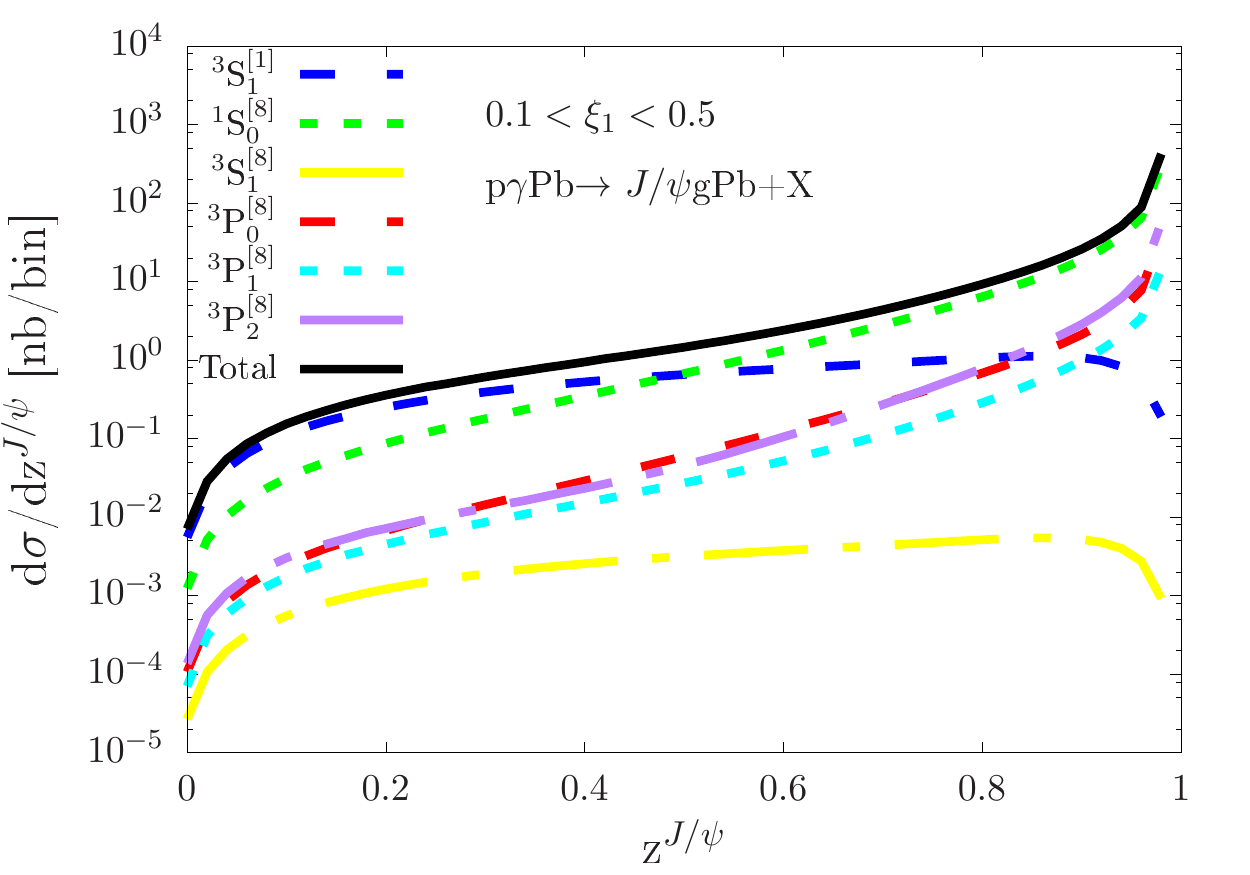}
   \includegraphics[height=4.8cm,width=4.4cm]{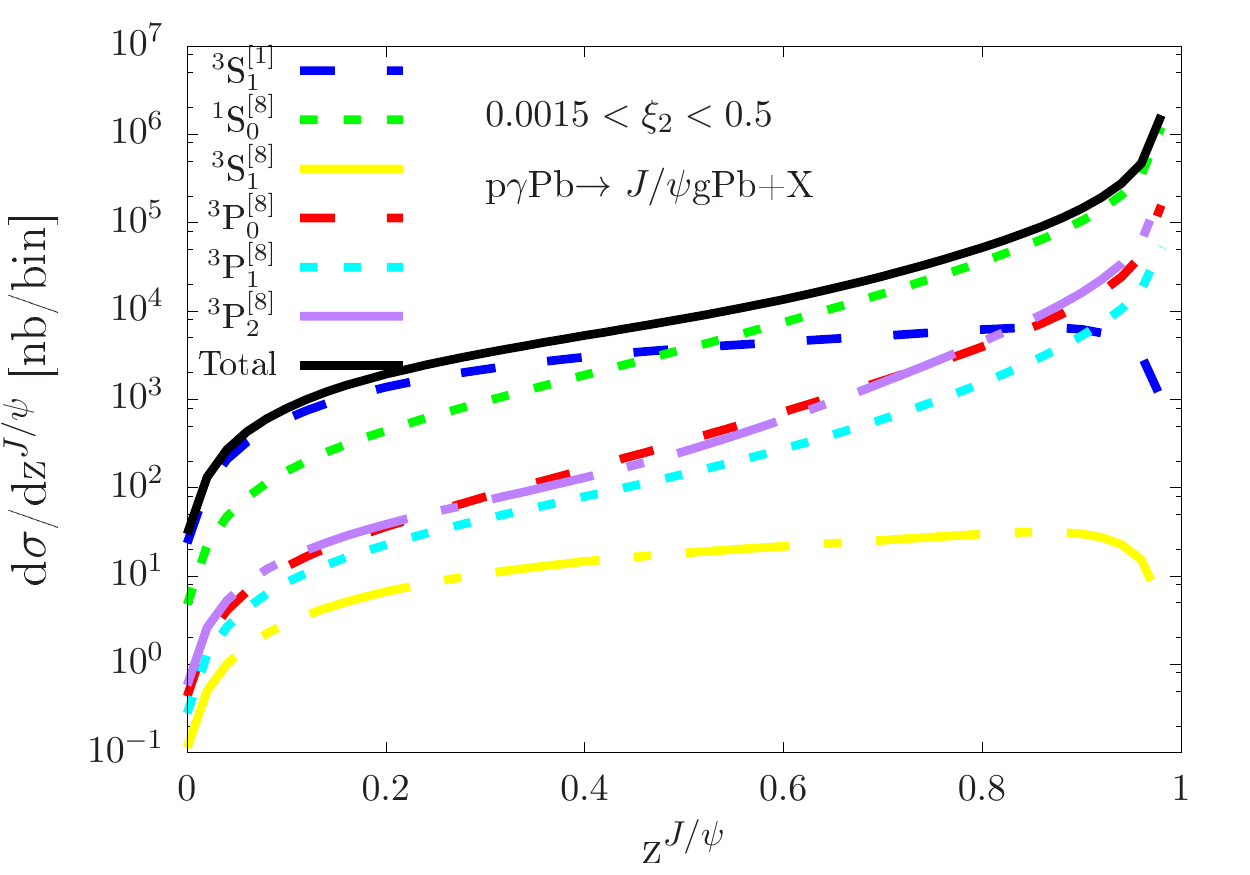}
   \includegraphics[height=4.8cm,width=4.4cm]{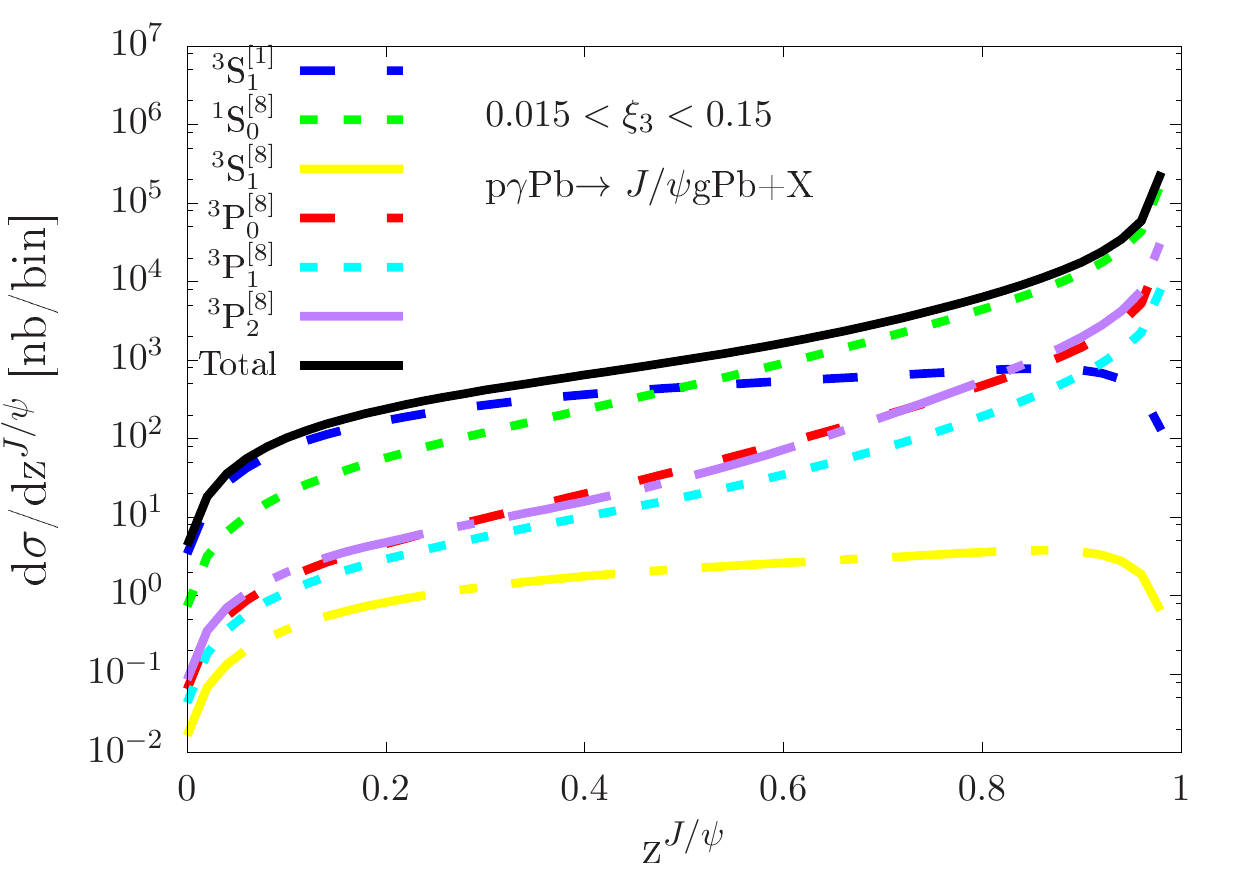}
   \includegraphics[height=4.8cm,width=4.4cm]{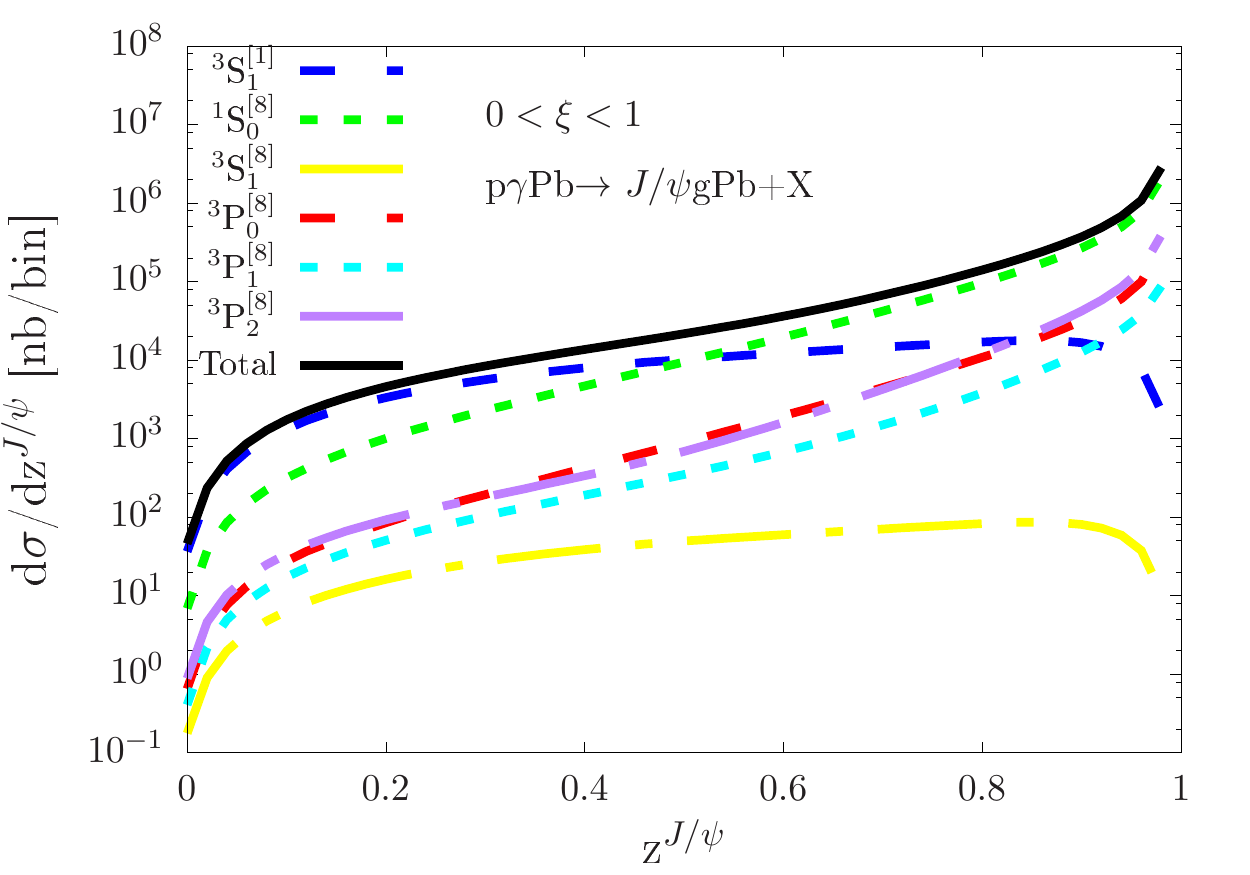}
   \includegraphics[height=4.8cm,width=4.4cm]{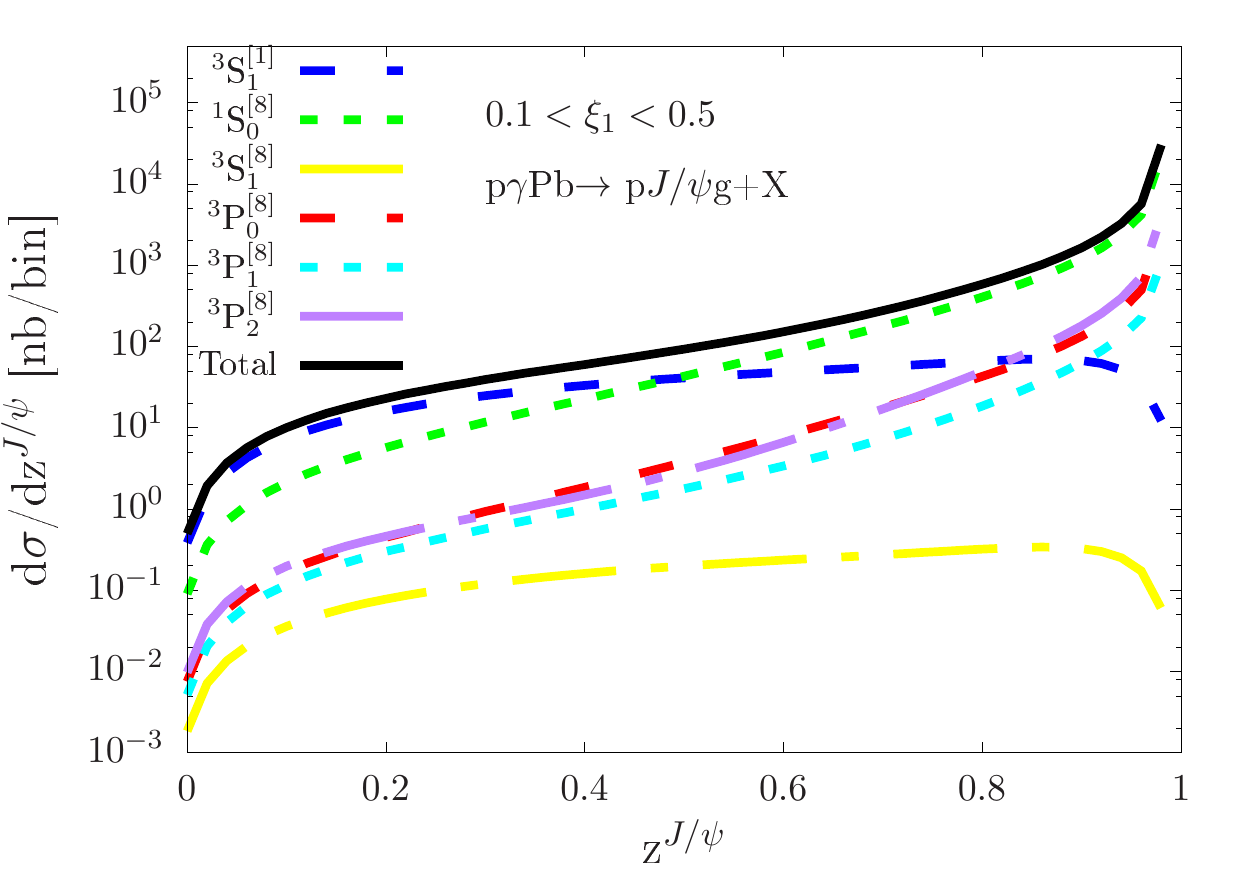}
   \includegraphics[height=4.8cm,width=4.4cm]{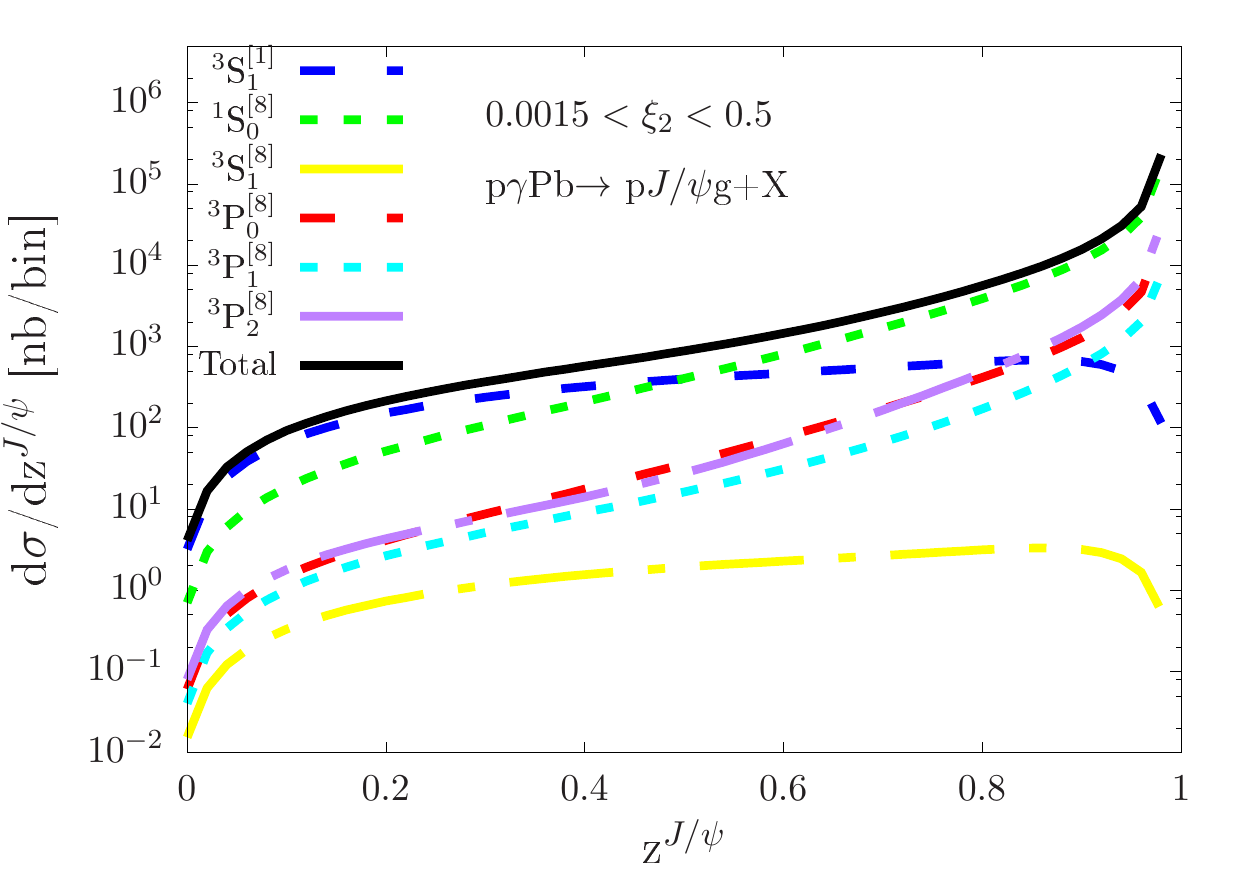}
   \includegraphics[height=4.8cm,width=4.4cm]{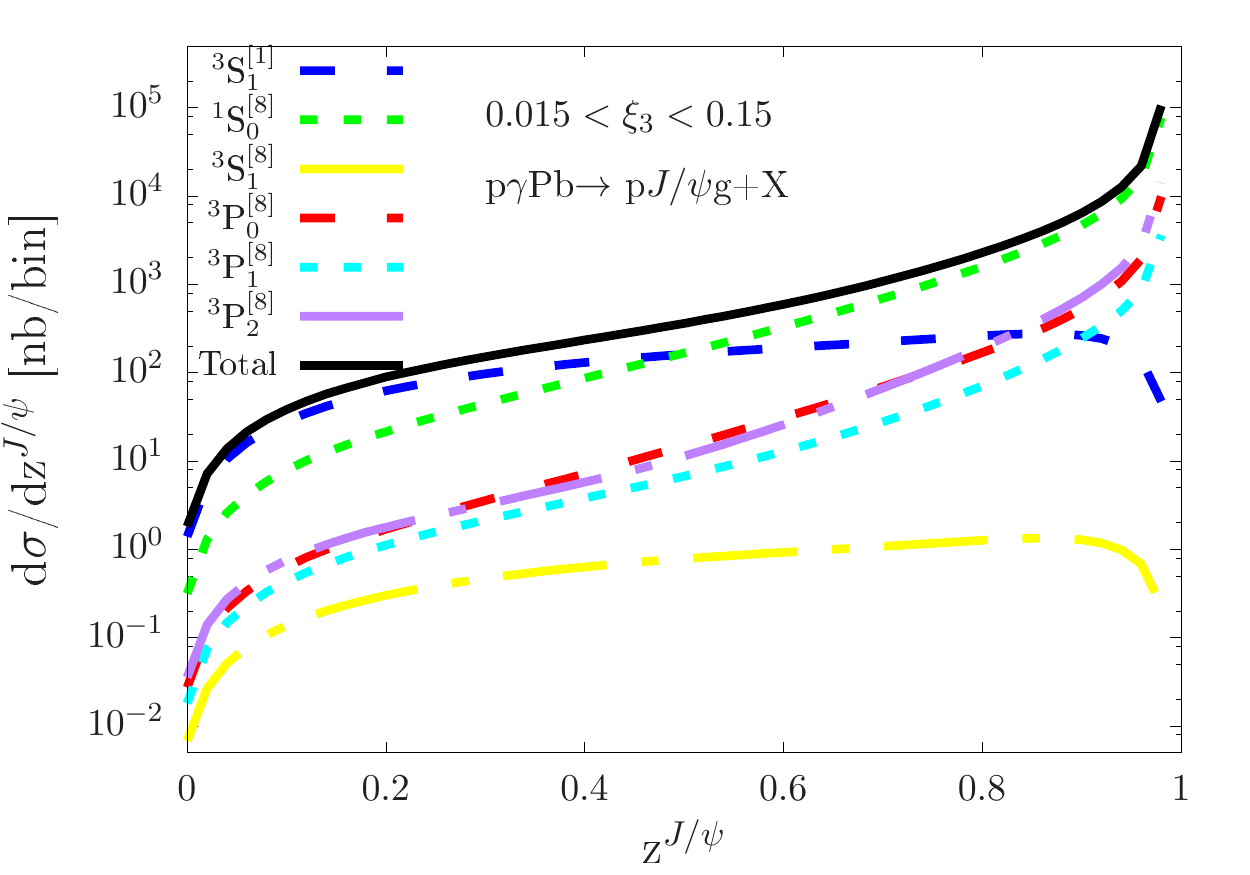}
   \includegraphics[height=4.8cm,width=4.4cm]{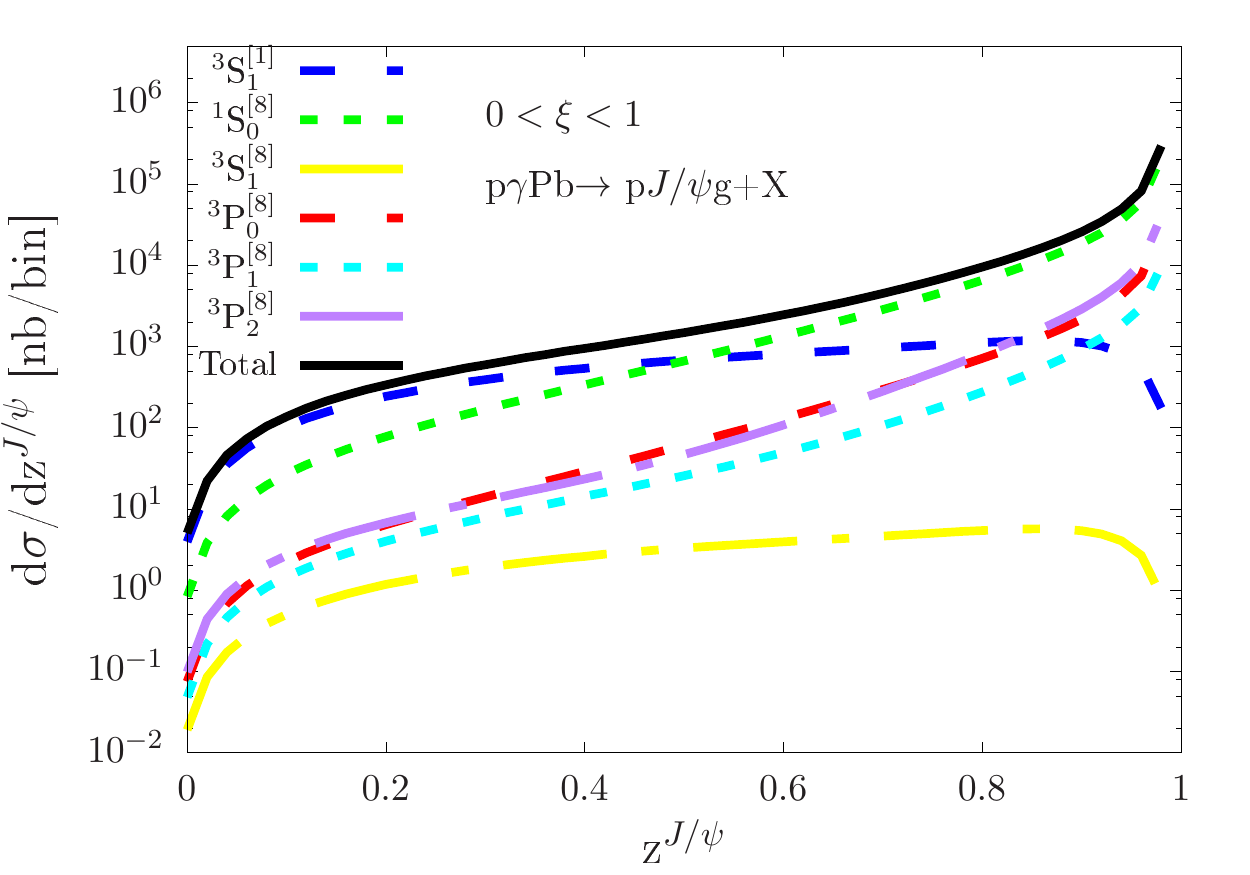}
   %\end{minipage}
   \caption{ \normalsize (color online) The $\rm z^{J/\psi }$ distributions for  the
    $\rm p\gamma Pb\to \mathcal{Q}gPb+X$  (top panel) and $\rm  p\gamma Pb\to p\mathcal{Q}g+X$  (bottom panel)  processes,
     and the contributions of the $\rm ^{3}S_{1}^{[1]}$ (blue dashed line),
   $\rm ^{1}S_{0}^{[8]}$ (green dotted line), $\rm ^{3}S_{1}^{[8]}$ (yellow dash dotted line),
   $\rm ^{3}P_{0}^{[8]}$ (red dashed line), $\rm ^{3}P_{1}^{[8]}$ (cyan dotted line), $\rm ^{3}P_{2}^{[8]}$ (purple dash dotted line) channels and
   Total (black solid line).}
\label{fig6:limits}
\end{figure}
The values of scaling variable $\rm z^{J/\psi}$ can be used for the separation of the elastic, inelastic and diffractive photoproduction events in $\rm z^{J/\psi}$  distributions. The 2$\to$1 subprocess cross section vanishes except $\rm z^{J/\psi}=1$ and contributes to $\rm J/\psi$ photoproduction in forward direction which means that the $\rm pp$, $\rm PbPb$, $\rm Pb p$ collisions in 2$\to$1 subprocess don't have the total contribution in the region $0<\rm z^{J/\psi}<0.9$ and nearly vertically increase at the endpoint $\rm z^{J/\psi}=1$.
However, the total contributions for 2$\to$2 subprocess increase along with $\rm z^{J/\psi}$ in the region $0<\rm z^{J/\psi}<0.9$ and drastically increase obliquely for $0.9<\rm z^{J/\psi}<1$. Here, the forward distribution is only regarded and the $\rm Pbp$ collision is divided into two. The total $\rm z^{J/\psi}$  distribution will be twofold larger in $\rm pp$ and $\rm PbPb$ collisions and the sum of the two in $\rm Pb p$ collisions if the backward distribution is involved. In Figs. \ref{fig5:limits} and \ref{fig6:limits}, we present our prediction for $\rm z^{J/\psi}$ distributions and it has been observed that the color-singlet channel $\rm ^{3}S_{1}^{[1]}$ and the color octet channel  $\rm ^{3}S_{1}^{[8]}$ begin to increase lightly and go down logarithmically  with the increase of  $\rm z^{J/\psi }$. The other remaining color octet channels also begin to increase lightly and drastically going up logarithmically along with  $\rm z^{J/\psi }$. The color-singlet channel gives the main contribution  in the  region of  $\rm 0 < z^{J/\psi }< 0.5$ while the color octet channel $\rm ^{1}S_{0}^{[8]}$ provides the main one in the region of  $\rm   z^{J/\psi }>0.5$.  The low contribution  to  the  $\rm z^{J/\psi}$  distribution comes from the color octet channel $\rm ^{3}S_{1}^{[8]}$ as $\rm z^{J/\psi}$ enlarges.
In our work, we have only considered the direct photon process produced  from proton or lead in the inelastic process, and our $\rm J/\psi $ photoproduction is direct which means that the quarkonium is directly produced in the hard scattering. The feeddown is one of the contributors to the $\rm J/\psi $ spectrum which is expected to originate from
higher chamonium states as well as B meson decays \cite{LindenLevy2009}. Some processes for instance the single and  double resolved $\rm J/\psi $ photoproduction channel can also contribute significantly at small $\rm z^{J/\psi }$ \cite{Kniehl1999,Klasen1998,Godbole2002}, and  color singlet model  is dominant over the photon-gluon fusion \cite{Ko1996} as in the direct photon process, whereas the elastic/diffractive process  gives contribution to the $\rm J/\psi $ photoproduction at large $\rm z^{J/\psi }$\cite{Butenschoen2010}. The contributions of resolved photoproduction and the diffractive production are in general excluded from experimental measurements due to the unavailability of the data, in order to make a significant comparison \cite{Aaron2010}. The feed-down contribution (15\%)  from an excited state $\rm \psi (2S)$ \cite{Chekanov2003} and decay contribution  (1\%) from $\chi_{c_{J}}$ states \cite{Butenschoen2010,Artoisenet2009} to $\rm J/\psi$ photoproduction are not included in our current work. The light quark contribution is expected to be negligible compared to the photon-gluon fusion process \cite{Ko1996}. The $\rm J/\psi$ photoproduction cross section to NLO confirmed that the corrections are considerable, enlarging towards large transverse momentum of $\rm J/\psi$ meson \cite{Kramer1996, Kramer1995, Artoisenet2009, Aaron2010}. For example, the NRQCD cross section of the $\rm J/\psi$ photoproduction to LO can be improved by next to leading order (NLO) corrections up to 115\% in the considered kinematic in electron proton collisions through $ \rm \gamma g$ fusion\cite{Butenschoen2010}. We have commented that NLO has been vital and will need to be considered in future works.

\begin{figure*}[htp]
\centering
   %\begin{minipage}[t]{4.0cm}
   \includegraphics[height=4.4cm,width=4.4cm,angle=0]{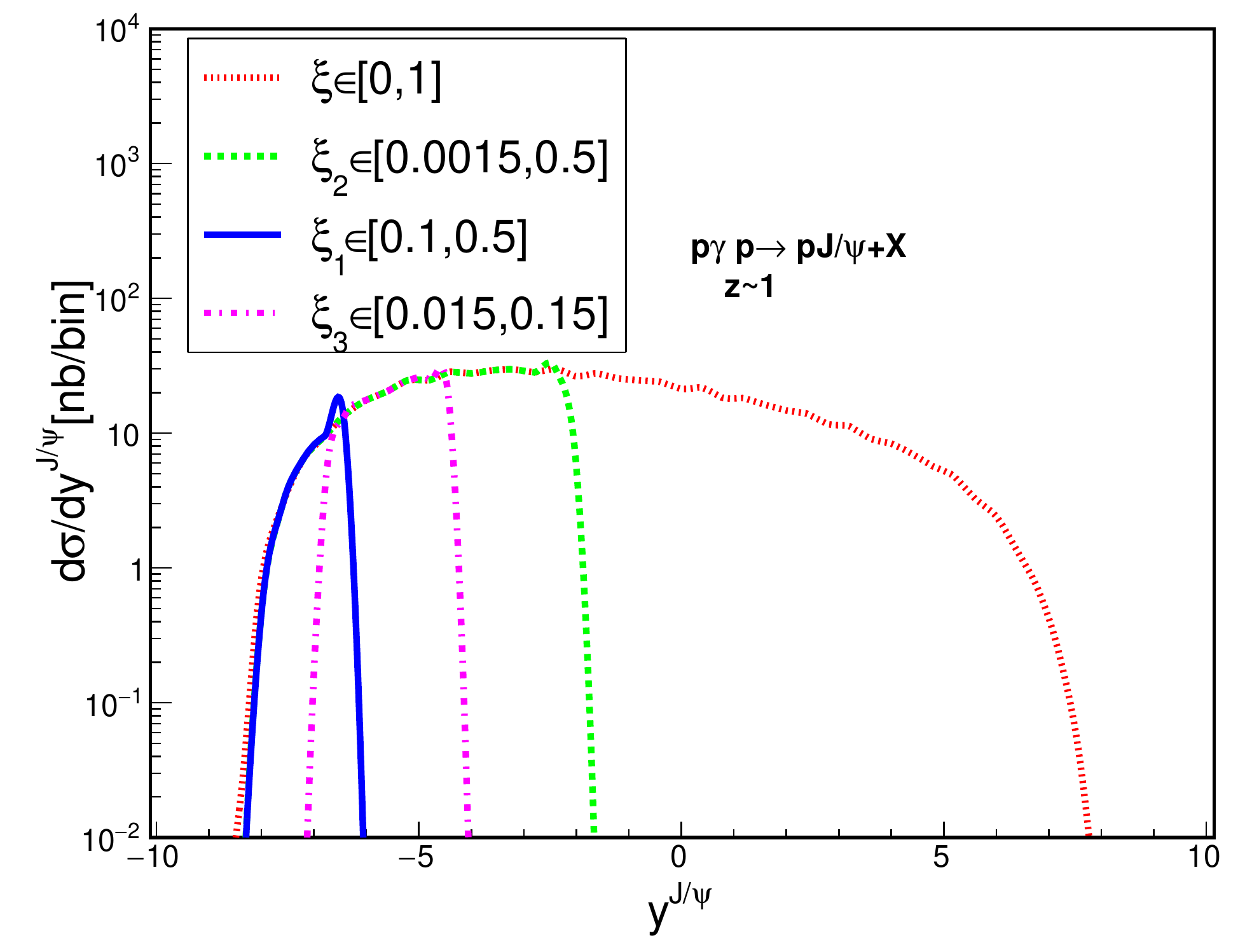}
   \includegraphics[height=4.4cm,width=4.4cm,angle=0]{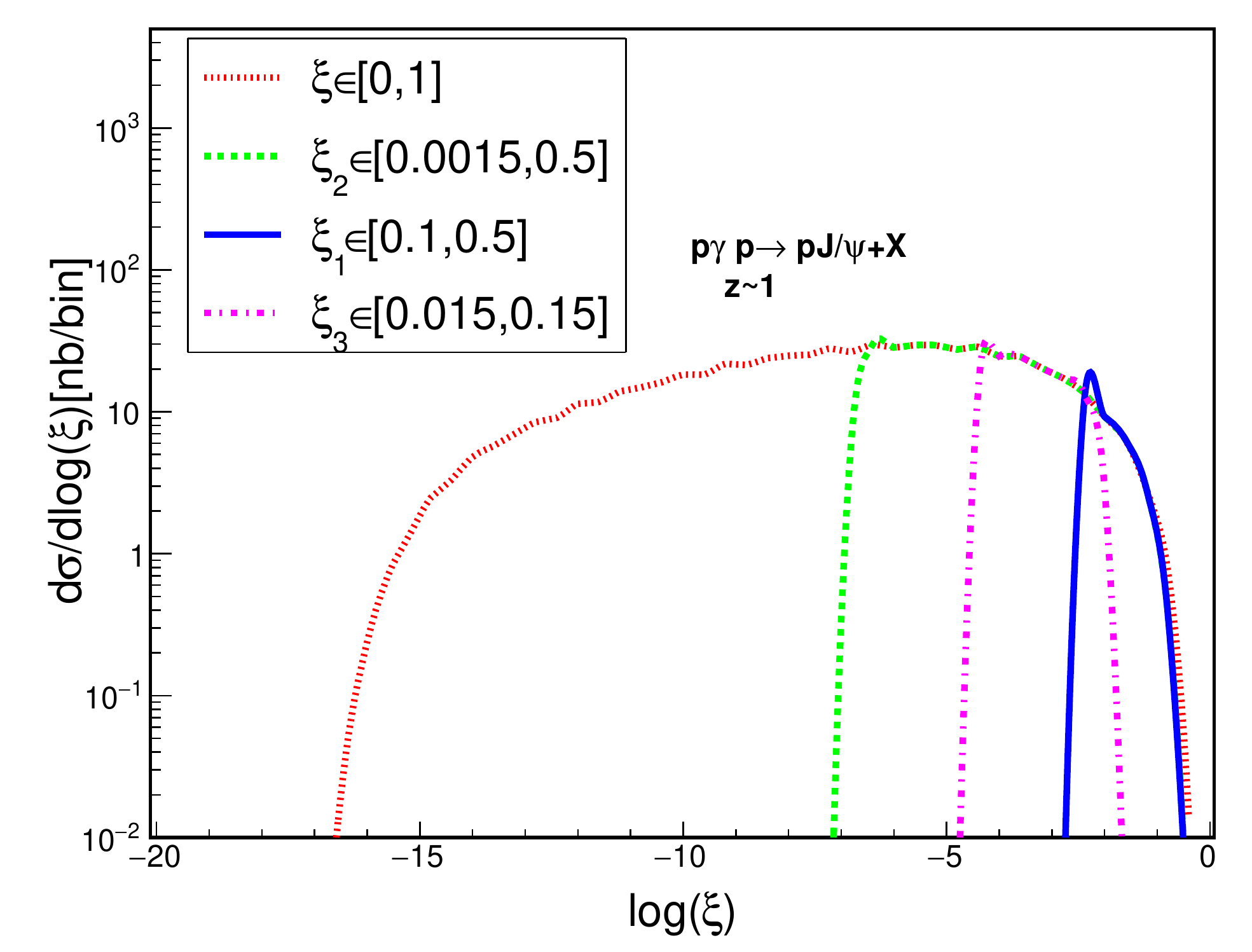}
   \includegraphics[height=4.4cm,width=4.4cm,angle=0]{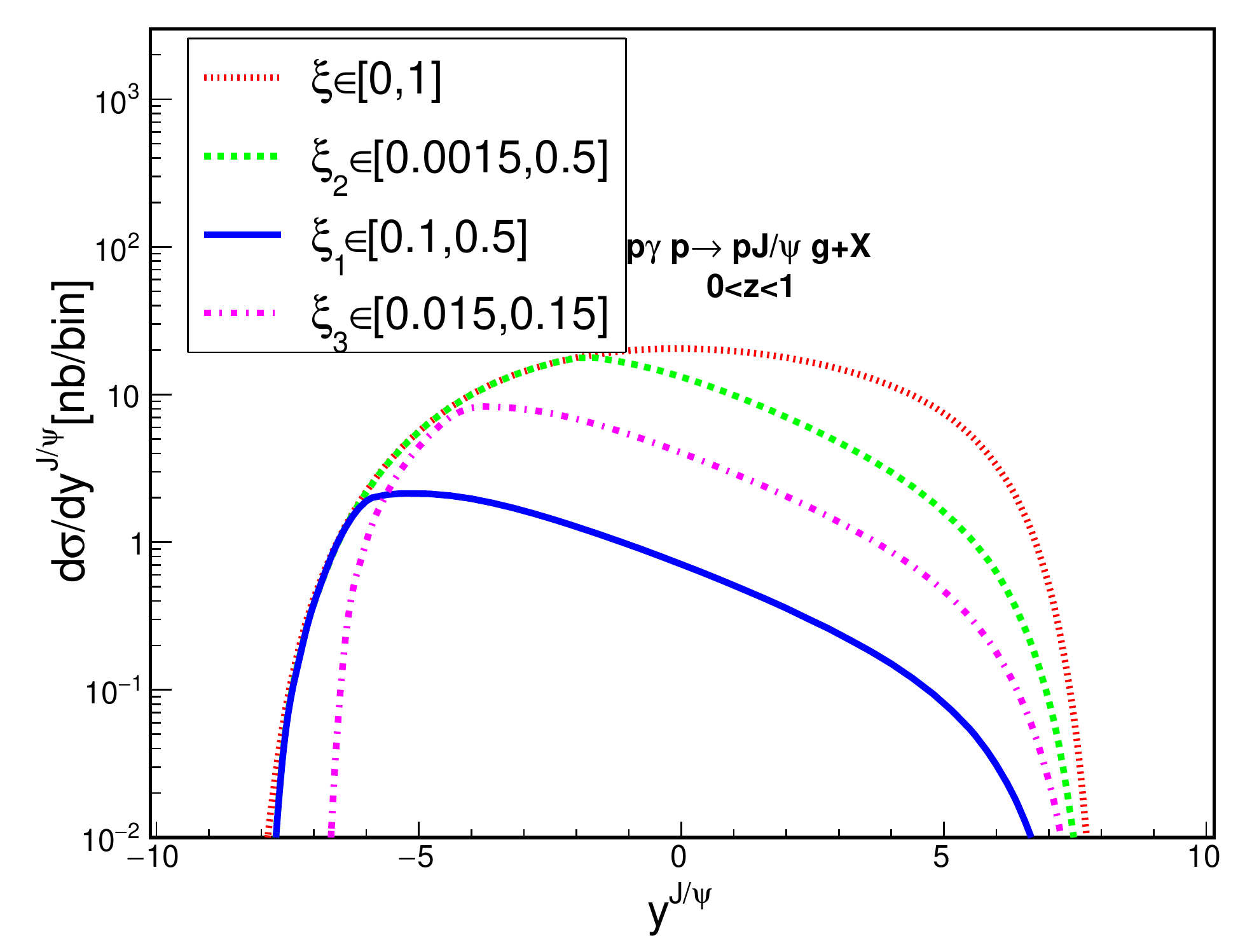}
   \includegraphics[height=4.4cm,width=4.4cm,angle=0]{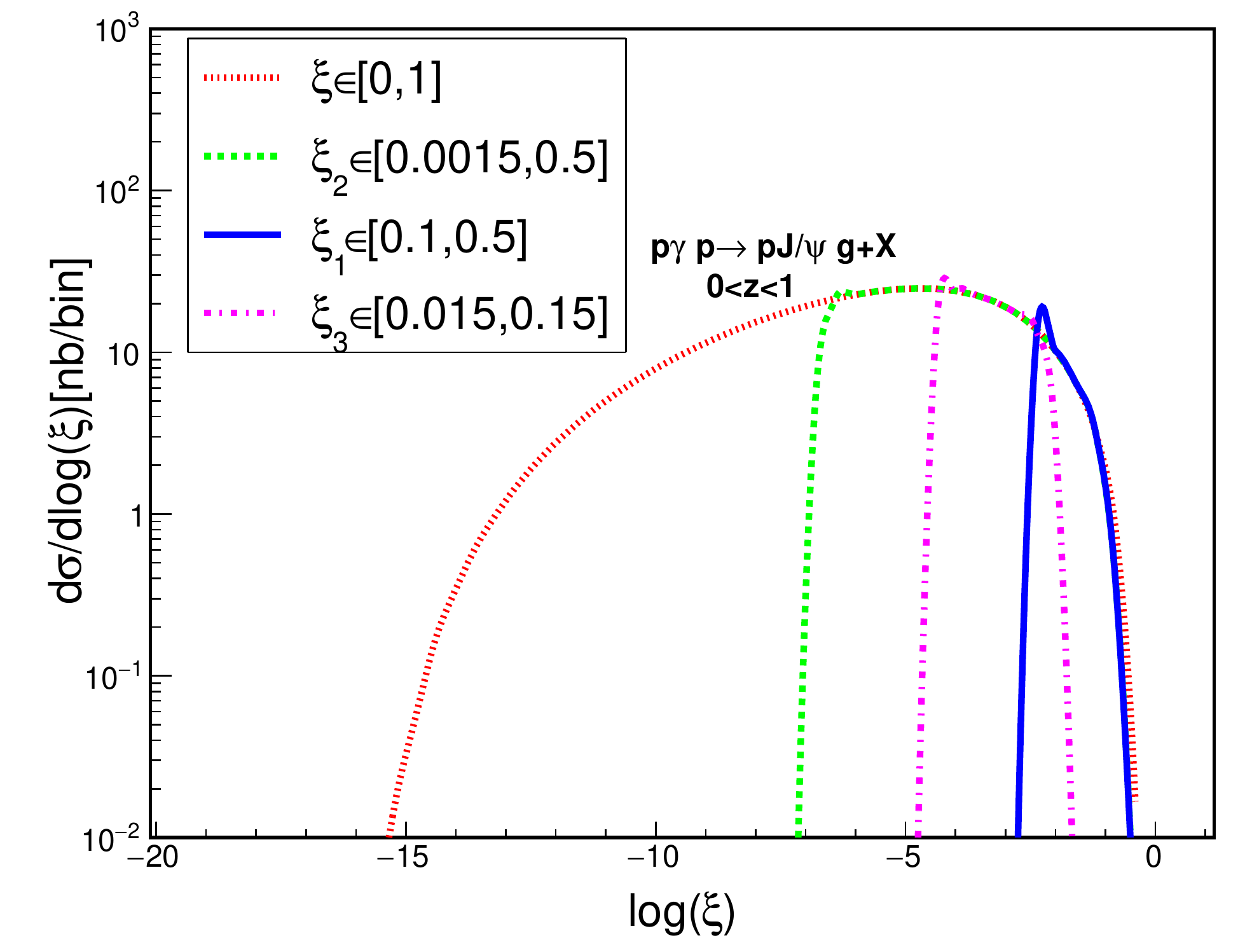}
   \includegraphics[height=4.4cm,width=4.4cm,angle=0]{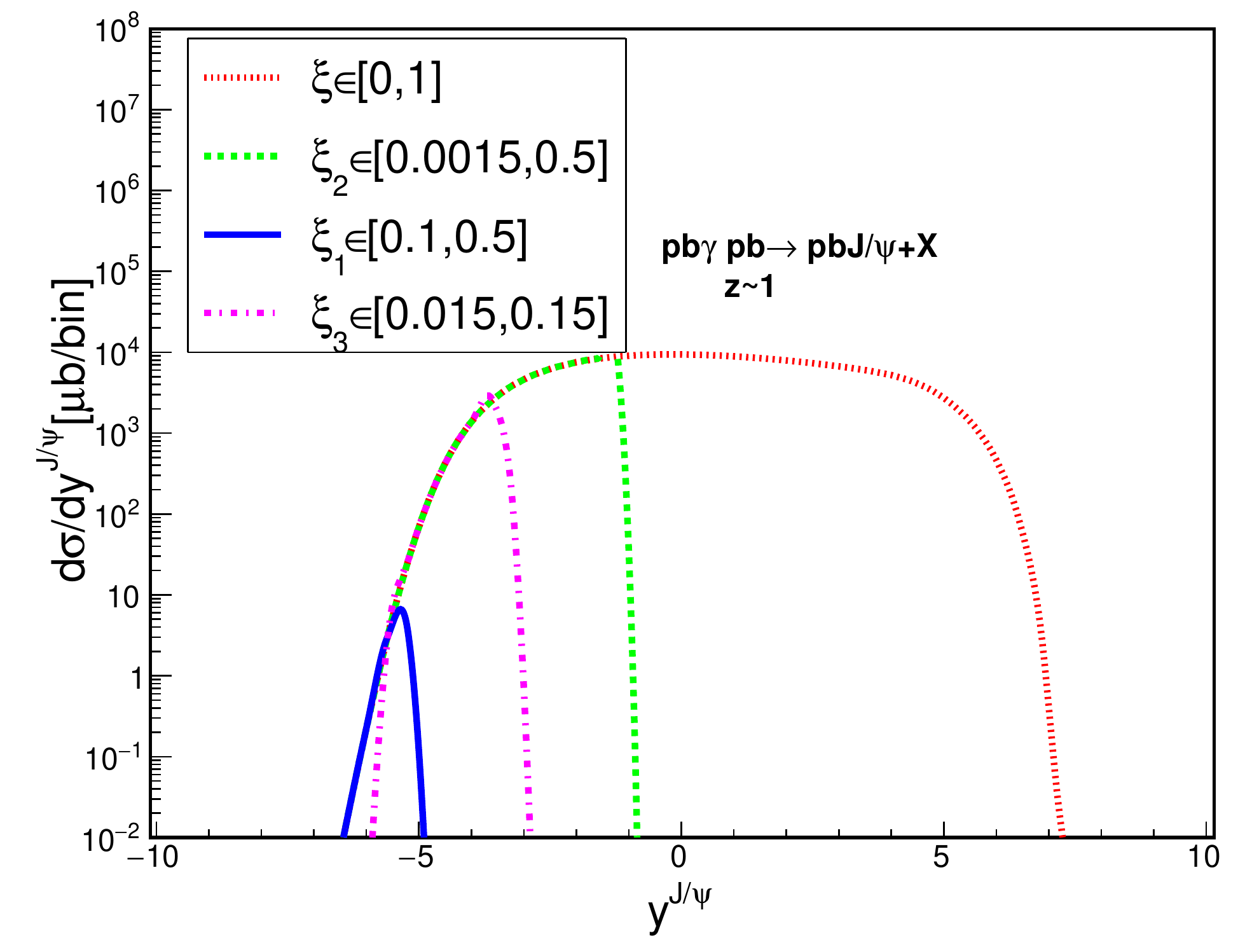}
   \includegraphics[height=4.4cm,width=4.4cm,angle=0]{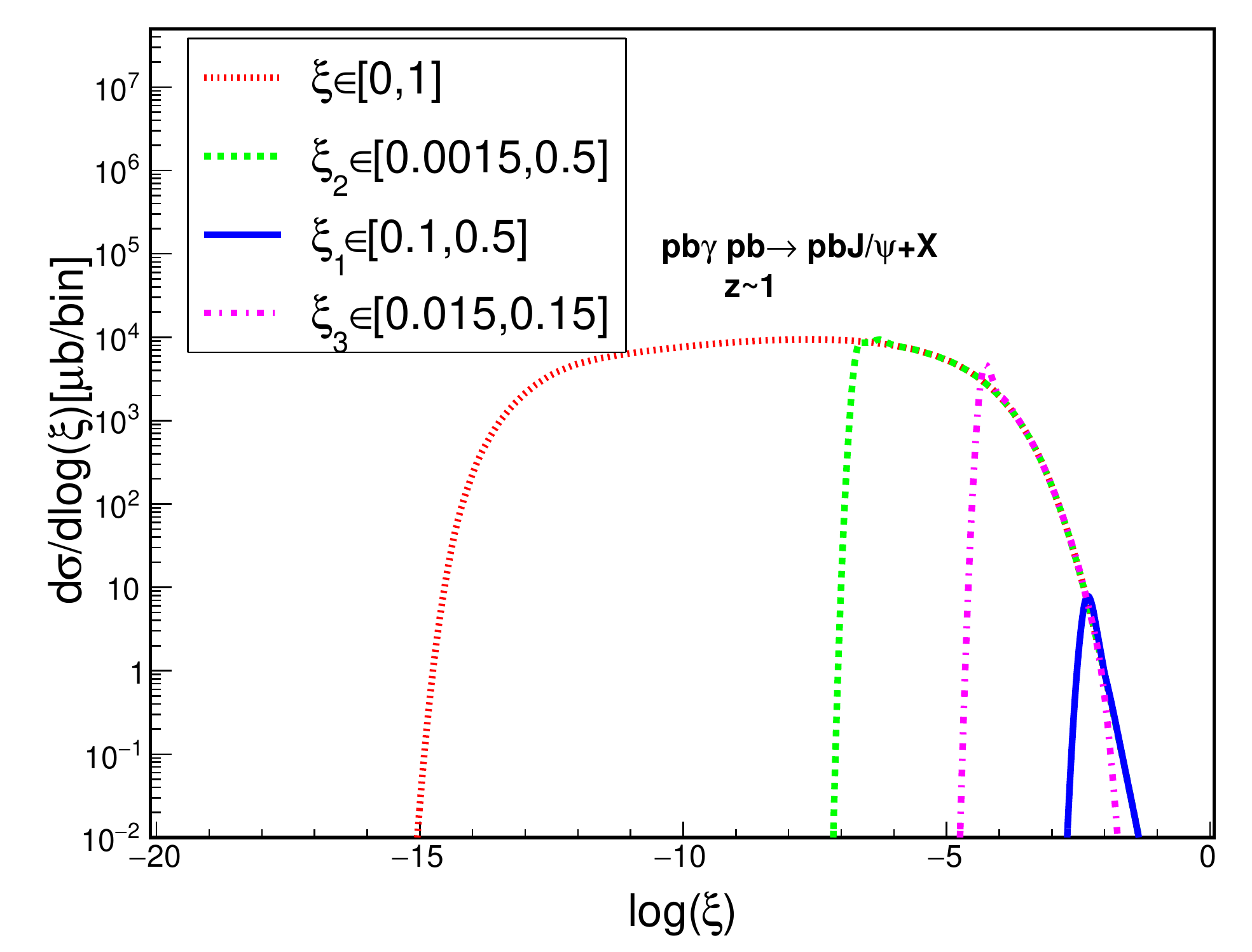}
   \includegraphics[height=4.4cm,width=4.4cm,angle=0]{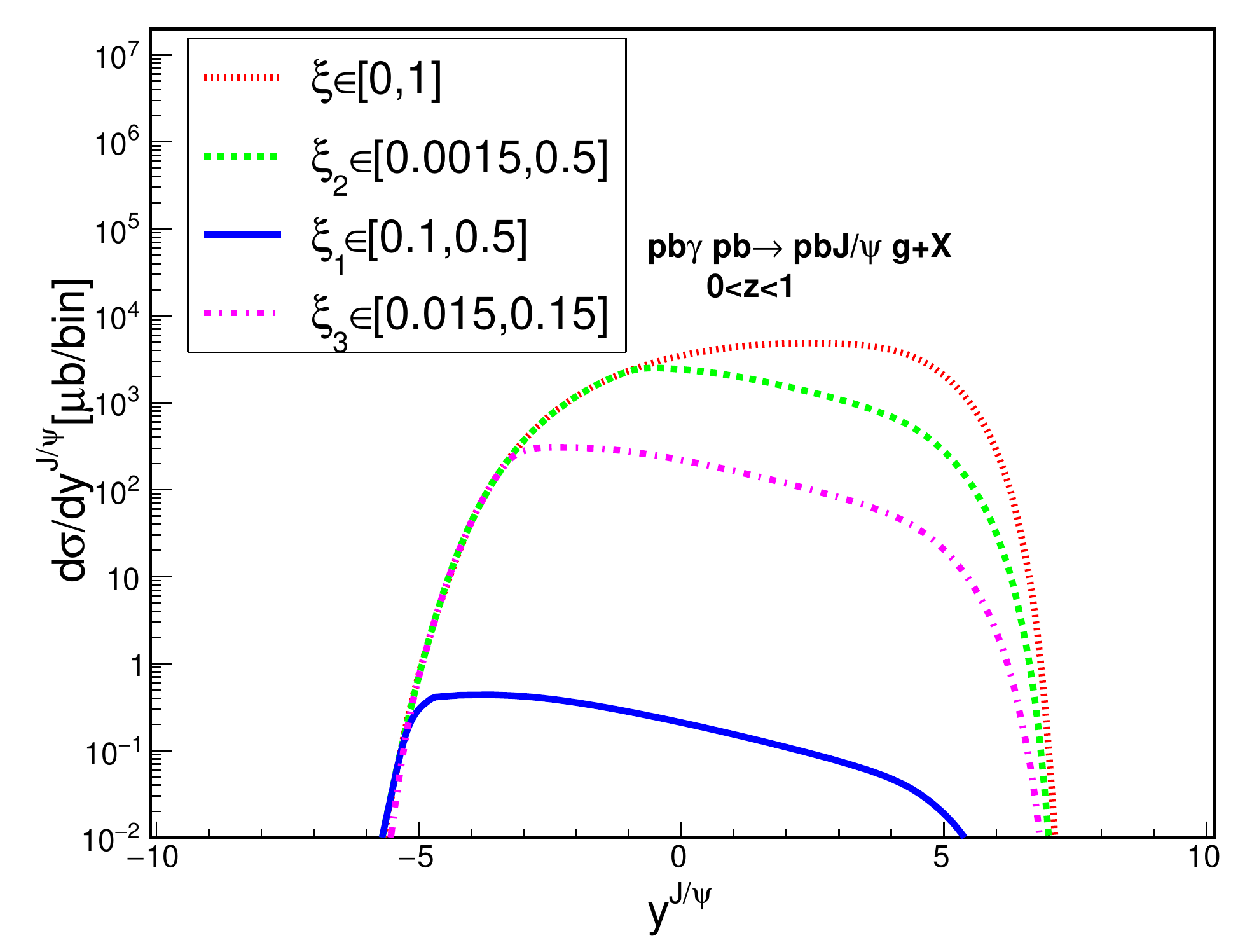}
   \includegraphics[height=4.4cm,width=4.4cm,angle=0]{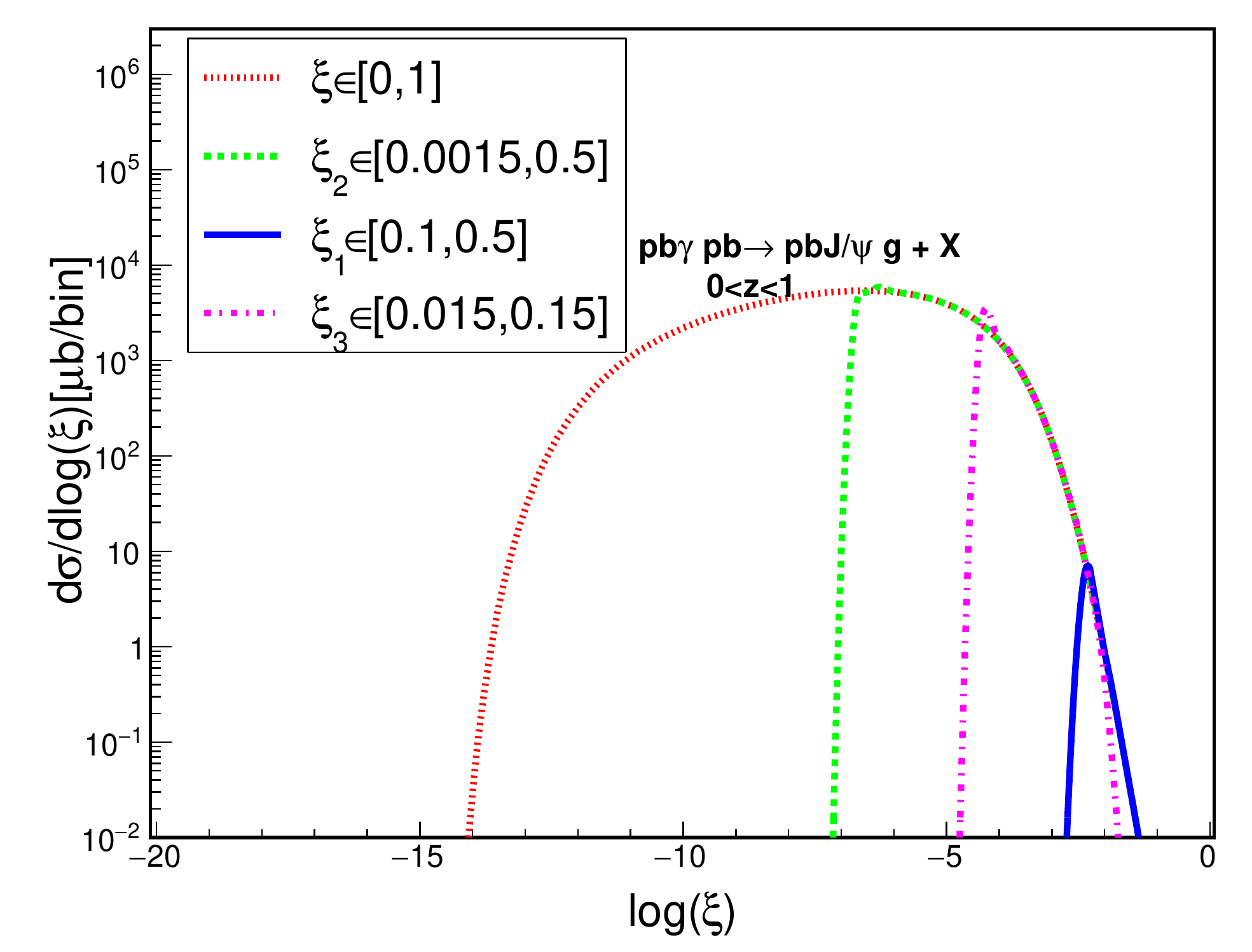}
   %\end{minipage}
   \caption{ \normalsize
  (color online)  The  $\rm y^{J/\psi}$ and $\rm \log \xi$ distributions
  for the $\rm p\gamma p\to pJ/\psi +X$ and $\rm p\gamma p\to pJ/\psi g+X$  processes (top panels)
  and $\rm Pb\gamma Pb\to PbJ/\psi +X$  and $\rm Pb\gamma Pb\to Pb J/\psi g +X$ processes (bottom panels)
  for the forward detector with $\xi$ (red dashed line), at the CMS-TOTEM forward detector with $\xi_{1}$ (green dashed line),
  at the CMS-TOTEM forward detector with $\xi_{2}$ (blue solid line), and
  at  the AFP-ATLAS forward detector with $\xi_{3}$ (magenta dash dotted line).}
\label{fig7:limits}
\end{figure*}
In Fig.\ref{fig7:limits}, we exhibit our predictions of the $\rm y^{J/\psi }$ and $\rm \log\xi$ distributions for the inelastic $\rm J/\psi$ photoproduction in coherent $\rm pp$ and $\rm PbPb$ collisions at LHC energies for four different forward detector acceptances.
The two first figures on top (bottom) left panel and two last figures on top (bottom) right panel represent the $\rm y^{J/\psi}$ and  $\rm \log\xi$ distributions for $\rm pp$ ($\rm PbPb$) collision in 2$\to$1 and 2$\to$2 subprocesses. Given the photon flux and colliding energy, the study of the $\rm J/\psi$ photoproduction distributions can be employed to constrain the photoproduction cross section.
In the case of $\rm pp$ and $\rm PbPb$ collisions, both incident protons and leads are sources of photons providing equally at forward and backward rapidities, respectively. The forward rapidity is defined by the negative $\rm \phi^{J/\psi}$ and the backward one  by the positive $\rm \phi^{J/\psi}$.  As a result, the total  $\rm y^{J/\psi }$ and $\rm \log\xi$ distributions at forward and backward are symmetric about midrapidity. The forward contribution is only taken into the consideration while total contributions would be two times larger. In the figures on the top (bottom) panels, the $\rm y^{J/\psi}$ and $\rm \log\xi$ distributions for $\rm pp$ ($\rm PbPb$) collision are computated considering the parametrization for the gluon distribution in the proton (lead). we observe  that the shapes of $\rm y^{J/\psi}$ and $\rm \log\xi$ distributions of 2$\to$1 subprocess are similar and symmetric. Contrariwise, the shapes of $\rm y^{J/\psi}$ and $\rm \log\xi$ distributions of 2$\to$2 subprocess are dissimilar and asymmetric due to the emission of gluon in the final state. The $\rm pp$ and $\rm PbPb$ contributions peak for the negative rapidities as shown in our figures due the the flux of photon in forward rapidities.

\begin{figure*}[htp]
\centering
   %\begin{minipage}[t]{4.0cm}
   \includegraphics[height=4.4cm,width=4.4cm,angle=0]{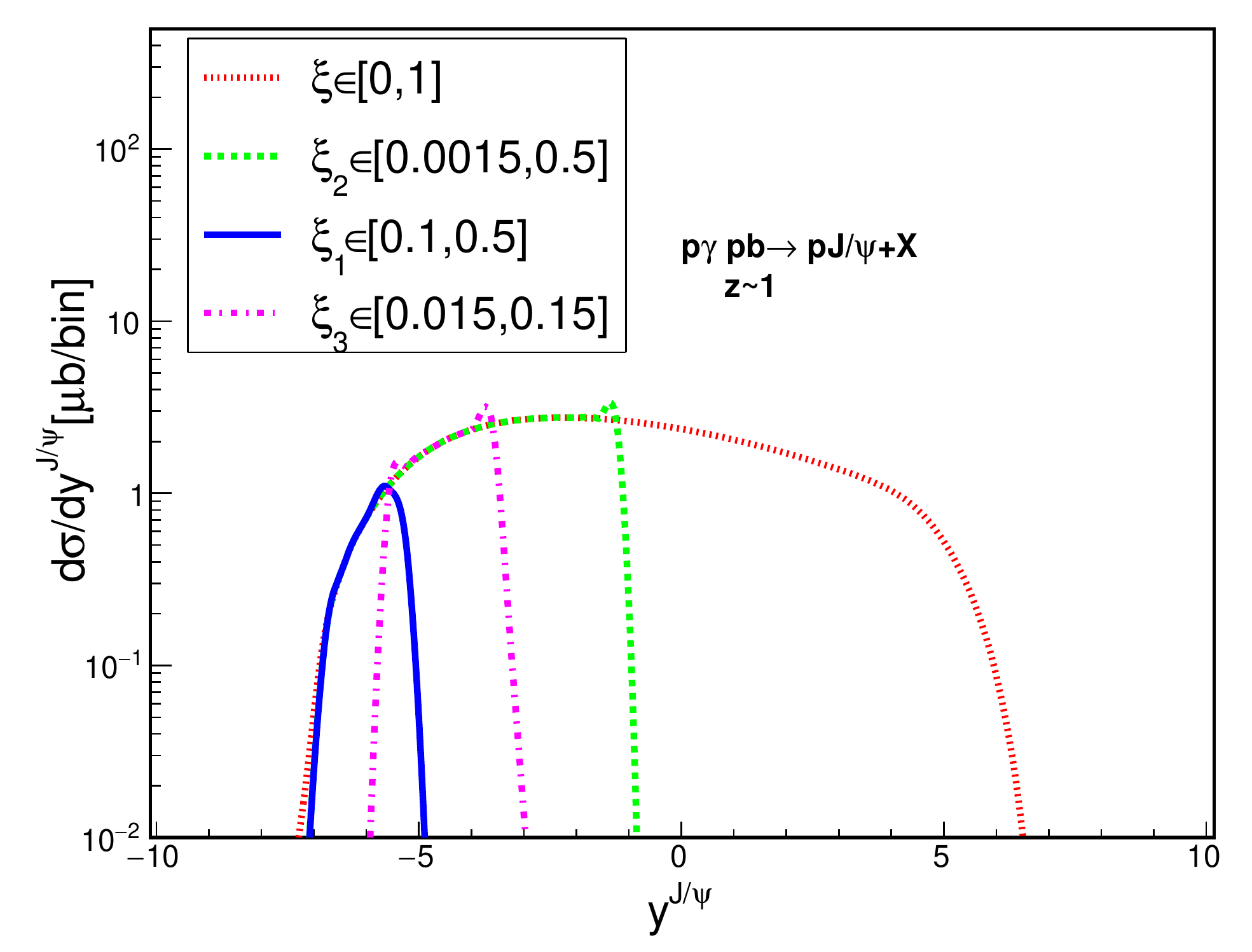}
   \includegraphics[height=4.4cm,width=4.4cm,angle=0]{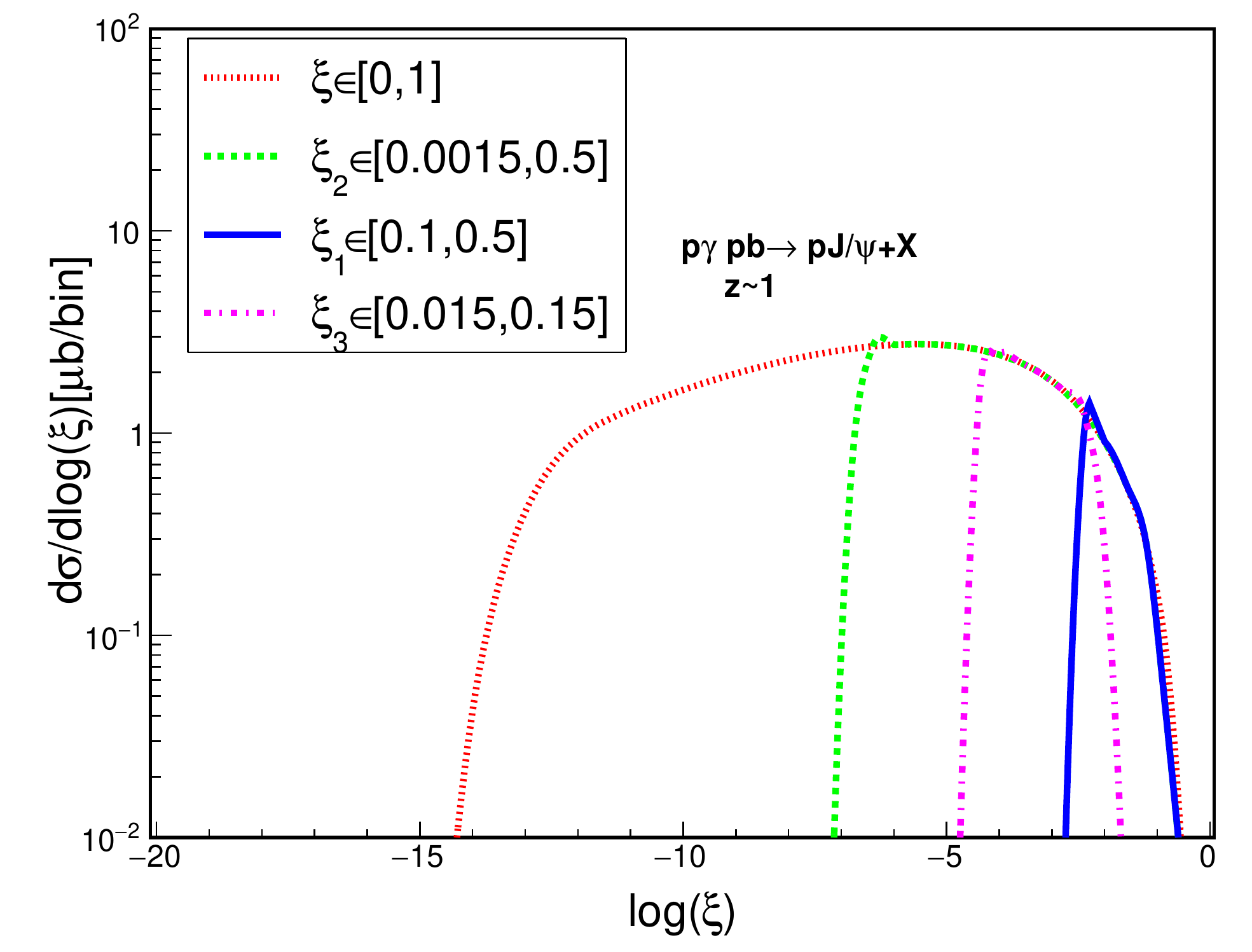}
   \includegraphics[height=4.4cm,width=4.4cm,angle=0]{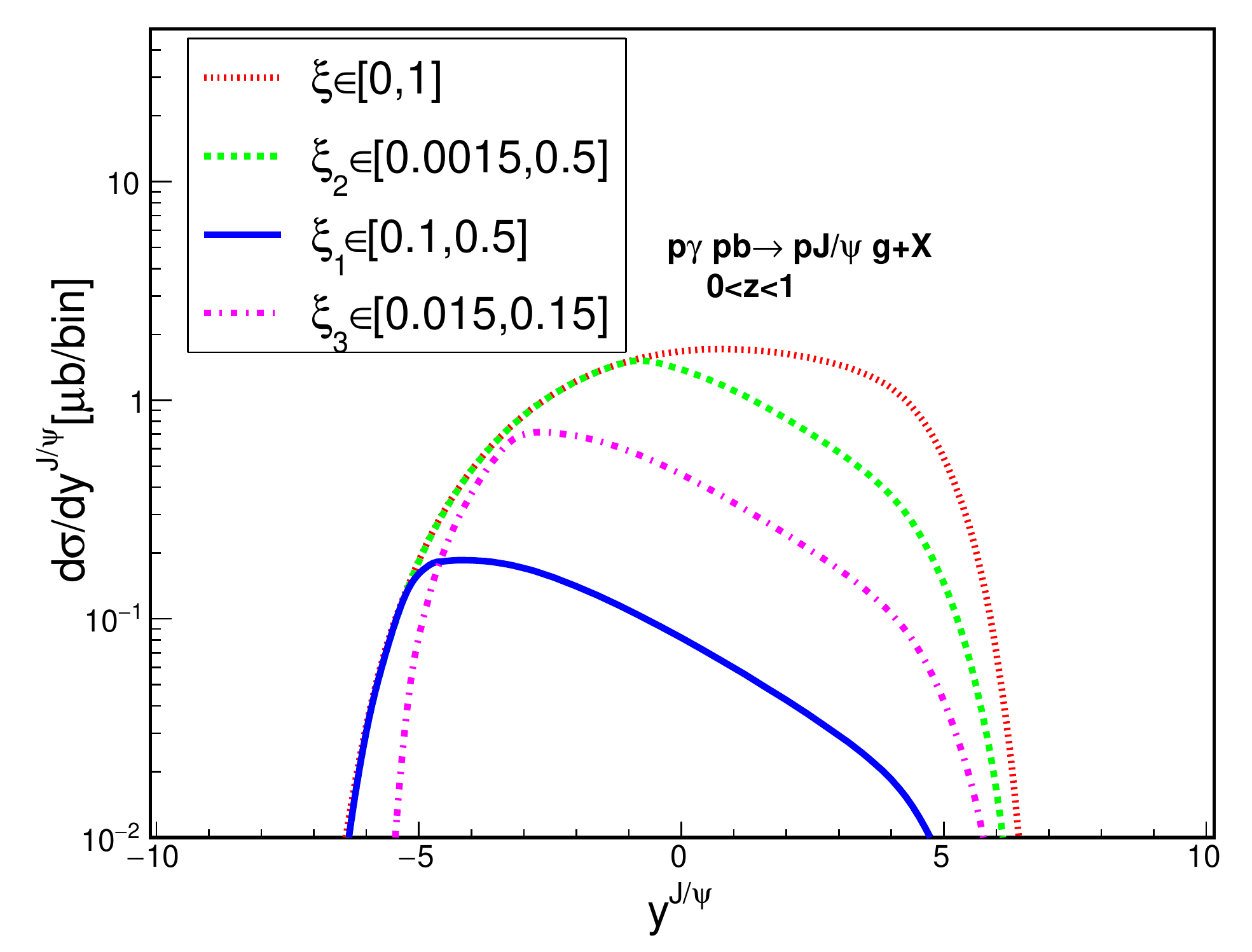}
   \includegraphics[height=4.4cm,width=4.4cm,angle=0]{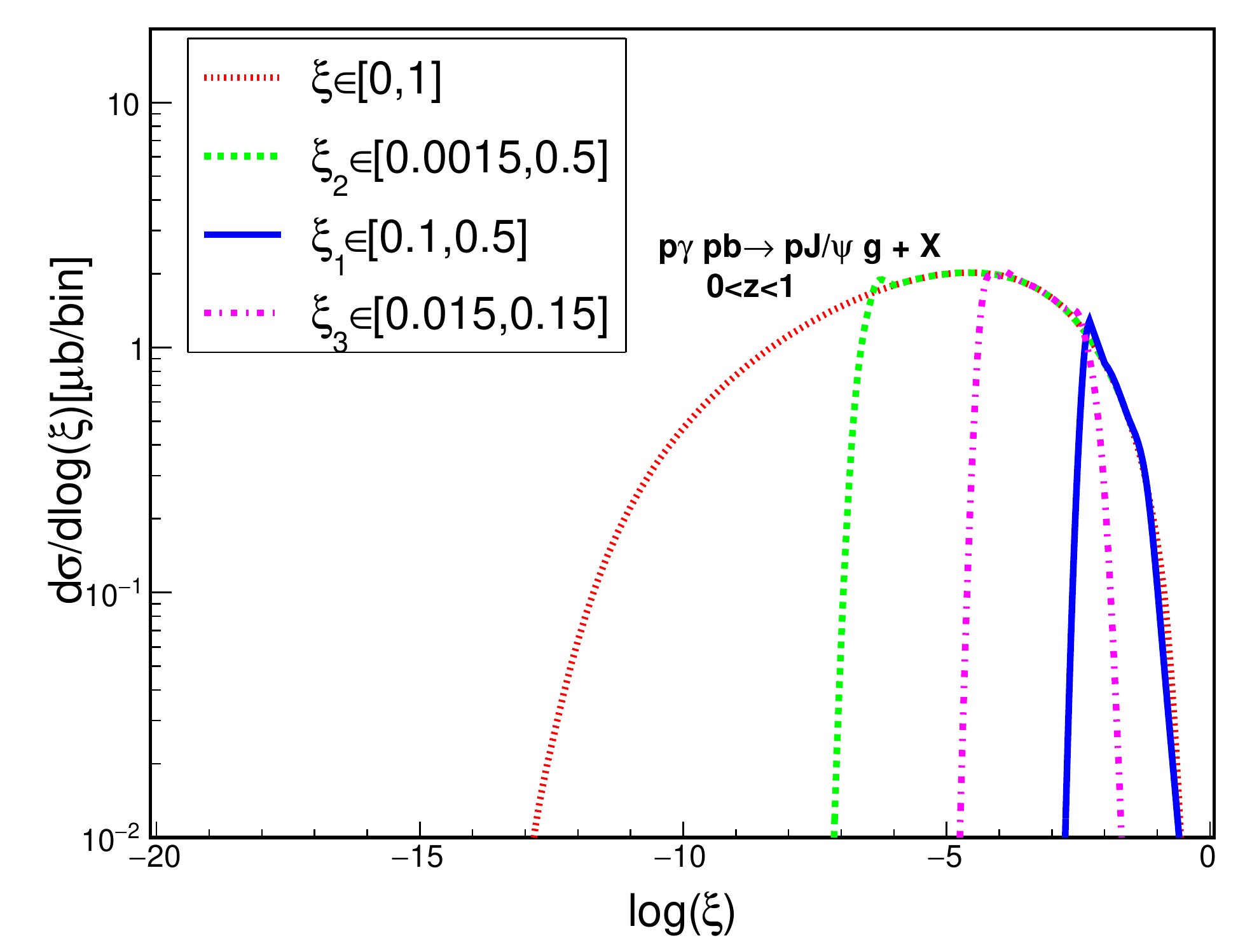}
   \includegraphics[height=4.4cm,width=4.4cm,angle=0]{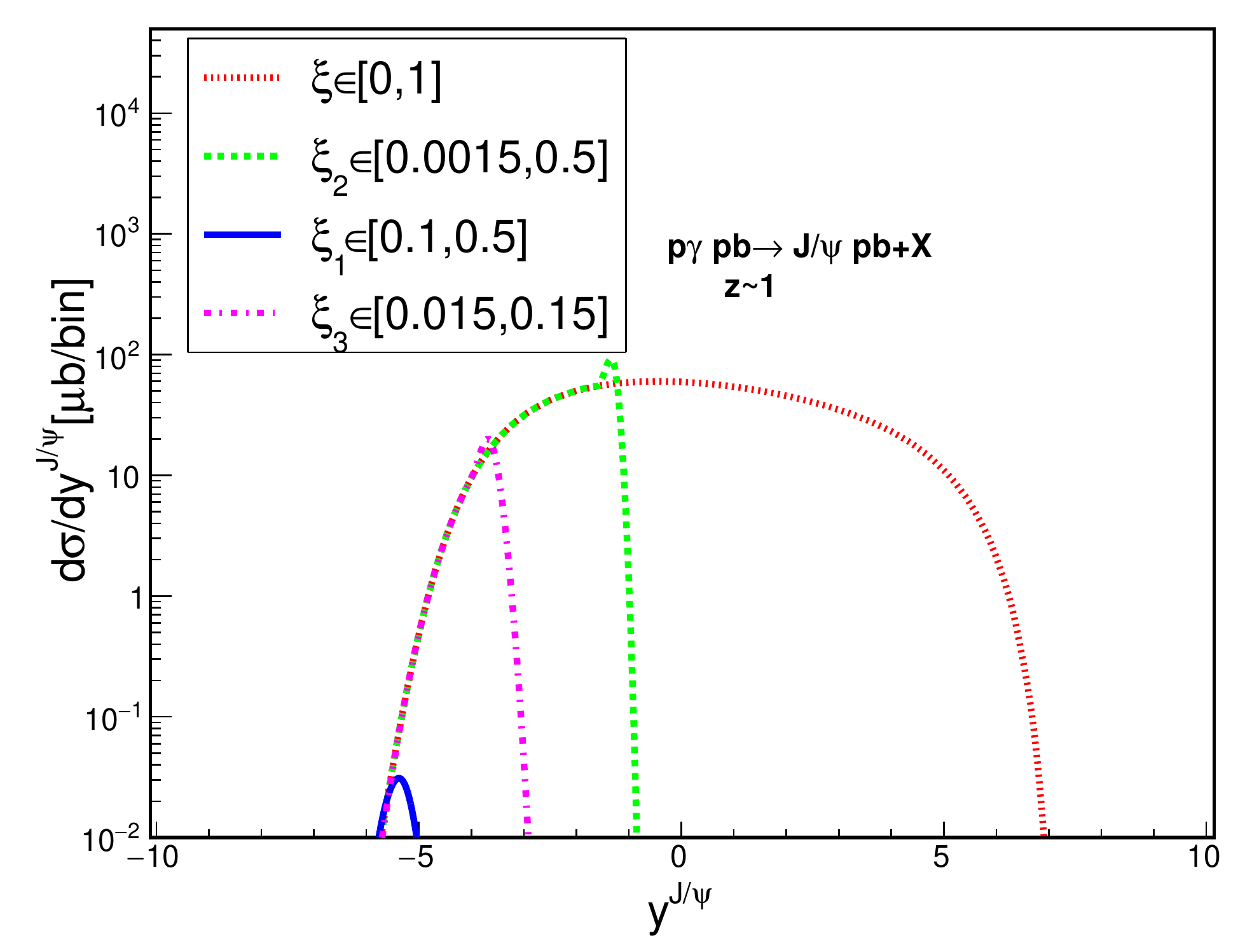}
   \includegraphics[height=4.4cm,width=4.4cm,angle=0]{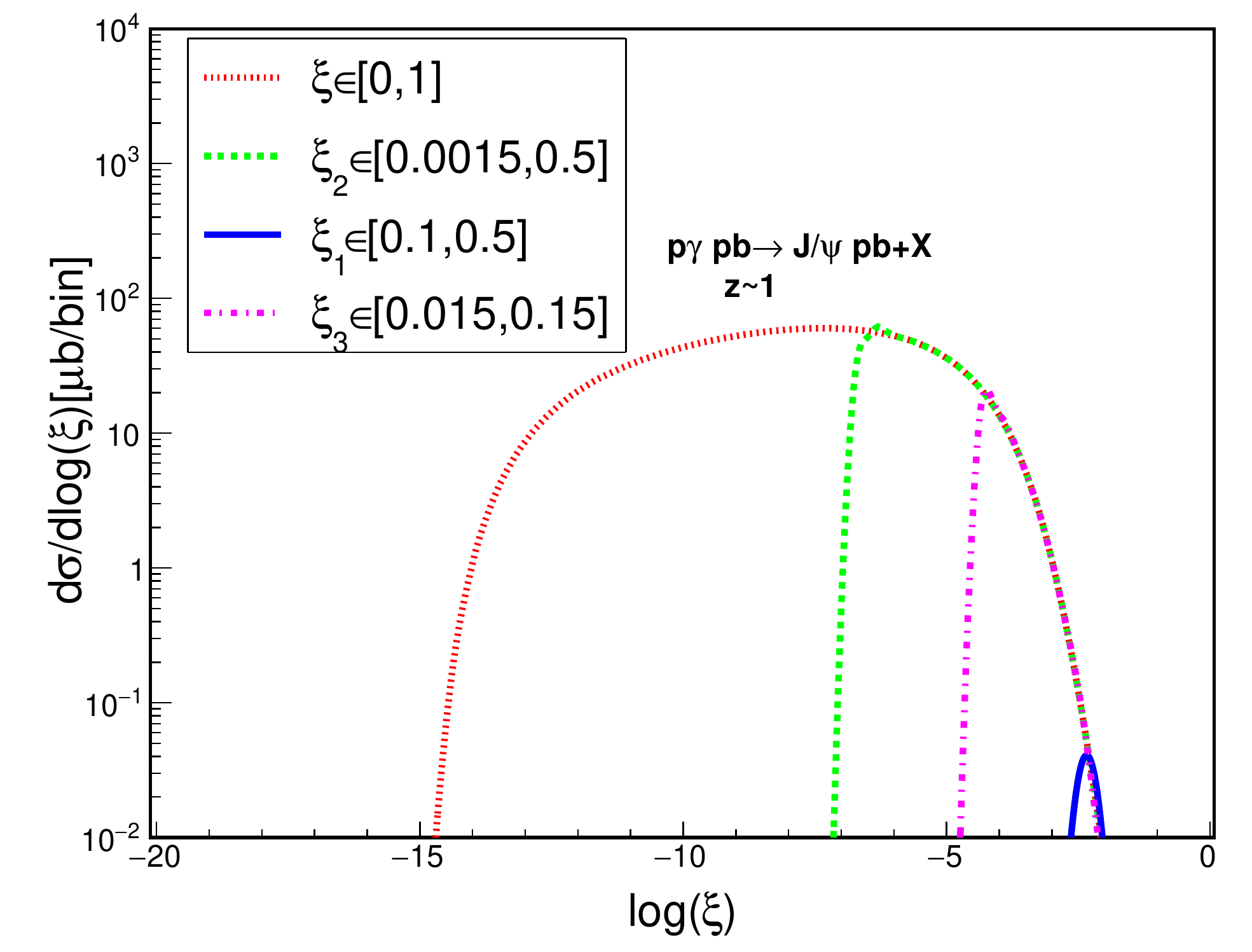}
   \includegraphics[height=4.4cm,width=4.4cm,angle=0]{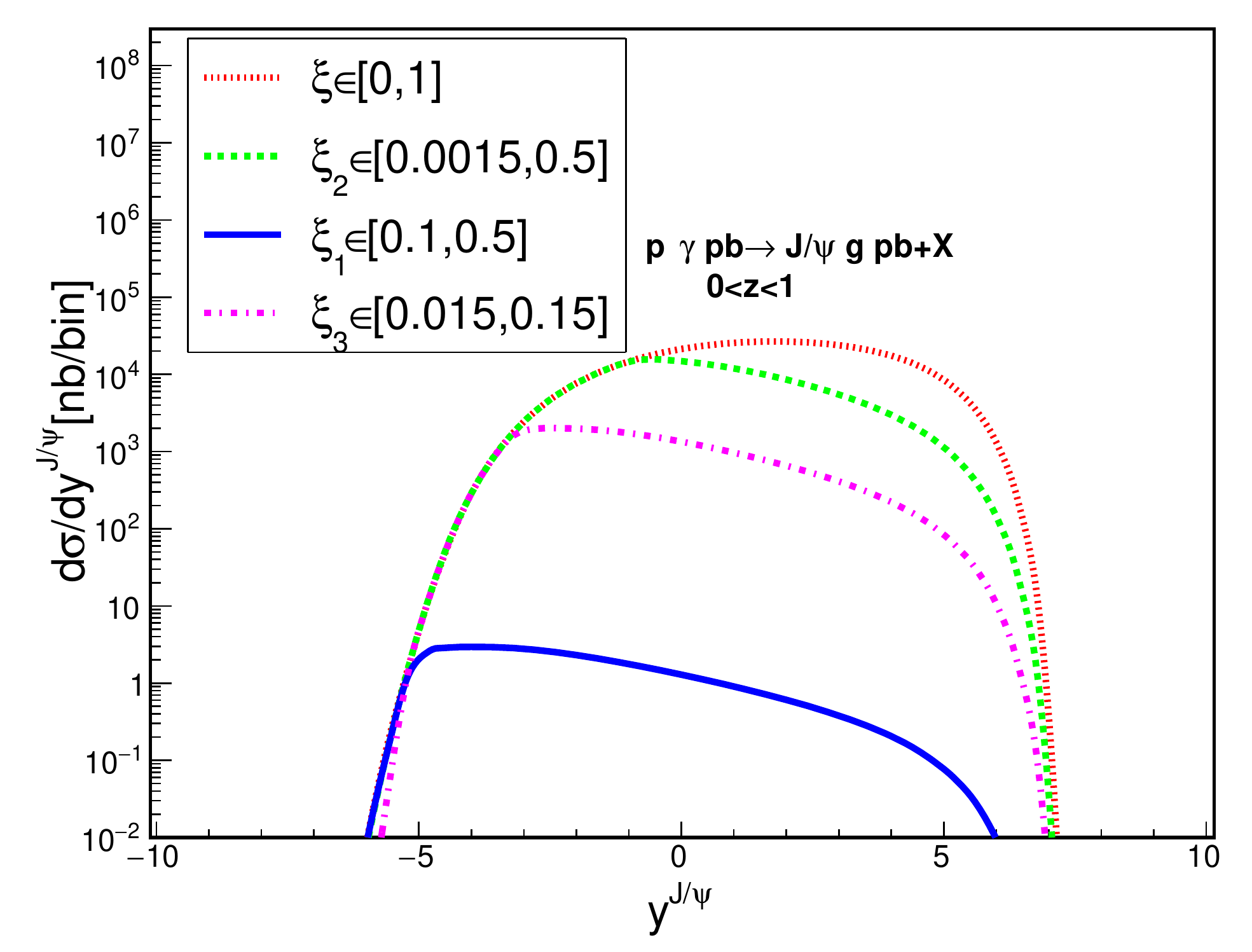}
   \includegraphics[height=4.4cm,width=4.4cm,angle=0]{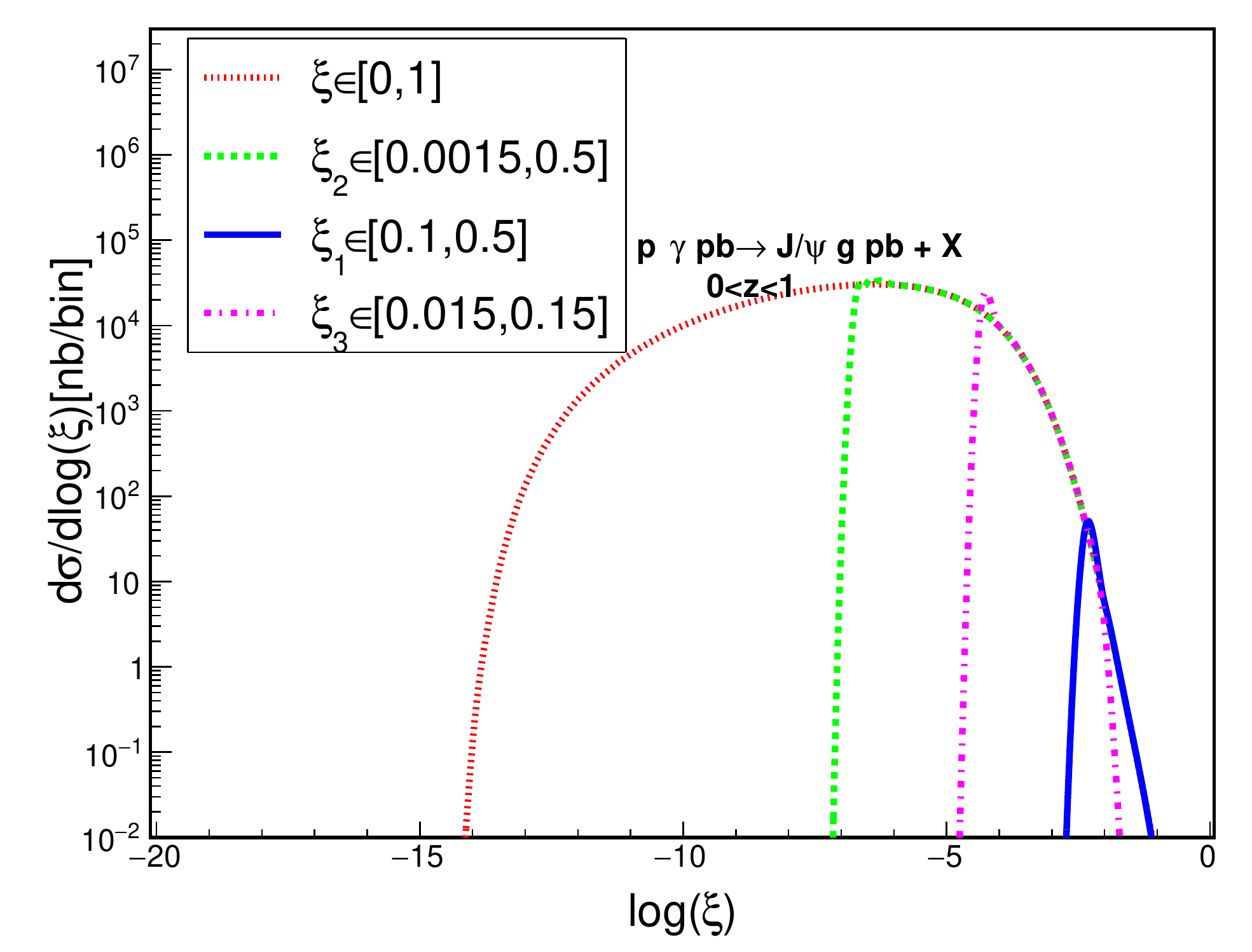}
   %\end{minipage}
   \caption{ \normalsize
  (color online)  The  $\rm y^{J/\psi}$ and $\rm \log \xi$ distributions with nuclear shadowing effect
  for the  $\rm p\gamma Pb\to p J/\psi+X $ and $\rm p\gamma Pb\to p J/\psi g +X$  processes ( top panels)
  and $\rm p\gamma Pb\to J/\psi Pb+X$ and $\rm p\gamma Pb\to J/\psi gPb+X$ processes (bottom panels)
  for the forward detector with $\xi$ (red dashed line), at the CMS-TOTEM forward detector with $\xi_{1}$ (green dashed line),
  at the CMS-TOTEM forward detector with $\xi_{2}$ (blue solid line), and
  at  the AFP-ATLAS forward detector with $\xi_{3}$ (magenta dash dotted line).}
\label{fig8:limits}
\end{figure*}
Fig.\ref{fig8:limits} shows the $\rm y^{J/\psi}$ and $\rm \log\xi$ distributions for $\rm pPb$ collision where $\rm p\gamma Pb\to p J/\psi(+g)+X$ means initial photon is emitted from proton while this proton remains undissociated in the final state ($\rm \gamma Pb$ interaction), and $\rm p\gamma Pb\to J/\psi Pb(+g)+X$ means the initial photon is emitted from lead while this lead remains undissociated ($\rm \gamma p$ interaction). The full contribution to pPb interation would be the sum of these two while here we split them and study them separately. The two first figures on top (bottom) left panel and two last figures on top (bottom) right panel represent the $\rm y^{J/\psi }$ and  $\rm \log\xi$ distributions for $\rm Pbp$ collision in 2$\to$1( 2$\to$2) subprocesses. As for $\rm Pbp$ collisions, the $\rm y^{J/\psi}$ and $\rm \log\xi$ distribution at forward and backward are asymmetric about midrapidity due to the differences of the fluxes and parton distribution function.
For example, in $\rm \gamma p$ interaction, the photon comes from the nuclei, with photon flux being proportional to $\rm Z^{2}$ and the photoproduction cross section is defined by the gluon density of the proton ($x\rm G_{g/p}(x,\mu_{f})$). While in the $\rm \gamma Pb$ ones the photon originates from the proton and the photoproduction cross section is determined by the gluon density of nuclei, which is improved by a factor of the order of $\rm A=208$  with reference to $x\rm G_{g/p}(x,\mu_{f})$. Consequently, the contributions of the $\rm \gamma p$ and $\rm \gamma Pb$ interactions are different and the contributions of the $\rm \gamma p$ interactions are larger than those of  $\rm \gamma Pb$ ones. The experimental disentanglement between the $\rm \gamma p$ and $\rm \gamma Pb$ interactions is technically possible by examining the final state
by tagging the intact hadron using forward detectors or/and by using the zero-degree calorimeters to veto very forward going neutral fragments.
The shapes of the $\rm y^{J/\psi}$ and $\rm \log\xi$ distributions for $\rm Pbp$ collisions in 2$\to$1 subprocess are similar and almost symmetric whereas for 2$\to$2 subprocess they are totally dissimilar and asymmetric. The dominant $\rm y^{J/\psi}$ and $\rm \log\xi$ distributions also arise from $\rm 0.0015<\xi_{2}<0.5$.

\section{SUMMARY}
\label{SUMMARY}

We examine in this paper the photoproduction of single $\rm J/\psi$ meson to leading order in the NRQCD framework at the LHC with forward detector acceptances. We look into both color-singlet and -octet contributions for different
Fock states in $\rm pp$, $\rm PbPb$ and $\rm pPb$ collision modes. The differential cross sections of $\rm z^{J/\psi}$, $\rm p_{T}^{J/\psi}$, $\rm y^{J/\psi}$ and $\rm \log\xi$ for $\rm J/\psi$ meson are presented and nuclear shadowing effects are considered. Our results show that the main contribution of the $\rm p_{T}^{J/\psi}$ distribution comes from the color-octet $\rm ^{1}S_{0}^{[8]}$ channel. As for $\rm z^{J/\psi}$ distribution, the main contribution arises from the color-singlet $\rm ^{3}S_{1}^{[1]}$ channel for small-$\rm z^{J/\psi}$ region and $\rm ^{1}S_{0}^{[8]}$ channel for large-$\rm z^{J/\psi}$ region. The smallest ones are from the $\rm ^{3}S_{1}^{[8]}$ in the whole range of $\rm z^{J/\psi }$ and $\rm p_{T}^{J/\psi }$. It has been found that the produced signal relies on the choice of forward detector acceptance, where in our case $\rm 0.0015<\xi_{2}<0.5$ is the one that keeps the largest contribution. The signal $\rm PbPb$ collision is boosted due to the enhancement of the photon flux for nuclei which is proportional to $\rm Z^2$, and gluon distribution of lead by a factor of order of $\rm A$. The large predicted values for the production rates in $\rm pp$ collisions together with in an upgraded $\rm pPb$ collisions indicating that the $\rm J/\psi$ photoproduction should be possible to investigate at the CERN LHC. It may provide more crucial and realistic route toward quarkonium production in inelastic process regime. Its trace at the LHC with forward detector acceptances may be obvious to investigate the color-singlet and color-octet mechanism and will be valuable for exploring the new area of heavy quarkonium production.

\begin{acknowledgments}
Tichouk thanks professors Gang Li and Mao Song for very useful disscussions. Hao Sun is supported by the National Natural Science Foundation of China (Grant No.11675033) and by the Fundamental Research Funds for the Central Universities (Grant No. DUT18LK27).
\end{acknowledgments}

\appendix
\section{LIST OF AMPLITUDE SQUARES}
\label{factorization formalism}

In the appendix, we give the list of the amplitude squares timed long distance matrix elements for the different partonic processes computed  with the FORM package \cite{1203.6543} of the following states: $\rm ^{3}S_{1}^{[1]}$, $\rm ^{1}S_{0}^{[8]}$, $\rm ^{3}S_{1}^{[8]}$, $\rm ^{3}P_{0}^{[8]}$, $\rm ^{3}P_{1}^{[8]}$, $\rm ^{3}P_{2}^{[8]}$ for the partonic processes $\rm \gamma g\to Q\overline{Q}[n]+g$ and $\rm ^{1}S_{0}^{[8]}$, $\rm ^{3}P_{0}^{[8]}$, $\rm ^{3}P_{2}^{[8]}$ for $\rm \gamma g\to Q\overline{Q}[n]$. The amplitude squares timed long distance matrix elements for $2\to 2$ partonic processes for different Fock state contributions are
\begin{eqnarray}
\rm &&\overline{\rm \sum \bigskip }\left\vert \mathcal{M}\left[\rm ^{2S+1}L_{J}^{[1,8]}\right]
\right\vert ^{2}=\frac{1}{\rm N_{col}N_{pol}} \overline{\sum}\left\vert\mathcal{A}_{S,L}\right\vert ^{2}
 \end{eqnarray}
\begin{eqnarray}
\rm
\overline{\sum \bigskip}\left\vert \mathcal{M}\left[ ^{3}S_{1}^{[1]}\right] \right\vert ^{2}=
\frac{32(4\pi)^{3}\alpha \alpha_{s}^{2}e_{c}^{2}((\hat{s}\hat{t}+\hat{t}\hat{u}+\hat{s}\hat{u})^{2}-M_{\mathcal{Q}}^{2}\hat{s}\hat{t}
\hat{u})M_{{J/\psi }}}{27(\hat{s}+\hat{t})^{2}(\hat{s}+\hat{u})^{2}(\hat{t}+\hat{u})^{2}}\rm \langle 0|\mathcal{O}_{1}^{J/\psi }[^{3}S_{1}]|0\rangle;
 \end{eqnarray}
\begin{eqnarray}
\rm
\overline{\sum \bigskip}\left\vert \mathcal{M}\left[ ^{1}S_{0}^{[8]}\right] \right\vert ^{2}=
\frac{3(4\pi)^{3}\alpha \alpha_{s}^{2}e_{c}^{2}\hat{s}\hat{u}(M_{{J/\psi }}^{8}+\hat{s}^{4}+\hat{t}^{4}+
\hat{u}^{4})}{M_{{J/\psi }}\hat{t}(\hat{s}+\hat{t})^{2}(\hat{s}+\hat{u})^{2}(\hat{t}+\hat{u})^{2}}\rm \langle 0|\mathcal{O}_{8}^{J/\psi }[^{1}S_{0}]|0\rangle;
 \end{eqnarray}
\begin{eqnarray}
\rm
\overline{\sum \bigskip }\left\vert  \mathcal{ M}\left[ ^{3}S_{1}^{[8]}\right] \right\vert ^{2}=\frac{20(4\pi )^{3}\alpha \alpha _{s}^{2}e_{c}^{2}((\hat{s}\hat{t}+\hat{t}\hat{u}+\hat{s}\hat{u})^{2}-M_{{J/\psi }}^{2}\hat{s}\hat{t}\hat{u})M_{{J/\psi }}}
{9(\hat{s}+\hat{t})^{2}(\hat{s}+\hat{u})^{2}(\hat{t}+\hat{u})^{2}}\rm \langle 0|\mathcal{O}_{8}^{J/\psi }[^{3}S_{1}]|0\rangle;
 \end{eqnarray}
\begin{eqnarray}
\rm
\overline{\sum \bigskip}\left\vert \mathcal{M}\left[ ^{3}P_{0}^{[8]}\right] \right\vert ^{2}&=&\rm
\frac{4(4\pi)^{3}\alpha \alpha _{s}^{2}e_{c}^{2}}{(\hat{s}+\hat{t})^{2}(\hat{s}+\hat{u})^{2}(\hat{t}+\hat{u})^{2}}[\frac{9M_{{J/\psi }}^{5}\hat{s}
\hat{u}}{\hat{t}}+\frac{\hat{s}\hat{u}\widehat{t}^{3}(2M_{{J/\psi }}^{4}+3\hat{t}M_{{J/\psi }}^{2}+\hat{s}\hat{u})^{2}}{M_{{J/\psi }}^{3}(\hat{s}+
\hat{t})^{2}(\hat{t}+\hat{u})^{2}}+ \\
&&\rm \nonumber\frac{\hat{s}^{3}\hat{u}(3\hat{s}^{2}M_{{J/\psi }}^{2}-\hat{t}\hat{
u}(2M_{{J/\psi }}^{2}-\hat{s}))^{2}}{M_{{J/\psi }}^{3}\hat{t}(\hat{t}+\hat{s})^{2}(\hat{s}+\hat{u})^{2}}+\frac{\hat{u}^{3}\hat{s}(\hat{s}
\hat{t}(2M_{{J/\psi }}^{2}-\hat{u})-3\hat{u}^{2}M_{{J/\psi }}^{2})^{2}}{M_{{J/\psi }}^{3}\hat{t}(
\hat{t}+\hat{s})^{2}(\hat{s}+\hat{u})^{2}}]\rm \langle 0|\mathcal{O}_{8}^{J/\psi}[^{3}P_{0}]|0\rangle;
 \end{eqnarray}
\begin{eqnarray}
\rm
\overline{\sum \bigskip }\left\vert  \mathcal{M}\left[ ^{3}P_{1}^{[8]}\right] \right\vert ^{2} &=&\rm
\frac{8(4\pi)^{3}\alpha \alpha _{s}^{2}e_{c}^{2}}{M_{{J/\psi }}^{3}(\hat{s}+\hat{t})^{4}(
\hat{s}+\hat{u})^{4}(\hat{t}+\hat{u})^{4}}[\hat{s}^{7}(
\hat{t}^{4}+2\hat{t}^{3}\hat{u}+4\hat{t}^{2}\hat{u}^{2}+2
\hat{t}\hat{u}^{3}+\hat{u}^{4})+\hat{s}^{6}(\hat{t}+\widehat{u})^{2} \\
&&\rm \nonumber\times (3\hat{t}^{3}+7\hat{t}^{2}\hat{u}+7\hat{t}\hat{u}^{2}-\hat{u}^{3})+\hat{s}^{5}(3\hat{t}^{6}+22\hat{t}^{5}
\hat{u}+60\hat{t}^{4}\hat{u}^{2}+76\hat{t}^{3}\hat{u}^{3}+36\hat{t}^{2}\hat{u}^{4}+ \\
&&\rm \nonumber4\hat{t}\hat{u}^{5}-\hat{u}^{6})+\hat{s}^{4}(\hat{t}+
\hat{u})(\hat{t}^{6}+12\hat{t}^{5}\hat{u}+46\hat{t}^{4}
\hat{u}^{2}+72\hat{t}^{3}\hat{u}^{3}+32\hat{t}^{2}\hat{u}^{4}+4\hat{t}\hat{u}^{5}+ \\
&&\rm \nonumber\hat{u}^{6})+2\hat{s}^{3}\hat{t}\hat{u}(\hat{t}^{6}+10
\hat{t}^{5}\hat{u}+38\hat{t}^{4}\hat{u}^{2}+59\hat{t}^{3}
\hat{u}^{3}+38\hat{t}^{2}\hat{u}^{4}+10\hat{t}\hat{u}^{5}+\hat{u}^{6})+ \\
&&\rm \nonumber2\hat{s}^{2}\hat{t}^{2}\hat{u}^{2}(\hat{t}+\hat{u})(\hat{t}^{4}+9\hat{t}^{3}\hat{u}+20\hat{t}^{2}\hat{u}
^{2}+10\hat{t}\hat{u}^{3}+2\hat{u}^{4})+\hat{s}\hat{t}^{3}\hat{u}^{3}(\hat{t}+\hat{u})^{2} \\
&&\rm \nonumber\times (2\hat{t}^{2}+9\hat{t}\hat{u}+2\hat{u}^{2})+
\hat{t}^{4}\hat{u}^{4}(\hat{t}+\hat{u})^{3}]\rm \langle 0|\mathcal{O}_{8}^{J/\psi }[^{3}P_{1}]|0\rangle;
 \end{eqnarray}
\begin{eqnarray}
\rm
\overline{\sum \bigskip }\left\vert  \mathcal{M}\left[ ^{3}P_{2}^{[8]}\right] \right\vert ^{2}&=&\rm
\frac{8(4\pi)^{3}\alpha \alpha _{s}^{2}e_{c}^{2}}{5M_{{J/\psi }}^{3}\hat{t}(\hat{s}+\hat{
t})^{4}(\hat{s}+\hat{u})^{4}(\hat{t}+\hat{u})^{4}}[12\hat{s}^{9}\hat{u}(\hat{t}+\hat{u})^{2}+12\hat{s}^{8}
\hat{u}(\hat{t}+\hat{u})(5\hat{t}^{2} +\\
&&\rm \nonumber7\hat{t}\hat{u}+4\hat{u}^{2})+\hat{s}^{7}(3\hat{t}^{5}+142\hat{t}^{4}\hat{u}+384\hat{t}^{3}\hat{u}^{2}+454%
\hat{t}^{2}\hat{u}^{3}+303\hat{t}\hat{u}^{4}+96\hat{u}^{5})+ \\
&&\rm \nonumber\hat{s}^{6}(\hat{t}+\hat{u})(9\hat{t}^{5}+202\hat{t}
^{4}\hat{u}+438\hat{t}^{3}\hat{u}^{2}+442\hat{t}^{2}\hat{u}^{3}+309\hat{t}\hat{u}^{4}+120\hat{u}^{5})+ \\
&&\rm \nonumber\hat{s}^{5}(9\hat{t}^{7}+198\hat{t}^{6}\hat{u}+736\hat{t}^{5}\hat{u}^{2}+1200\hat{t}^{4}\hat{u}^{3}+1184\hat{t}^{3}\hat{u}^{4}+860\hat{t}^{2}\hat{u}^{5}+ \\
&&\rm \nonumber429\hat{t}\hat{u}^{6}+96\hat{u}^{7})+\hat{s}^{4}(\hat{t}+\hat{u})(3\hat{t}^{7}+100\hat{t}^{6}\hat{u}+450\hat{t}%
^{5}\hat{u}^{2}+720\hat{t}^{4}\hat{u}^{3}+ \\
&&\rm \nonumber688\hat{t}^{3}\hat{u}^{4}+496\hat{t}^{2}\hat{u}^{5}+255\hat{t}\hat{u}^{6}+48\hat{u}^{7})+2\hat{s}^{3}\hat{u}(11%
\hat{t}^{8}+114\hat{t}^{7}\hat{u}+ \\
&&\rm \nonumber362\hat{t}^{6}\hat{u}^{2}+585\hat{t}^{5}\hat{u}^{3}+600\hat{t}^{4}\hat{u}^{4}+440\hat{t}^{3}\hat{u}^{5}+227\hat{
t}^{2}\hat{u}^{6}+66\hat{t}\hat{u}^{7}+ \\
&&\rm \nonumber6\hat{u}^{8})+2\hat{s}^{2}\hat{t}\hat{u}^{2}(\hat{t}+
\hat{u})^{2}(19\hat{t}^{5}+76\hat{t}^{4}\hat{u}+104\hat{t}^{3}\hat{u}^{2}+84\hat{t}^{2}\hat{u}^{3}+ \\
&&\rm \nonumber48\hat{t}\hat{u}^{4}+12\hat{u}^{5})+\hat{s}\hat{t}^{2}\hat{u}^{3}(\hat{t}+\hat{u})^{2}(22\hat{t}^{4}+59\hat{t}^{3}\hat{u}+ \\
&&\rm \nonumber58\hat{t}^{2}\hat{u}^{2}+36\hat{t}\hat{u}^{3}+12\hat{u}^{4})+3(\hat{t}+\hat{u})^{3}\hat{t}^{5}\hat{u}
^{4}]\rm \langle 0|\mathcal{O}_{8}^{J/\psi }[^{3}P_{2}]|0\rangle.
\end{eqnarray}
The amplitude squares timed long distance matrix elements for $\rm 2\to 1$ partonic processes for different Fock state contributions are
\begin{eqnarray}
\rm
\overline{\sum \bigskip }\left\vert \mathcal{M}\left[ ^{1}S_{0}^{[8]}\right] \right\vert ^{2}=\frac{(4\pi
)^{2}\alpha\alpha _{s}e_{c}^{2} }{2M_{{J/\psi }}}\rm \langle 0|\mathcal{O}_{8}^{J/\psi }[^{1}S_{0}]|0\rangle;
\end{eqnarray}
\begin{eqnarray}
\rm
\overline{\sum \bigskip }\left\vert \mathcal{ M}\left[ ^{3}P_{0}^{[8]}\right] \right\vert ^{2}=\frac{6(4\pi
)^{2}\alpha \alpha _{s}e_{c}^{2}}{M_{{J/\psi }}^{3}}\rm \langle 0|\mathcal{O}_{8}^{J/\psi }[^{3}P_{0}]|0\rangle;
\end{eqnarray}
\begin{eqnarray}
\rm
\overline{\sum \bigskip }\left\vert  \mathcal{ M}\left[ ^{3}P_{2}^{[8]}\right] \right\vert ^{2}=\frac{8(4\pi
)^{2}\alpha\alpha _{s}e_{c}^{2} }{5M_{{J/\psi }}^{3}}\rm \langle 0|\mathcal{O}_{8}^{J/\psi }[^{3}P_{2}]|0\rangle.
\end{eqnarray}

\end{document}